\documentclass[11pt,a4paper]{article} 
\pdfoutput=1 
\usepackage{jcappub}

\usepackage{mathrsfs}
\DeclareMathAlphabet{\mathpzc}{OT1}{pzc}{m}{it}

\usepackage[english]{babel}

\usepackage{amsmath}
\usepackage{amssymb}

\usepackage{mathrsfs}
\DeclareMathAlphabet{\mathpzc}{OT1}{pzc}{m}{it}

\usepackage{graphicx}

\usepackage{cancel} 

\newcommand{\Z}{\mathbb{Z}} 

\newcommand{\T}[1]{#1^T} 
\newcommand{\hc}[2][]{#2^{\dagger #1}} 
\newcommand{\abs}[1]{|#1|} 
\newcommand{\vev}[1]{\langle #1 \rangle} 

\renewcommand{\Re}{\mathrm{Re}}
\renewcommand{\Im}{\mathrm{Im}}

\title{Minimal semi-annihilating \\ $\Z_{N}$ scalar dark matter}

\author[a]{Genevi\`{e}ve B\'{e}langer,}
\author[b]{Kristjan Kannike,}
\author[c]{Alexander Pukhov}
\author[b]{and Martti Raidal}

\affiliation[a]{LAPTH, Universit\'e de Savoie,\\ CNRS, B.P.110, F-74941 Annecy-le-Vieux Cedex, France}
\affiliation[b]{National Institute of Chemical Physics and Biophysics, \\
R\"{a}vala 10, Tallinn, Estonia}
\affiliation[c]{Skobeltsyn Institute of Nuclear Physics,\\ Moscow State University, Moscow 119991, Russia}

\emailAdd{belanger@lapth.cnrs.fr}
\emailAdd{kristjan.kannike@cern.ch}
\emailAdd{pukhov@lapth.cnrs.fr}
\emailAdd{martti.raidal@cern.ch}
 
\abstract{
We study the dark matter from an inert doublet and a complex scalar singlet  stabilized by $\Z_{N}$ symmetries. This field content is the minimal one that allows dimensionless semi-annihilation couplings for $N > 2$. 
We consider explicitly the $\Z_3$ and $\Z_4$ cases and take into account constraints from perturbativity, unitarity, vacuum stability, necessity for the electroweak $\Z_{N}$ preserving vacuum to be the global minimum, electroweak precision tests, upper limits from direct detection  and properties of  the Higgs boson. Co-annihilation and semi-annihilation of dark sector particles as well as dark matter conversion significantly modify the cosmic abundance and direct detection phenomenology.
}

\arxivnumber{14xx.xxxx}

\begin{document}

\maketitle

\section{Introduction}
\label{sec:intro}

The $\Lambda$CDM model that explains 22\% of the Universe's energy density with non-baryonic collisionless cold dark matter (DM) 
has turned out to give an excellent description of the Universe at large scales~\cite{Ade:2013zuv}. The most popular candidates
for the DM are weakly interacting massive particles (WIMPs) \cite{Jungman:1995df,Bertone:2004pz,Bergstrom:2000pn} that are stable due to an imposed discrete symmetry. 
A large class of  models beyond the standard model (SM), such as supersymmetric models~\cite{Nilles:1983ge,Haber:1984rc},  correctly predict the observed DM 
abundance as a thermal relic density of WIMPs.

At the same time, there is an increasing number of experimental and observational hints
that the WIMP paradigm may not be realised in Nature in its simplest form. 
The negative results in searches for DM direct~\cite{Aprile:2011hi,Aprile:2012nq,Akerib2013111} and indirect detection~\cite{Cirelli:2010xx} severely constrain the simplest DM models.
The not yet conclusive cosmological observations (see \cite{Weinberg:2013aya} for a review) suggest that the DM density profiles in the centres of galaxies and in dwarf galaxies,
 and the masses of the biggest satellite halos, may significantly deviate from the results of $N$-body simulations. 
Those hints may suggest that the DM freeze-out processes are non-standard, and the DM interactions with baryons and with other DM particles 
may be more complicated than the simplest models predict. In addition, the dark sector may have complicated dynamics with more than 
one DM component. In light of those results studies of non-standard DM dynamics in non-minimal models are  well motivated.

The discovery of the Higgs boson at the Large Hadron Collider (LHC) \cite{Aad:2012tfa,Chatrchyan:2012ufa} has proven that 
 scalar particles play an important r\^{o}le in fundamental physics. Since the nature of DM is yet unknown, scalar DM models 
 are among the best motivated DM scenarios~\cite{McDonald:1993ex,Barger:2007im,Barger:2008jx,Burgess:2000yq,Gonderinger:2009jp,Cai:2011kb,Deshpande:1977rw,Ma:2006km,Barbieri:2006dq,LopezHonorez:2006gr,Belanger:2012zr,Belanger:2012vp}. The latest studies show that the SM scalar potential is very close to the critical bound \cite{Bezrukov:2012sa,Degrassi:2012ry,Buttazzo:2013uya,Masina:2012tz}. Scalar DM couplings to the Higgs boson, the so-called Higgs portal~\cite{Patt:2006fw,Chu:2011be,Djouadi:2011aa,Djouadi:2012zc}, 
 can stabilise the SM Higgs potential via its contribution to the running of Higgs quartic self-coupling~\cite{Gonderinger:2009jp,Kadastik:2011aa,Chen:2012faa,Cheung:2012nb,Gonderinger:2012rd,Chao:2012mx,Gabrielli:2013hma} or via singlet threshold effects~\cite{Lebedev:2012zw,EliasMiro:2012ay,Hambye:2013dgv}.
 The scalar DM framework is also suitable for constructing DM models based on Abelian $\Z_N$ or non-Abelian (discrete) symmetries  \cite{Batell:2010bp,DeMontigny:1993gy,Agashe:2010gt,DEramo:2010ep,Martin:1992mq,Agashe:2010tu,Belanger:2012vp,Ma:2007gq,Belanger:2012zr,Ivanov:2012hc,Lovrekovic:2012bz,DEramo:2012rr,Ko:2014nha} that have non-standard freeze-out processes, such as semi-annihilations~\cite{Hambye:2009fg,Hambye:2008bq,Arina:2009uq,DEramo:2010ep}, that modify the predictions for the DM abundance and for its interactions with matter.
 Due to the new type of processes the relations between DM annihilation cross sections and spin-independent scattering cross section
 with nuclei are modified, explaining the present negative results.

The aim of this work  is to perform a comprehensive study of $\Z_3$ and $\Z_4$ scalar DM models with semi-annihilation by scanning systematically over their full parameter space.
We consider models presented in~\cite{Belanger:2012vp} with scalar sectors that comprise, in addition to the Higgs doublet, gauge singlet and doublet scalars. The $\Z_{4}$ model may have more than one species of stable DM. 
The $\Z_{3}$ singlet model~\cite{Ma:2007gq,Belanger:2012zr} and the inert doublet model \cite{Deshpande:1977rw,Ma:2006km,Barbieri:2006dq,LopezHonorez:2006gr} with a $Z_2$ symmetry are just limiting cases of this general framework.
Since the new semi-annihilation modes, $\mathrm{DM}+\mathrm{DM} \to \mathrm{DM} + \mathrm{SM}$, modify the DM freeze-out, our aim is to study the impact of the non-standard physics on DM direct detection and on the LHC Higgs phenomenology in those models. In particular, we study the possible deviations of the Higgs boson
decay mode to two photons, $h\to \gamma\gamma,$ from the SM prediction. 
As for the singlet model we found that the main constraint from Higgs physics comes from the upper bound on the invisible width that rules out the light DM scenarios. We also discuss the possibility of having direct signals from two different DM candidates. 

The layout of our paper is the following. We formulate $\Z_N$ symmetric models and study their field content in Section~\ref{sec:ZN}.
The scalar potentials that give rise to semi-annihilations are presented in Section~\ref{sec:z:3:4:semiannih}.
We list the various experimental and theoretical constraints on those models in Section~\ref{sec:constraints}. 
The results of our study for $\Z_3$ models are presented in Section~\ref{sec:Z:3} and for $\Z_4$ models  in Section~\ref{sec:Z:4}.
We conclude in Section~\ref{sec:concl}. One loop $\beta$-functions for renormalisation of our models are presented in Appendix~\ref{se:z3:z4:1-loop:beta:functions}.

\section{Conditions on $\Z_{N}$ charges and potential terms}
\label{sec:ZN}

\subsection{$\Z_{N}$ symmetries}

A field $\phi$ with $\Z_{N}$ charge $X_{\phi}$ transforms under a $\Z_{N}$ group as $\phi \to \omega^{X_{\phi}} \phi$, where $\omega^N = 1$, that is $\omega = \exp (i 2 \pi/N)$. Since the addition of charges is modulo $N$, the possible values of $\Z_{N}$ charges can be restricted to $0, 1, \ldots, N-1$ without loss of generality. 
Of course, for $N > 2$, the field $\phi$ has to be complex to be charged under $\Z_{N}$. In general the field $\phi$ transformation has to be complex unless $N/X_\phi=2$ when N is even.

A $\Z_{N}$ symmetry can arise as a discrete gauge symmetry from breaking a $U(1)_{X}$ gauge group with a scalar, whose $X$-charge is $N$ \cite{Krauss:1988zc,Martin:1992mq}. From the phenomenological point of view, however, it may be impossible to distinguish different top-down assignments of discrete charges to fields, since they can yield the same low energy scalar potential. For larger values of $N$, the conditions the $\Z_{N}$ symmetry impose on the Lagrangian approximate the original $U(1)$ symmetry for two reasons. For a given field content, the number of possible renormalisable Lagrangian terms is limited and will be exhausted for some small finite $N$, although they will appear in different combinations for different values of $N$. In addition, if the $\Z_{N}$ symmetry results from breaking a $U(1)_{X}$, the discrete charges of particles cannot be arbitrarily large, because that would make the model nonperturbative. The discrete charges of fields will equal their $U(1)$ charges if the latter are smaller than $N$, and the scalar potential will be the same as in the unbroken $U(1)$ in this case.

For $\Z_{2}$, the DM can have only one component, because the only possible discrete charge $X_{\text{DM}} = 1$ to keep DM from decaying into SM fields with $X = 0$. The same is true for $\Z_{3}$: although the discrete charges can take values $0, 1, 2$, one has $2 = -1 \mod 3$, so the dark sector particles with $X = 2$ are just the antiparticles of those with $X = 1$. For both $\Z_{4}$ and $\Z_{5}$, DM can have two components. In general, for $\Z_{N}$, it can have $\lfloor N/2 \rfloor$ components.

\subsection{Field content of the minimal model}

$\Z_N$ symmetries with $N>2$ can lead to new phenomena such as semi-annihilation \cite{DEramo:2010ep,Hambye:2009fg,Hambye:2008bq,Arina:2009uq} and dark matter conversion \cite{Liu:2011aa,Belanger:2011ww,Adulpravitchai:2011ei}. The simplest such model is the $\Z_3$ singlet scalar dark matter \cite{Ma:2007gq,Belanger:2012zr} where the cubic term of the singlet produces semi-annihilation -- missing in the well-studied $\Z_2$ case of the complex singlet \cite{McDonald:1993ex,Barger:2007im,Barger:2008jx,Burgess:2000yq,Gonderinger:2009jp}. 

In general, however, there must be more than one neutral particle in the dark sector to give rise to different behaviour, in particular to multicomponent DM. The scalar sector of the minimal dark matter model with both semi-annihilation and DM conversion contains, in addition to the standard model Higgs boson $H_{1}$, one extra scalar doublet $H_{2}$ -- similar to the well-known inert doublet \cite{Deshpande:1977rw,Ma:2006km,Barbieri:2006dq,LopezHonorez:2006gr} -- and one extra complex scalar singlet $S$. 
Note that for such a field content  even a $\Z_2$ symmetry yields qualitatively novel features  concerning dark matter phenomenology, electroweak symmetry breaking and collider phenomenology as compared to the inert doublet model~\cite{Kadastik:2009cu,Kadastik:2009dj,Kadastik:2009ca,Kadastik:2009gx,Huitu:2010uc,Cohen:2011ec,Biswas:2013nn}. 

Because the only scalar field in the Standard Model is a doublet, the new doublet is essential to write quartic semi-annihilation terms such as the $\lambda_{S12}$ term in eq.~\eqref{eq:V:Z:3}. The presence of the singlet is as essential, since it is impossible to allow only the $\lambda_6 \abs{H_1}^2 \hc{H_1} H_2$ term of the two Higgs doublet model (2HDM) for semi-annihilation without also allowing the $\lambda_7 \abs{H_2}^2 \hc{H_1} H_2$ term which mixes the two doublets.%
\footnote{Here and below we use the standard 2HDM symbols for the interaction terms of the doublets.}

\subsection{Constraints on charge assignments}

To allow the SM Yukawa terms of the Higgs $H_{1}$, and to keep the DM stable, the discrete charges must satisfy certain requirements. On one hand, the $H_1$ Yukawa terms with $u$- and $d$-type quarks and charged leptons must separately have zero discrete charge modulo $N$. From a low energy point of view, however,  we can simply set the charges of all standard model fields to zero. On the other hand, we want to forbid the $|H_{1}|^{2} S$  and other terms that lead to mixing,%
\footnote{In principle either $H_{1}$ and $H_{2}$, or $H_{1}$ and $S$ could mix, leaving the other dark sector particle to be the DM. However, in the models we study we demand no mixing with $H_{1}$ to allow for a richer dark sector phenomenology.}
together with Yukawa couplings for $H_2$. Therefore, the discrete charges must satisfy
\begin{equation}
  X_{1} = 0, \enspace  X_{2} > 0, \enspace  X_{S} > 0.
\label{eq:constraints:V:simpler}
\end{equation}

All possible scalar potentials contain a common piece $V_{0}$ since under any $\Z_{N}$ and for any charge assignment each field can be paired with its Hermitian conjugate to form an invariant:
\begin{equation}
\begin{split}
  V_{0}&=  \mu_{1}^{2} \abs{H_{1}}^{2} + \lambda_{1} \abs{H_{1}}^{4} 
  + \mu_{2}^{2} |H_{2}|^{2} + \lambda_{2} |H_{2}|^{4} + \mu_{S}^{2} |S|^{2} + \lambda_{S} |S|^{4} \\
  &+ \lambda_{S1} |S|^{2} |H_{1}|^{2}
  + \lambda_{S2} |S|^{2} |H_{2}|^{2} + \lambda_{3} |H_{1}|^{2} |H_{2}|^{2} 
  + \lambda_{4} (H_{1}^{\dagger} H_{2}) (H_{2}^{\dagger} H_{1}).
\end{split}
\label{eq:V:c}
\end{equation}

\subsection{Doublet-like DM}

In case of mixing angle, $\theta$, between the neutral components of the doublet and the singlet, dark matter can be either doublet-like or singlet-like.
In the first case, if DM is degenerate with the next-to-lightest stable particle (NLSP) it can only contribute to a small fraction of the relic density $\Omega_{\text{DM}}$ because of the large coannihilation contribution due to the $Z H^0 A^0$ coupling, where $H^{0}$ and $A^{0}$ are the real and imaginary neutral components of $H_{2}$. If dark matter is singlet-like, the coannihilation cross section is suppressed by an additional factor of $\sin^2\theta$ . In this case even full degeneracy of DM and NLSP does not affect the relic density significantly.

In the case of the doublet-like DM, there are three ways to lift the degeneracy between the DM and the NLSP particle (we keep $X_{1}$ for generality):
\begin{enumerate}
\item A nonzero $\mu_{S}^{\prime 2} S^2$ term (together with a $\mu_{SH} \hc{S}\hc{H_{1}} H_{2}$ or $\mu'_{SH} S \hc{H_{1}} H_{2}$ mixing term to convey the mass gap from $S$ to $H_{2}$). 
If the $N$ in the $\Z_{N}$ is odd, the term is not allowed due to $X_{S} > 0$. For an even $N$, $X_{S} = N/2$ is possible and, in addition, $X_{2} = N/2 + X_{1}$ (or $1 \leftrightarrow2$) is needed. 
\item A $\lambda_{5} (\hc{H_{1}} H_{2})^{2}$ term. A $\lambda_{5}$ term can only be allowed for even $N$, because it transforms as $\omega^{2 (X_1 - X_2)}$ and as $X_1 \neq X_2$, the exponent $2 (X_1 - X_2)$ cannot be zero modulo an odd number such as $3$.
\item  \emph{Both} the $\mu_{SH} \hc{S} \hc{H_{1}} H_{2}$ and $\mu'_{SH} S \hc{H_{1}} H_{2}$ terms. It is only possible to have both these terms for an even $N$ with $X_{S} = N/2$ and $X_{2} = N/2 + X_{1}$ (or vice versa). In this case the $\mu_{S}^{\prime 2}$ and $\lambda_{5}$ terms are allowed as well.
\end{enumerate}
None of these work for odd $N$, in which case the only possibility is singlet-like DM.

The semi-annihilation terms $\lambda_{S12} \hc{H_{1}} H_{2} S^2$ and $\lambda_{S21} \hc{H_{2}} H_{1} S^2$ are forbidden if both $\mu_{SH}$ and $\mu'_{SH}$ terms are allowed, nor can they coexist with the $\mu_{S}^{\prime 2}$ term. Thus, for doublet-like DM with semi-annihilation, the only option to lift degeneracy between the DM and NLSP particle is a $\lambda_5$ term.

\subsection{Detour into $SO(10)$}

The discrete $\Z_N$ symmetry that stabilises dark matter may arise from breaking a $U(1)_X$ subgroup of the gauge group of a grand unified theory (GUT) \cite{Pati:1974yy,Georgi:1974sy}. One of the simplest examples is $SO(10)$ \cite{Fritzsch:1974nn} which is broken down to the SM gauge group as $SO(10) \to SU(5) \times U(1)_X \to SU(5) \times \Z_N$. There are two ways to combine the two $U(1)$ subgroups of $SO(10)$ into hypercharge $U(1)_Y$ and $U(1)_X$: standard $SU(5)$ \cite{Georgi:1974sy} and flipped $SU(5)$ \cite{Barr:1981qv,Derendinger:1983aj,Barr:1988yj}.

The SM matter fields and a heavy neutrino singlet can be put in a $\bf 16$ of $SO(10)$, and the SM Higgs field  in the $\bf 10$ of $SO(10)$. The minimal choice of representation to embed the complex singlet and the new doublet in is a scalar $\bf 16$. Under $SU(5) \times U(1)_X$  the representations decompose as
\begin{align}
{\bf 16} &= {\bf 10}_{1}^{16} + {\bf \bar{5}}_{-3}^{16} + {\bf 1}_{5}^{16}, \\
{\bf 10} &= {\bf 5}_{-2}^{10} + {\bf \bar{5}}_{2}^{10}.
\end{align}

In standard $SU(5)$, the relation to the SM fields is
\begin{equation}
{\bf \bar{5}}_{-3}^{16}
= 
\T{
\begin{pmatrix}
d_{1}^{c} & d_{2}^{c} & d_{3}^{c} & \nu & e\end{pmatrix}_{L}
},
\quad
\begin{pmatrix}
\begin{pmatrix}
u \\ d
\end{pmatrix}_{L}
&
u_{L}^{c}
&
e_{L}^{c}
\end{pmatrix}
\in
{\bf 10}_{1}^{16},
\quad
{\bf 1}_{5}^{16} = \nu_{L}^{c},
\end{equation}
where we have suppressed colour indices on fields in the $\bf 10$.

The down-type and lepton Yukawa interactions are ${\bf 10}_{1}^{16} {\bf \bar{5}}_{-3}^{16} {\bf \bar{5}}_{2}^{10}$ and the up-type Yukawa couplings are ${\bf 10}_{1}^{16} {\bf 10}_{1}^{16} {\bf 5}_{-2}^{10}$. The SM Higgs is $H_{1} \in {\bf \bar{5}}_{2}^{10}$.

$S$ is the scalar analogue of the neutrino singlet in ${\bf 1}_{5}^{16}$ and $\hc{H_{2}}$ is the scalar analogue of the lepton doublet in ${\bf \bar{5}}_{-3}^{16}$. Therefore the $U(1)$-charges of the scalar sector are $X_{S} = 5$, $X_{1} = 2$ and $X_{2} = 3$ (these are equal, modulo $N$, to the discrete $\Z_{N}$ charges of the fields).

In flipped $SU(5)$, the relation to the SM fields is
\begin{equation}
{\bf \bar{5}}_{-3}^{16}
= 
\T{
\begin{pmatrix}
u_{1}^{c} & u_{2}^{c} & u_{3}^{c} & \nu & e 
\end{pmatrix}_{L}
},
\quad
\begin{pmatrix}
\begin{pmatrix}
u \\ d
\end{pmatrix}_{L}
&
d_{L}^{c}
&
\nu_{L}^{c}
\end{pmatrix}
\in
{\bf 10}_{1}^{16},
\quad
{\bf 1}_{5}^{16} = e_{L}^{c}.
\end{equation}

The down-type Yukawa interactions are ${\bf 10}_{1}^{16} {\bf 10}_{1}^{16} {\bf 5}_{-2}^{10}$, the lepton Yukawa interactions are ${\bf 1}_{5}^{16} {\bf \bar{5}}_{-3}^{16} {\bf 5}_{-2}^{10}$ and the up-type Yukawa couplings are ${\bf 10}_{1}^{16} {\bf \bar{5}}_{-3}^{16} {\bf \bar{5}}_{2}^{10}$. The SM Higgs is $H_{1} \in {\bf 5}_{-2}^{10}$. Note that the doublet must be flipped too with respect to standard $SU(5)$.

$S$ is the scalar analogue of the neutrino singlet in ${\bf 10}_{1}^{16}$ and $\hc{H_{2}}$ is the scalar analogue of the lepton doublet in ${\bf \bar{5}}_{-3}^{16}$. The $U(1)$-charges of the scalar sector are $X_{S} = 1$, $X_{1} = -2$ and $X_{2} = 3$, which are equal, modulo $N$, to the discrete $\Z_{N}$ charges of the fields.

\section{$\Z_N$  potentials}
\label{sec:z:3:4:semiannih}

\subsection{The $\Z_{2}$ potential}

For the sake of completeness, we include the unique scalar potential symmetric under $\Z_{2}$:
\begin{equation}
\begin{split}
    V &= V_{0} + \frac{\mu_{S}^{\prime 2}}{2} ( S^{2}
    + S^{\dagger 2} ) + \frac{\mu_{S H}}{2} (\hc{S}\hc{H_{1}} H_{2} + S \hc{H_{2}} H_{1}) \\
    &+ \frac{\mu'_{S H}}{2} (S\hc{H_{1}} H_{2} + \hc{S} \hc{H_{2}} H_{1}) 
    + \frac{\lambda_{5}}{2} \left[(\hc{H_{1}} H_{2})^{2} + (\hc{H_{2}} H_{1})^{2} \right]
    \\
    &+ \frac{ \lambda'_{S}}{2} (S^{4} + S^{\dagger 4})
    + \frac{ \lambda''_{S} }{2} \abs{S}^{2} (S^{2} + S^{\dagger 2})
    \\
    &+ \frac{ \lambda'_{S1} }{2}
    \abs{H_{1}}^{2} (S^{2} + S^{\dagger 2} )
    + \frac{ \lambda'_{S2} }{2}
    \abs{H_{2}}^{2} (S^{2} + S^{\dagger 2} ).
\end{split}
\label{eq:V:Z2}
\end{equation}
For the given dark sector of $H_{2}$ and $S$, scalar potentials for higher $\Z_{N}$ that do not contain semi-annihilation terms will be equivalent to the potential \eqref{eq:V:Z2} with some terms set to zero.

The $\Z_{2}$ potential \eqref{eq:V:Z2} in the case of $SO(10)$ GUT in which some interactions are suppressed was studied in detail in \cite{Kadastik:2009cu,Kadastik:2009dj,Kadastik:2009ca,Kadastik:2009gx,Huitu:2010uc}. Both $H_{2}$ and $S$ are odd under  $\Z_{2}$ but only one of them is dark matter.

\subsection{$\Z_3$ potential with semi-annihilation}
\label{sec:z:3:semiannih}

A $\Z_{3}$ potential that induces semi-annihilation processes is 
\begin{equation}
\begin{split}
  V_{\Z_{3}} &= V_{0} + \frac{\mu''_{S}}{2} (S^{3} + S^{\dagger 3}) 
  + \frac{\lambda_{S12}}{2} (S^{2} H_{1}^{\dagger}   H_{2} + S^{\dagger 2} H_{2}^{\dagger} H_{1}) \\
  &+ \frac{\mu_{SH}}{2} (S H_{2}^{\dagger} H_{1} + S^{\dagger} H_{1}^{\dagger} H_{2}),
\end{split}
\label{eq:V:Z:3}
\end{equation}
invariant under e.g. the $\Z_{3}$ charges $X_{1} = 0, X_{2} = X_{S} = 1$.
Another such potential is obtained from eq.~(\ref{eq:V:Z:3}) by substituting $S \to S^\dagger$ (with $\mu_{SH} \to \mu'_{SH}$ and $\lambda_{S12} \to \lambda_{S21}$). From a low energy point of view, the two potentials are indistinguishable. Both the standard $SU(5)$, in which case the $\Z_{3}$ charges of fields are $X_{S} = 2, X_{1} = 2, X_{2} = 0$, and flipped $SU(5)$, with $X_{S} = 1, X_{1} = 1, X_{2} = 0$, yield the potential obtained by $S \to S^\dagger$ in \eqref{eq:V:Z:3}.

Our $\Z_{3}$ and $\Z_{4}$ lagrangians are invariant under hypercharge symmetry $H_{1} \to e^{i \phi_Y} H_{1}$ and $H_{2} \to e^{i \phi_Y} H_{2}$.
We can use a hypercharge rotation of the doublets \cite{Dreiner:2005rd} to satisfy $X_{1} = 0$ and $X_{2} > 0$. For example, for standard $SU(5)$, we can rotate the charges to $X_{S} = 1$, $X_{1} = 0$, $X_{2} = -1 \equiv 2 \mod 3$, which upon replacing $H_{1}$ and $H_{2}$ with their conjugates gives $X_S = 1$, $X_{1} = 0$ and $X_{2} = 1$, which gives the $\Z_{3}$ scalar potential \eqref{eq:V:Z:3}.


The $\mu_{SH}$ term in \eqref{eq:V:Z:3} induces mixing between the neutral components of $H_2$ and $S$.
In terms of the mass eigenstates  $x_1$, $x_2$, we have 
\begin{equation}
  H_{2} = \begin{pmatrix}
    -i H^{\pm} \\
    x_{1} \sin{\theta} + x_{2} \cos{\theta}
  \end{pmatrix},
  \quad
  S= x_1 \cos{\theta}  - x_2 \sin{\theta}.
\end{equation}   
Considering the masses $M_{h}$, $M_{H^{\pm}}$, $M_{x_{1}}$, $M_{x_{2}}$
and the mixing angle $\theta$ as free parameters of the model, we have
\begin{align}
  \mu_S^2   &= M_{x_1}^2 \cos^2{\theta} + M_{x_2}^2 \sin^2{\theta}
  - \lambda_{S1} \frac{v^{2}}{2}, \\
  \mu_{SH}  &= -4 (M_{x_2}^2-M_{x_1}^2) 
  \frac{\cos{\theta} \sin{\theta}}{\sqrt{2} v }, \\
  \mu_{1}^{2} &= -\frac{M_{h}^{2}}{2}, \\
  \mu_2^2   &= M_{H^{\pm}}^{2} - \lambda_3 \frac{v^2}{2},
  \label{eq:Z3:def:lambda:mu2} \\
    \lambda_1 &= \frac{1}{2} \frac{M_{h}^{2}}{v^{2}}, \label{eq:Z3:def:lambda:1} \\
    \lambda_{4} &= \left( M_{x_{1}}^{2} \sin^{2}{\theta} 
    + M_{x_{2}}^{2} \cos^{2}{\theta} - M_{H^{\pm}}^{2} \right) \frac{2}{v^{2}}.\label{eq:Z3:def:lambda:4}
\end{align}

\subsection{$\Z_4$ potential with semi-annihilation}
\label{sec:z:4:semiannih}

The only potential for $\Z_{4}$ that contains semi-annihilation terms%
\footnote{
The other four scalar potentials can formally be obtained from the $\Z_2$-invariant potential (\ref{eq:V:Z2}) by setting all the new terms added to $V_{0}$ to zero, with the exception of the
1) $\lambda'_S$, $\mu_{SH}$ (this is the potential that emerges from $SO(10)$ for both standard and flipped $SU(5)$),
2) $\lambda'_S$, $\mu'_{SH}$,
3) $\mu'_S$, $\lambda'_S$, $\lambda''_S$, $\lambda'_{S1}$, $\lambda'_{S2}$,
4) $\mu'_S$, $\lambda'_S$, $\lambda''_S$, $\lambda'_{S1}$, $\lambda'_{S2}$, $\lambda_{5}$, 
  $\mu_{SH}$, $\mu'_{SH}$ terms.
}
 is
\begin{equation}
\begin{split}
V_{\Z_{4}} &= V_{0} + \frac{\lambda'_{S}}{2} (S^{4} + S^{\dagger 4}) + 
\frac{\lambda_{5}}{2} \left[(H_{1}^{\dagger} H_{2})^{2} + (H_{2}^{\dagger} H_{1})^{2} \right] \\
&+ \frac{\lambda_{S12}}{2} (S^{2} H_{1}^{\dagger} H_{2} + S^{\dagger 2} H_{2}^{\dagger} H_{1}) + \frac{\lambda_{S21}}{2} (S^{2} H_{2}^{\dagger} H_{1} + S^{\dagger 2} H_{1}^{\dagger} H_{2}),
\end{split}
\label{eq:pot:3:Z:4}
\end{equation}
invariant under e.g. the assignment of $\Z_{4}$ charges $X_{1} = 0, X_{2} = 2, X_{S} = 1$.
The dark sector particles do not mix with each other, because $S$ and $H_{2}$ have
different $\Z_4$ charges. As a result this model has two dark sectors with the complex scalar $S$ in the first one, and the second one comprising the charged Higgs boson $H^{\pm}$ and the real scalars $H^{0}$ and $A^{0}$. Any of the neutral particles with a non-zero $\Z_4$ charge can be a dark matter candidate. 
Considering the masses of the scalars $M_{h}^{2}$, $M_{H^{\pm}}$, $M_{S}$, $M_{H^0}$ and $M_{A^0}$ as independent
parameters, we have 
\begin{align}
  \mu_S^2 &= M_S^2 - \lambda_{S1} \frac{v^2}{2},\\
   \mu_{1}^{2} &= -\frac{M_{h}^{2}}{2}, \\
  \mu_2^2 &= M_{H^{\pm}}^{2} - \lambda_3 \frac{v^2}{2}, \\ 
  \lambda_{1} &= \frac{1}{2} \frac{M_{h}^{2}}{v^{2}}, \label{eq:Z4:def:lambda:1} \\
  \lambda_{4} &= \left( \frac{M_{A^0}^2 + M_{H^0}^2 }{2} -  M_{H^{\pm}}^{2} \right) \frac{2}{v^{2}},
  \label{eq:Z4:def:lambda:4} \\
  \lambda_{5} &= \frac{M_{H^0}^2-M_{A^0}^2}{v^2} \label{eq:Z4:def:lambda:5}.
\end{align}

\section{Experimental and theoretical constraints}
\label{sec:constraints}

\subsection{Perturbativity}
\label{sec:perturbativity}

There are several possible definitions of perturbativity constraints on scalar quartic couplings. In \cite{Cheung:2012nb}, for example, it was required that the contribution of each coupling $\lambda_{i}$ to its own $\beta$-function is less than unity so that the couplings do not run too fast. We follow \cite{Lerner:2009xg}, comparing the couplings $\lambda_{i}$ in the potential to the vertices in the Feynman rules for mass eigenstates. Barring accidental cancellations, the vertex factors have to be smaller than $4 \pi$ to ensure that the one-loop level quantum corrections are smaller than the tree level contributions. For example, the quartic singlet self-interaction term $\lambda_{S} \abs{S}^{4}$ yields the vertex factor $i \, 4 \lambda_{S}$. Demanding that $4 \lambda_{S} < 4 \pi$ gives $\lambda_{S} < \pi$. If a quartic coupling $\lambda_{i}$ in the potential occurs in several vertices, we choose the strongest bound.

\subsection{Perturbative unitarity}
\label{sec:unitarity}

At high energy, the tree-level scalar-scalar scattering matrix is dominated by the quartic contact interaction terms. The $s$-wave scattering amplitudes should not exceed the perturbative unitarity limit for this partial wave, requiring that the eigenvalues of the $S$-matrix $\mathcal{M}$ must be smaller than the unitarity bound given by
\begin{equation}
  \abs{\Re \mathcal{M}} < \frac{1}{2}.
\end{equation}

The unitarity bounds of the 2HDM were first studied in \cite{Kanemura:1993hm,Akeroyd:2000wc}. We will extend the formalism of \cite{Ginzburg:2003fe,Ginzburg:2005dt} for the 2HDM  to states containing the singlet $S$. The initial states are classified according to their total hypercharge $Y$ ($0$, $1$ or $2$), weak isospin $\sigma$ ($0$, $\frac{1}{2}$ or $1$) and discrete $\Z_N$ charge $X$. The $\Z_{3}$ unitarity bounds are reducible to those of the $\Z_{4}$ case, and therefore we present only the latter.

For the sake of brevity, we list only the two sets of initial states which differ from the 2HDM initial states given in \cite{Ginzburg:2003fe,Ginzburg:2005dt}.
The full set of possible initial states with hypercharge $Y = 1$ and $\sigma = \frac{1}{2}$ is
\begin{equation}
  H_{2} \hc{S}, H_{1} S, H_{1} \hc{S}, H_{2} S.
\end{equation}
The full set of possible initial states with hypercharge $Y = 0$ and $\sigma = 0$ is
\begin{equation}
 \frac{1}{\sqrt{2}} \hc{H_{1}} H_{1}, 
 \frac{1}{\sqrt{2}} \hc{H_{2}} H_{2}, 
 \hc{S} S, 
 \frac{1}{\sqrt{2}} S^{2}, 
 \frac{1}{\sqrt{2}} S^{\dagger 2}, 
 \frac{1}{\sqrt{2}} \hc{H_{1}} H_{2}, 
 \frac{1}{\sqrt{2}} \hc{H_{2}} H_{1},
\end{equation}
where the first three states have discrete $\Z_{4}$ charge $X = 0$ and the last four states have $X = 2$.

We do present all the scattering matrices and bounds on their eigenvalues for the $\Z_{4}$ model. They reduce to the $\Z_{3}$ case with $\lambda'_{S} = 0$, $\lambda_{5} = 0$, $\lambda_{S21} = 0$.
The scattering matrices are
\begin{align}
  8 \pi S_{Y=2,\sigma=1} &= 
  \begin{pmatrix}
    2 \lambda_{1} & \lambda_{5} & 0 \\
    \lambda_{5}^{*} & 2 \lambda_{2} & 0 \\
    0 & 0 & \lambda_{3} + \lambda_{4}
  \end{pmatrix},
  &
  8 \pi S_{Y=2,\sigma=0} &= \lambda_{3} - \lambda_{4},
  \\
  8 \pi S_{Y=1,\sigma=\frac{1}{2}} &= 
  \begin{pmatrix}
    \lambda_{S2} & \lambda_{S21} & 0 & 0 \\
    \lambda_{S21}^{*} & \lambda_{S1} & 0 & 0 \\
    0 & 0 & \lambda_{S1} & \lambda_{S12} \\
    0 & 0 & \lambda_{S12}^{*} & \lambda_{S2}
  \end{pmatrix},
  &
  8 \pi S_{Y=0,\sigma=1} &= 
  \begin{pmatrix}
    2 \lambda_{1} & \lambda_{4} & 0 & 0 \\
    \lambda_{4} & 2 \lambda_{2} & 0 & 0 \\
    0 & 0 & \lambda_{3} & \lambda_{5}^{*} \\
    0 & 0 & \lambda_{5} & \lambda_{3}
  \end{pmatrix},
\end{align}
\begin{equation}
    8 \pi S_{Y=0,\sigma=0} = 
  \begin{pmatrix}
    6 \lambda_{1} & 2 \lambda_{3} + \lambda_{4} & \sqrt{2} \lambda_{S1} & 0 & 0 & 0 & 0 \\
    2 \lambda_{3} + \lambda_{4} & 6 \lambda_{2} & \sqrt{2} \lambda_{S2} & 0 & 0 & 0 & 0 \\
    \sqrt{2} \lambda_{S1} & \sqrt{2} \lambda_{S2} & \lambda_{S} & 0 & 0 & 0 & 0 \\
    0 & 0 & 0 & \lambda_{S} & \lambda'_{S} & \lambda_{S21} & \lambda_{S12} \\
    0 & 0 & 0 & \lambda'_{S} & \lambda_{S} & \lambda_{S12} & \lambda_{S21} \\
    0 & 0 & 0 & \lambda_{S21} & \lambda_{S12} & \lambda_{3} + 2 \lambda_{4} & 3 \lambda_{5}^{*}   
    \\
    0 & 0 & 0 & \lambda_{S21} & \lambda_{S21} & 3 \lambda_{5} & \lambda_{3} + 2 \lambda_{4} \\
  \end{pmatrix}.
\end{equation}

The eigenvalues $\Lambda_{Y\sigma i}^{X}$ of the above scattering matrices (where $i = \pm \text{ or }1,2,3$) can be written as
\begin{align}
  \Lambda_{21\pm}^{0} &= \lambda_{1} + \lambda_{2} \pm \sqrt{(\lambda_{1} - \lambda_{2})^{2} + \abs{\lambda_{5}}^{2}},
  \\
  \Lambda_{21}^{2} &= \lambda_{3} + \lambda_{4},
  \\
  \Lambda_{20}^{2} &= \lambda_{3} - \lambda_{4},
  \\
  \Lambda_{1\frac{1}{2}\pm}^{0,1} &= \frac{1}{2} \left( \lambda_{S1} + \lambda_{S2}
  \pm \sqrt{ (\lambda_{S1} - \lambda_{S2})^{2} + 4 \abs{\lambda_{S21}}^{2}} \right),
  \\
  \Lambda_{1\frac{1}{2}\pm}^{2} &= \frac{1}{2} \left( \lambda_{S1} + \lambda_{S2}
  \pm \sqrt{ (\lambda_{S1} - \lambda_{S2})^{2} + 4 \abs{\lambda_{S12}}^{2}} \right),
  \\
  \Lambda_{01\pm}^{0} &= \lambda_{1} + \lambda_{2} \pm \sqrt{(\lambda_{1} - \lambda_{2})^{2} + \lambda_{4}^{2}},
  \\
  \Lambda_{01\pm}^{2} &= \lambda_{3} \pm \abs{\lambda_{5}},
  \\
  \abs{ \Lambda_{00\, 1,2,3}^{0} } &\leqslant \frac{1}{3} \left( 6 \lambda_1 + 6 \lambda_2 + \lambda_S + 2 \, [ 36 (\lambda_{1}^{2} - \lambda_{1} \lambda_{2} + \lambda_{2}^{2}) \phantom{^{\frac{1}{2}}} 
 \right. \label{eq:Lambda000} \\ 
 & \left. - 6 (\lambda_{1} + \lambda_{2}) \lambda_{S} + \lambda_{S}^{2} 
 + 3 (2 \lambda_{3} + \lambda_{4})^{2} + 6 (\lambda_{S1}^{2} + \lambda_{S2}^{2}) ]^{\frac{1}{2}} \right), \notag
 \\
 \Lambda_{00+\pm}^{2} &= \frac{1}{2} 
 \left( \lambda_{S} + \lambda'_{S} + \lambda_{3} + 2 \lambda_{4} + 3 \lambda_{5} 
 \phantom{ \sqrt{ (\lambda'_{S})^{2} } }
 \right. \\
 & \left. \pm \sqrt{ (\lambda_{S} + \lambda'_{S} -\lambda_{3} - 2 \lambda_{4} - 3 \lambda_{5})^{2} + 4 (\lambda_{S12} + \lambda_{S21})^{2} } \right), \notag
 \\
  \Lambda_{00-\pm}^{2} &= \frac{1}{2} 
 \left( \lambda_{S} - \lambda'_{S} + \lambda_{3} + 2 \lambda_{4} - 3 \lambda_{5} 
 \phantom{ \sqrt{ (\lambda'_{S})^{2} } }
 \right. \\
 & \left. \pm \sqrt{ (-\lambda_{S} + \lambda'_{S} + \lambda_{3} + 2 \lambda_{4} - 3 \lambda_{5})^{2} + 4 (\lambda_{S12} - \lambda_{S21})^{2} } \right). \notag
\label{eq:Z4:semiannih:unitarity}
\end{align}
Since $\Lambda_{00\, 1,2,3}^{0}$ are too cumbersome to be presented in full, in \eqref{eq:Lambda000} we have given an upper bound on their absolute values by applying Samuelson's inequality \cite{samuelson} to the characteristic equation (in our numerical calculations we use the exact eigenvalues). The inequality arises from the observation that a collection of $n$ points is within $\sqrt{n - 1}$ standard deviations of their mean. For the polynomial $a_{n} x^{n} + a_{n-1} x^{n-1} + \ldots + a_{1} x + a_{0}$ with only real roots, the roots lie in the interval bounded by
\begin{equation}
  x_{\pm} = -\frac{a_{n-1}}{n a_{n}} \pm \frac{n-1}{n a_{n}} \sqrt{a_{n-1}^{2} - \frac{2 n}{n-1} a_{n} a_{n-2}}.
\end{equation}

\subsection{Vacuum stability}
\label{sec:vacuum:stability}

In order to have a finite minimum of the potential energy, the scalar potential has to be bounded below, especially in the limit of large field values. The quadratic and cubic terms are negligible in this limit and therefore it suffices to consider only the quartic terms to find the constraints of vacuum stability. To ensure that the quartic potential is bounded below, we write the matrix of quartic interactions in a basis of non-negative field variables and demand this matrix to be copositive  \cite{Kannike:2012pe}.

The Higgs doublet bilinears can be parameterised as \cite{Ginzburg:2004vp}
\begin{equation}
  \abs{H_{1}}^{2} = r_{1}^{2}, \quad \abs{H_{2}}^{2} = r_{2}^{2}, 
  \quad \hc{H_{1}} H_{2} = r_{1} r_{2} \rho e^{i \phi}.
  \label{eq:V:SM:ID:param}
\end{equation}
The parameter $\abs{\rho} \in [0,1]$ as implied by the Cauchy inequality $0 \leqslant \abs{\hc{H_{1}} H_{2}} \leqslant \abs{H_{1}} \abs{H_{2}}$. The singlet can be written in the polar form as $S = s e^{i \phi_S}$.

Then the quartic part of the $\Z_4$-symmetric potential \eqref{eq:pot:3:Z:4} takes the form 
\begin{equation}
\begin{split}
  V_{\Z_4} &\supset \lambda_1 r_1^4 + \lambda_2 r_2^4 
  + \left[ \lambda_3 + (\lambda_4 + \lambda_5 \cos 2 \phi) \rho^2 \right]r_1^2 r_2^2 
  + \left( \lambda_S + \lambda'_S \cos 4 \phi_S \right) s^4
  \\
  &+ s^2 \Big( \lambda_{S1} r_1^2 + \lambda_{S2} r_2^2 
  + \rho \left[ \lambda_{S12} \cos{(\phi + 2 \phi_S)} 
  + \lambda_{S21} \cos{(\phi - 2 \phi_S)} \right] r_1 r_2 \Big) 
\end{split}
\label{eq:quartic:pot:Z4}
\end{equation}

We discuss only the $\Z_4$ case as its vacuum stability conditions reduce to the $\Z_{3}$ case with $\lambda'_{S} = 0$, $\lambda_{5} = 0$, $\lambda_{S21} = 0$ and further $\cos (\phi + 2 \phi_S) = -1$ so that the $\lambda_{S12}$ term can always be chosen negative in \eqref{eq:V:Z:3}. 
The parameters $r_1^2$, $r_2^2$ and $s^2$ are non-negative and can be used as a basis for the matrix of quartic couplings. If the potential has semi-annihilation terms, then it contains terms with $r_{1} r_{2}$ besides $r_{1}^{2}$, $r_{2}^{2}$. This leads to an ambiguity in the matrix because of $r_1^2 r_2^2 = (r_1 r_2)^2$. In that case we define $r_{1} = r \cos{\gamma}$, $r_{2} = r \sin{\gamma}$, where $0 \leqslant \gamma \leqslant \frac{\pi}{2}$ is a free parameter. In the vacuum stability conditions, the potential must be minimised with respect to all free parameters such as $\phi$, $\phi_S$ or $\gamma$.

For the $\Z_4$ case the necessary and sufficient vacuum stability conditions are
\begin{align}
  \lambda_{1} > 0, \enspace \lambda_{2} > 0, \enspace  \lambda_{S} - \abs{\lambda'_{S}} &> 0 \\
  \lambda_{3} + 2 \sqrt{\lambda_{1} \lambda_{2}} &>0, \\
  \lambda_{3} + \lambda_{4} - \abs{\lambda_{5}} + 2 \sqrt{\lambda_{1} \lambda_{2}} &>0, 
  \label{eq:Z4:vac:stab:cond}
\end{align}
and \emph{either} the $\lambda_{S1}$ and $\lambda_{S2}$ terms dominate the semi-annihilation terms to make the term proportional to $s^2$ in \eqref{eq:quartic:pot:Z4} positive,
\begin{equation}
  \lambda_{S1} \geqslant 0, \quad \lambda_{S2} \geqslant 0, \quad 2 \sqrt{\lambda_{S1} \lambda_{S2}} > \abs{\lambda_{S12}} + \abs{\lambda_{S21}},
  \label{eq:simple:Z:4:semiannih:vac:stab:cond}
\end{equation}
\emph{or} in the $(r^2, s^2)$ basis the general vacuum stability condition
\begin{equation}
\sqrt{\Lambda_{11} \Lambda_{22}} + \Lambda_{12}^2 > 0,
\label{eq:Z:4:semiannih:vac:stab:cond}
\end{equation}
where
\begin{align}
\Lambda_{11} &= \lambda_{1} \cos^4 \gamma + (\lambda_3 + (\lambda_4 + \lambda_5 \cos 2 \phi) \rho^2) \cos^2 \gamma \sin^2 \gamma + \lambda_2 \sin^4 \gamma, 
  \\
  \Lambda_{22} &= \lambda_S + \lambda'_S \cos 4 \phi_S,
  \\
  \Lambda_{12} &= \frac{1}{2} \Big[\lambda_{S1} \cos^2 \gamma 
  + \rho ( \lambda_{S21} \cos (\phi - 2 \phi_S)
  + \lambda_{S12} \cos (\phi + 2 \phi_S) ) \cos \gamma \sin \gamma  \\
  &+ \lambda_{S2} \sin^2 \gamma) \Big], \notag
\end{align}
has to hold for all values of the parameters in the ranges $0 \leqslant \gamma \leqslant \frac{\pi}{2}$, $0 \leqslant \abs{\rho} \leqslant 1$, $0 \leqslant \phi \leqslant 2 \pi$ and $0 \leqslant \phi_S \leqslant 2 \pi$. (In fact, the vacuum stability conditions for the $\Z_{3}$ case can also be found analytically by minimising the condition \eqref{eq:Z:4:semiannih:vac:stab:cond} with respect to the free parameters. However, the resulting expressions are extremely complicated.)

Note that in the special case $\gamma = 0, \frac{\pi}{2}$, the vacuum stability condition 
\begin{align}
  \lambda_{S1} + 2 \sqrt{\lambda_{1} (\lambda_{S} - \abs{\lambda'_{S}})} &>0, \label{eq:vac:stab:S:H1} \\
    \lambda_{S2} + 2 \sqrt{\lambda_{2} (\lambda_{S} - \abs{\lambda'_{S}})} &>0,
\end{align}
arise when two field directions at a time are non-zero ($h$ or $H^0,A^0$ and $S$) as was the case in \eqref{eq:Z4:vac:stab:cond}.

\subsection{Globality of the $\Z_{N}$-symmetric vacuum}
\label{sec:vacuum:globality}

The $\Z_{N}$-symmetric, EW-breaking vacuum has to be the global minimum of the scalar potential.%
\footnote{We do not consider metastability, but expect corrections to the allowed values of parameters to be of the order of $10\%$ as in \cite{Belanger:2012zr}.} To check whether the correct vacuum is a global minimum, the solutions or a given point in the parameter space will be substituted back in the scalar potential and compared to each other. Below we give a classification of the stationary points of the scalar potentials with the field content $H_1$, $H_2$, $S$.

In the general 2HDM there are three possible forms of vacua due to $SU(2)$ invariance. Via $SU(2)_{L} \times U(1)_{Y}$ transformations, the vacuum expectation values can always be reduced to the form
\begin{equation}
\vev{H_{1}} = 
  \begin{pmatrix}
  0
  \\
  \frac{v_{1}}{\sqrt{2}}
  \end{pmatrix},
  \quad
  \vev{H_{2}} = 
  \begin{pmatrix}
  v_{+} 
  \\
  \frac{v_{2}}{\sqrt{2}} e^{i \theta}
  \end{pmatrix},
  \label{eq:vevs:doublet}
\end{equation}
where $v_{1,2,+}$ are real and $\theta = 0$ when $v_+ \neq 0$ (see \cite{Branco:2011iw}).
%
In the classification of the extrema, we extend the notation used for the inert doublet model in \cite{Ginzburg:2010wa} by adding the index $S$ to the symbol when  a singlet VEV is present (if only the singlet VEV breaks $\Z_{N}$, we prefix the name of the vacuum by `non-').

\begin{table}[htdp]
\caption{Classification of the stationary points of the model.}
\begin{center}
\begin{tabular}{ccccc}
\textbf{Vacuum} & $v_1$ & $v_2$ & $v_+$ & $v_S$ \\
Fully symmetric $\text{EW}$ & & & & \\
EW-symmetric $\text{EW}_S$ & & & & \checkmark \\
Inert $\text{I}_1$ & \checkmark & & & \\
Non-inert $\text{I}_{1S}$ & \checkmark & & & \checkmark \\
Inert-like $\text{I}_{2}$ & & \checkmark & & \\
Non-inert-like $\text{I}_{2S}$ & & \checkmark & & \checkmark \\
Mixed $\text{M}$ & \checkmark & \checkmark & & \\
Mixed $\text{M}_S$ & \checkmark & \checkmark & & \checkmark \\
Charged $\text{CB}$ & \checkmark & \checkmark & \checkmark & \\
Charged $\text{CB}_S$ & \checkmark & \checkmark & \checkmark & \checkmark \\
CP-violating $\text{CP}_S$ & \checkmark & \checkmark & & \checkmark \\
\end{tabular}
\end{center}
\label{tab:vacua}
\end{table}%

In the fully symmetric vacuum ($\text{EW}$) all VEVs are zero, while in the EW-symmetric $\text{EW}_{S}$, $\Z_{N}$ is broken by the singlet VEV. In the inert vacuum $\text{I}_{1}$, only the SM Higgs $H_{1}$ gets a VEV and the neutral component of $H_{2}$ can be DM candidate -- this is the vacuum which we require to be the global minimum of the potential. $\Z_{N}$ is broken by $v_{S}$ in the \textit{non-inert} $\text{I}_{1S}$. In the inert-like $\text{I}_{2}$, the r\^{o}les of $H_{1}$ and $H_{2}$ are reversed \cite{Ginzburg:2010wa}, except for the Yukawa couplings of $H_{1}$ that break $\Z_{N}$ at loop level; in the \textit{non-inert-like} $\text{I}_{2S}$, $\Z_{N}$ is already broken at tree level by $v_{S}$. In the mixed vacuum $\text{M}$ both doublets get a VEV, and in $\text{M}_{S}$ the singlet gets a VEV too.  To have a charge-breaking vacuum, all the $v_{1}$, $v_{2}$ and $v_{+}$ have to be nonzero and $v_{S}$ can be zero ($\text{CB}$) or not ($\text{CB}_{S}$). In a CP-violating extremum $\text{CP}_{S}$ there is a phase of $\theta$ between the VEVs of $H_{2}$ and $H_{1}$.

In the 2HDM, the existence of a normal extremum $\text{I}_1$, $\text{I}_2$ or $\text{M}$ that is a minimum implies that it has lower potential energy than the charge-breaking or CP-violating vacua \cite{Ferreira:2004yd, Barroso:2005sm}. It is not necessarily true when the model is extended with the singlet, since in a vacuum with a singlet VEV the effective doublet mass terms can be different. Also note that the CP-violating vacuum $\text{CP}_S$ does not exist in the pure inert doublet model and is enabled only by $v_{S} \neq 0$: the singlet VEV generates the necessary mixing term for $H_{1}$ and $H_{2}$.

Actual calculations are much simplified if in the scalar potential the doublet bilinears $\hc{H_{i}} H_{j}$ are expressed in terms of gauge orbit variables \cite{Maniatis:2006fs} (see also \cite{Ivanov:2006yq,Ivanov:2007de}). All the bilinears can be arranged into the Hermitian $2 \times 2$ matrix 
\begin{equation}
  K = 
  \begin{pmatrix}
    \hc{H_{1}} H_{1} & \hc{H_{2}} H_{1} \\
    \hc{H_{1}} H_{2} & \hc{H_{2}} H_{2}
  \end{pmatrix},
\end{equation}
which is decomposed as 
\begin{equation}
  K_{ij} = \frac{1}{2} (K_{0} \delta_{ij} + K_{a} \sigma^{a}_{ij}),
\end{equation}
where $\sigma^{a}$ are the Pauli matrices. The four real gauge orbit variables are
\begin{equation}
  K_{0} = \hc{H_{i}} H_{i}, \quad K_{a} = (\hc{H_{i}} H_{j}) \sigma^{a}_{ij}, \quad a = 1,2,3.
\label{eq:K:i:H:i}
\end{equation}
Positive semidefiniteness of the matrix $K$ implies the `future light cone' conditions
\begin{equation}
  K_{0} \geqslant 0, \quad K_{0}^{2} - K_{1}^{2} - K_{2}^{2} - K_{3}^{2} \geqslant 0.
  \label{eq:K:lightcone}
\end{equation}
Inverting \eqref{eq:K:i:H:i}, the potential can be written in terms of $K_\mu$. Each term in the potential is at most quadratic in $K_\mu$ which reduces the degree of the minimisation equations. Of the solutions to the equations, only the stationary points that satisfy \eqref{eq:K:lightcone} are physical.

In terms of the doublet VEVs \eqref{eq:vevs:doublet}, the VEVs of the gauge orbit variables are given by
\begin{equation}
  \vev{K_0} = \frac{v_1^2 + v_2^2 + v_+^2}{2}, \enspace \vev{K_1} = v_1 v_2 \cos{\theta}, \enspace \vev{K_1} = v_1 v_2 \sin{\theta}, \enspace \vev{K_3} = \frac{v_1^2 - v_2^2 - v_+^2}{2}.
\end{equation}

One can see that the condition that the doublet VEVs preserve $\Z_{N}$ is
\begin{equation}
  K_{1} = K_{2} = 0 \text{\quad and \quad} K_{0} = K_{3}.
\end{equation}

The vacua $\text{EW}$ and $\text{EW}_{S}$ are in the tip $K_{\mu} = 0$ of the doublet light cone. If we choose $\mu_{1}^{2} = -M_{h}^{2}/2$, then the SM Higgs mass is always negative at the tip and this point is by construction never a minimum, but a saddle point.

The charge-breaking vacua are inside the forward light cone:
\begin{equation}
  K_{0} > 0, \quad K_{0}^{2} - K_{1}^{2} - K_{2}^2 - K_{3}^2 > 0.
  \label{eq:stats:point:fully:broken:SM}
\end{equation}
This point is a minimum if in the basis $(K_{0}, K_{1}, K_{2}, K_{3}, \Re S, \Im S)$, the leading principal minors of the Hessian are all positive.

The vacua $\text{I}_{1}$, $\text{I}_{1S}$, $\text{I}_{2}$, $\text{I}_{2S}$, $\text{M}$, $\text{M}_{S}$ and $\text{CP}_{S}$, where the full electroweak gauge group $SU(2)_{L} \times U(1)_{Y}$ is broken into $U(1)_{\text{EM}}$, are on the null surface of the future light cone:
\begin{equation}
  K_{0} > 0, \quad K_{0}^{2} - K_{1}^{2} - K_{2}^{2} - K_{3}^{2} = 0.
  \label{eq:stats:point:partially:broken:SM}
\end{equation}
The latter condition is enforced by adding to the potential a Lagrange multiplier term
\begin{equation}
  V_{u} = - u \, (K_{0}^{2} - K_{1}^{2} - K_{2}^{2} - K_{3}^{2}).
\end{equation}
The inequality $K_{0} > 0$ in \eqref{eq:stats:point:fully:broken:SM} has to be checked separately.
This point is a minimum if $u > 0$ and the last five leading principal minors of the bordered Hessian in the basis $(u, K_{0}, K_{1}, K_{2}, K_{3}, \Re S, \Im S)$ are all negative.

The vacuum expectation values can be calculated in analytical form in most of the stationary points. For the extrema where the VEVs of both the doublets and the singlet are all non-zero, we solve the minimisation equations numerically with the PHCpack equation solver \cite{DBLP:journals/toms/Verschelde99}. 

The solutions can then be substituted back in the scalar potential, and the global minimum be found by the smallest value. We require the inert vacuum $\text{I}_{1}$ to be the global minimum. In addition, in order for the stationary point to be a (local) minimum, the scalar masses or eigenvalues of the Hessian matrix at the stationary point have to be positive.

This requirement limits the size of $\mu''_{S}$ in the $\Z_{3}$ model as in \cite{Belanger:2012zr} and requires $\mu_{S}^{2}, \mu_{2}^{2} > 0$. For example, for the $\Z_{3}$ model $\mu_2^2 > 0$ translates via \eqref{eq:Z3:def:lambda:mu2} into $M_{H^{\pm}}^{2} > \lambda_3 \frac{v^2}{2}$.

To ensure that we are in the inert and not in the inert-like vacuum, we must have $V_{\text{I}_{1}} < V_{\text{I}_{2}}$ or
\begin{equation}
  -\frac{\mu_{1}^{4}}{4 \lambda_{1}} < -\frac{\mu_{2}^{4}}{4 \lambda_{2}} 
  \quad \text{or} \quad M_{h}^{2} \frac{v^{2}}{2} > \frac{1}{\lambda_{2}} \left( M_{H^{\pm}}^{2} - \lambda_3 \frac{v^2}{2} \right)^{2}.
\end{equation}

\subsection{Electroweak precision tests}
\label{sec:EWPT}

The measurements of electroweak precision data put strong constraints on physics beyond the SM. The latest electroweak fit by the Gfitter group \cite{Baak:2012kk} gives for the oblique parameters $S$ and $T$ the central values
\begin{equation}
  S = 0.03 \pm 0.10, \quad T = 0.05 \pm 0.12,
\end{equation}
with a correlation coefficient of $+0.89$.

To calculate electroweak precision parameters $S$ and $T$, we use the results for general models with doublets and singlets \cite{Grimus:2008nb,Grimus:2007if}. The usual loop functions are defined as
\begin{equation}
  F(I, J) = \frac{I + J}{2} - \frac{I J}{I - J} \ln{\frac{I}{J}},
\end{equation}
with $F(I,I) = 0$ in the limit of $J \to I$, and
\begin{equation}
\begin{split}
  G \left( I, J, Q \right) &=
  - \frac{16}{3} + \frac{5 \left( I + J \right)}{Q}
  - \frac{2 \left( I - J \right)^2}{Q^2}
  \\
  &+ \frac{3}{Q} \left[ \frac{I^2 + J^2}{I - J}
- \frac{I^2 - J^2}{Q}
+ \frac{\left( I - J \right)^3}{3 Q^2} \right]
\ln{\frac{I}{J}}
+ \frac{r}{Q^3}\, f \left( t, r \right),
\end{split}
\end{equation}
where
\begin{equation}
  t \equiv I + J - Q 
  \quad \text{and} \quad
  r \equiv Q^2 - 2 Q (I + J) + (I - J)^2,
\end{equation}
\begin{equation}
\begin{split}
  f \left( t, r \right) \equiv \left\{ \begin{array}{lcl}
{
\sqrt{r}\, \ln{\left| \frac{t - \sqrt{r}}{t + \sqrt{r}} \right|}
} & \Leftarrow r > 0,
\\*[3mm]
0 & \Leftarrow r = 0,
\\*[2mm]
{
2\, \sqrt{-r}\, \arctan{\frac{\sqrt{-r}}{t}}
} & \Leftarrow  r < 0.
\end{array} \right.
\end{split}
\end{equation}

In terms of these functions, the non standard model contributions to the $S$ and $T$ parameters for the $\Z_{3}$ invariant potential \eqref{eq:V:Z:3} are
\begin{equation}
\begin{split}
  \Delta S &= \frac{1}{24 \pi} \left[ (2 s_{W}^2 -1)^2 \, G(M_{H^{\pm}}^{2},M_{H^{\pm}}^{2},M_{Z}^{2}) 
  + \cos^{4} \theta \, G(M_{x_{2}}^{2},M_{x_{2}}^{2},M_{Z}^{2}) \right. \\ 
  & + 2 \sin^{2} \theta \cos^{2} \theta \, G(M_{x_{1}}^{2},M_{x_{2}}^{2},M_{Z}^{2}) 
  + \sin^{4} \theta \, G(M_{x_{1}}^{2},M_{x_{1}}^{2},M_{Z}^{2}) \\
  & \left. + 2 \sin^{2} \theta \ln M_{x_{1}}^{2} + 2 \cos^{2} \theta \, \ln M_{x_{2}}^{2} 
  - 2 \ln M_{H^{\pm}}^{2} \right] \\
  &\approx \frac{1}{24 \pi}
  \left[ (2 s_{W}^2 -1)^2 \, G(M_{H^{\pm}}^{2},M_{H^{\pm}}^{2},M_{Z}^{2}) 
  + G(M_{x_{2}}^{2},M_{x_{2}}^{2},M_{Z}^{2}) \right. \\
  &\left. + 2 \ln M_{x_{2}}^{2} - 2 \ln M_{H^{\pm}}^{2} \right],
\end{split}
\end{equation}
and
\begin{equation}
\begin{split}
  \Delta T &= \frac{1}{16 \pi^{2} \alpha v^{2}} \left[ \sin^{2} \theta \, F(M_{H^{\pm}}^{2}, M_{x_{1}}^{2})  
  + \cos^{2} \theta \, F(M_{H^{\pm}}^{2}, M_{x_{2}}^{2})  \right. \\
  &\left. - \sin^{2} \theta \cos^{2} \theta \, F(M_{x_{1}}^{2}, M_{x_{2}}^{2})  \right] \\
  &\approx \frac{1}{16 \pi^{2} \alpha v^{2}} F(M_{H^{\pm}}^{2}, M_{x_{2}}^{2}) \approx \frac{1}{16 \pi^{2} \alpha v^{2}} \frac{2}{3} \frac{(M_{H^{\pm}}^{2} - M_{x_{2}}^{2})^{2}}{M_{x_{2}}^{2} + M_{H^{\pm}}^{2}},
\end{split}
\end{equation}
where in the limit of vanishing $\theta$ only the middle term survives. We see that the $T$ parameter is in general positive and that $M_{H^{\pm}}$ cannot be too different from $M_{x_{2}}$.

For the $\Z_{4}$ invariant potential \eqref{eq:pot:3:Z:4}, the complex singlet does not mix with the neutral components of the doublet and does not contribute to the EWPT at 1-loop level. The $S$ parameter is 
\begin{equation}
\begin{split}
  \Delta S &= \frac{1}{24 \pi} \left[ (2 s_{W}^{2} -1)^{2} G(M_{H^{\pm}}^{2},M_{H^{\pm}}^{2},M_{Z}^{2}) 
  + G(M_{H^{0}}^{2},M_{A^{0}}^{2},M_{Z}^{2}) \right. \\
  & \left. +\ln M_{A^{0}}^{2} + \ln M_{H^{0}}^{2} - 2 \ln M_{H^{\pm}}^{2} \right].
\end{split}
\end{equation}
  For the $T$ parameter, we reproduce the result of \cite{Barbieri:2006dq}:
\begin{equation}
  \Delta T = \frac{1}{32 \pi^{2} \alpha v^{2}} \left[ F(M_{H^{\pm}}^{2}, M_{H^{0}}^{2}) + F(M_{H^{\pm}}^{2}, M_{A^{0}}^{2}) - F(M_{H^{0}}^{2}, M_{A^{0}}^{2}) \right].
\end{equation}

\subsection{LEP limits}
\label{sec:LEP}

The results of precision measurements at the Large Electron-Positron Collider (LEP) exclude decays of the SM $Z$ and $W^{\pm}$ bosons into invisible particles, requiring \cite{Gustafsson:2007pc,Cao:2007rm} $M_{H^{\pm}} + M_{H^{0},A^{0}} > M_{W^{\pm}}$ and $M_{H^{0}} + M_{A^{0}}, 2 M_{H^{\pm}} > M_{Z}$. The searches for charginos and neutralinos have allowed to derive two additional indirect bounds: $M_{H^{\pm}} > 70-90$~GeV \cite{Pierce:2007ut} and exclusion of masses in the region \cite{Lundstrom:2008ai}
\begin{equation}
  M_{H^{0}} < 80~\text{GeV} ~\land~ M_{A^{0}} < 100~\text{GeV} ~\land~ M_{A^{0}} - M_{H^{0}} > 8~\text{GeV}.
\end{equation}

\subsection{Higgs diphoton signal and invisible decays}
\label{sec:higgs}

In these models the Higgs couplings are SM-like, except for the radiatively generated diphoton coupling which can receive a contribution from the charged Higgs. Modifications to the $h \to \gamma \gamma$ rate are expected to be large only for a light charged Higgs,
as in the inert doublet model \cite{Arhrib:2012ia,Posch:2010hx,Cao:2007rm,Borah:2012pu,Swiezewska:2012eh}.
 
The fit to the latest experimental data from TeVatron \cite{Aaltonen:2013kxa}, ATLAS \cite{Aad:2013wqa,ATLAS-CONF-2013-108,ATLAS-CONF-2013-079,ATLAS-CONF-2013-010,ATLAS-CONF-2013-009,ATLAS-CONF-2013-014} and CMS \cite{cms:2013:bosonic,cms:2013:comb:1,cms:2013:comb:2,cms:2013:fermions,CMS-PAS-HIG-12-050,CMS-PAS-HIG-13-001,Chatrchyan:2013mxa,Chatrchyan:2013iaa,CMS-PAS-HIG-13-004,cms:2013:tautau:update,Chatrchyan:2013zna,CMS-PAS-HIG-13-006,CMS-PAS-HIG-13-009}
 gives for the diphoton rate
\begin{equation}
  R_{\gamma\gamma} = 1.06 \pm 0.10,
\end{equation}
if all the other rates are fixed to their SM values \cite{Giardino:2013bma}.

Furthermore there is the possibility of invisible Higgs decays with the 125 GeV Higgs decaying into the singlet or scalar/pseudoscalar components of the doublet: the invisible branching ratio is $\text{BR}_{\text{inv}} < 0.24$ at $95\%$~C.L. \cite{Giardino:2013bma, Belanger:2013xza}. In the $\Z_3$ model this range is most likely ruled out by direct detection as seen previously  in the $\Z_{3}$ singlet dark matter model \cite{Belanger:2012zr}), while a large invisible rate can be generated in the $Z_4$ model as will be discussed below.


\subsection{Cosmic density of dark matter}
\label{sec:cosmic:density}

\label{sec:cosmic:density}
The PLANCK collaboration has recently released results for the cosmological parameters, in particular for the DM relic density~\cite{Ade:2013zuv}. When averaged with the WMAP-9 year data ~\cite{Hinshaw:2012aka}, it leads to the very precise value

\begin{equation}
\Omega h^2= 0.1199\pm  0.0027.
\end{equation}
We will use the 3$\sigma$ range below.

To compute the relic density in the $\Z_3$ model, we use micrOMEGAs\verb|_3.5| which takes into account all annihilation, coannihilation and semi-annihilation  channels \cite{Belanger:2013oya}. Final states with virtual gauge bosons that can be present  in (co-)annihilation and semi-annihilation  processes are also included.

In the $\Z_4$ model there can be two dark matter candidates:  the singlet with $X=1$  and the lightest component of the doublet, $H^{0},A^{0}$, with $X=2$. To compute the relic density, we use the generalized equations for the abundance $Y_i=n_i/s$:
\begin{eqnarray}
\label{z4eq1}
3H\frac{dY_1}{ds}&=&\sigma_v^{1100} \left(Y_1^2-\overline{Y}_1^2
\right) 
+   \sigma_v^{1120} \left(Y_1^2-Y_2 \frac{\overline{Y}_1^2}{\overline{Y}_2} 
\right) +\sigma_v^{1122}
\left( Y_1^2-Y_2^2 \frac{\overline{Y}_1^2}{\overline{Y}_2^2}\right),\\
\label{z4eq2}
3H\frac{dY_2}{ds}&=&\sigma_v^{2200} \left(Y_2^2-\overline{Y}_2^2
\right)
  -\frac{1}{2} \sigma_v^{1120} \left(Y_1^2-Y_2
   \frac{\overline{Y}_1^2}{\overline{Y}_2} 
\right) 
+ \frac{1}{2}  \sigma_v^{1210} Y_1 \left(Y_2-\overline{Y}_2 \right) \nonumber\\
&&+\sigma_v^{2211}
\left( Y_2^2-Y_1^2 \frac{\overline{Y}_2^2}{\overline{Y}_1^2}\right),
\end{eqnarray}
where  $\sigma_v^{abcd}$ stands for the thermally
averaged  cross section $\langle \sigma v\rangle$ for the reactions $ab\rightarrow cd$,  $a,b,c,d=0,1,2$ represent any particle with a given $X$ (SM particles have $X=0$), and
$\overline{Y}_a$ are the equilibrium abundances.
In $\sigma_v^{abcd}$ all annihilation and coannihilation processes are taken into account as well as annihilation into virtual gauge bosons. Semi-annihilation processes include all those where 2 DM particles annihilate into one DM and one standard particle, specifically 
$\sigma_v^{1110}$, $\sigma_v^{2220}$, $\sigma_v^{1120}$ and $\sigma_v^{1210}$. 
The method of solution for these equations as implemented in micrOMEGAs is described in \cite{Belanger:2012vp} and \cite{micromegas:in:prep}.

The abundances $Y_1$ and $Y_2$ will be modified by the interactions between the two dark sectors. After the light DM  freezes-out interactions such as $\mathpzc{h}\mathpzc{h}\rightarrow \mathpzc{l}\mathpzc{l}$ lead to a decrease of the abundance of the heavy component $\mathpzc{h}$ and to an increase in the light component $\mathpzc{l}$.
Such is also the case for semi-annihilation processes of the type $\mathpzc{h}\mathpzc{h}\rightarrow \mathpzc{l} 0$ (or its reverse $\mathpzc{h}0\rightarrow \mathpzc{l}\mathpzc{l}$).
The semi-annihilation process $12\rightarrow 10$ has no influence on $Y_1$ while leading to a decrease of $Y_2$. 
Note that this process is always kinematically open unless the only SM particle in the final state is heavier than $H^{0},A^{0}$. 
Finally, semi-annihilations involving only particles of a given sector always lead to a decrease of the abundance of the corresponding dark matter.

\subsection{Dark matter direct detection}
\label{sec:direct:detection}

The best upper limit on the spin independent (SI) scattering cross section on nuclei has been obtained by the LUX experiment \cite{Akerib2013111}: $\sigma_{\text{SI}} < 7.6 \times 10^{-46}~{\rm cm}^2$ for $M_{\text{DM}}= 33$~GeV. We also show the results from XENON100 (2012) \cite{Aprile:2012nq}.
Future detectors such as SuperCDMS(SNOLAB) \cite{supercdms}, XENON1T \cite{Aprile:2012zx} or  LZD \cite{2011arXiv1110.0103M} will increase the sensitivity by one to four orders of magnitude.

To compute the model predictions for the SI cross section we use micrOMEGAs\verb|_3.5|, and we assume that both DM and anti-DM have the same local density. In the $\Z_{3}$ model,
  the DM candidate $x_1$ a mixture of the complex doublet and singlet scalar and  there can be large differences in the scattering rate on protons and neutrons.  To compare directly with the  experimental limits,  which are obtained assuming isospin conservation, we compute the normalised cross section of DM on a point-like nucleus (that we take to be xenon)
 \begin{equation}
\sigma_{\rm SI}^{x \text{Xe}}= \frac{4 \mu_x^2}{\pi}\frac{\left( Z f_p+ (A-Z)f_n\right)^2}{A^2 }
\end{equation}
where $x$ denotes the DM candidate, $\mu_x$ the reduced mass, $f_p,f_n$ the amplitudes for protons and neutrons,  
and  the average over $x$ and $x^*$ is assumed implicitly.

In the $\Z_4$ model  with two dark matter candidates, we also compute the normalised cross section on xenon for each dark matter candidate after rescaling by the relative density of each component.

\subsection{Dark matter indirect detection}
\label{sec:indirect:detection}

The annihilation of DM in the Milky Way halo can lead to excesses in the cosmic ray fluxes  (photons, positrons, antiprotons) that
provide indirect evidence for DM. The measurements of the gamma-ray flux from Dwarf Spheroidal Galaxies by the Fermi-LAT satellite  provide the most stringent constraint for light DM annihilating into $bb$ or $\tau\tau$ \cite{Ackermann:2013yva}.  The canonical cross-section, $\langle \sigma v\rangle \approx 3\times 10^{-26}~{\rm cm}^3/{\rm s}$  is ruled out for  DM masses below 30 GeV~\cite{Ackermann:2011wa}.
For heavier DM, the  measurements of the antiproton flux from PAMELA~\cite{Adriani:2010rc}, for the MED set of propagation parameters~\cite{Donato:2003xg} have roughly the same sensitivity as Fermi-LAT, with a  limit of $\sigma v\approx 10^{-25}~{\rm cm}^3/{\rm s}$ for a DM mass of 200 GeV in the $bb,\tau\tau$ or $WW$ channels~\cite{Cirelli:2013hv}. Since in our models the light masses are severely constrained by the Higgs invisible width we will in the numerical analysis consider only the PAMELA limit from antiprotons and we will assume the MED set of propagation parameters. 

The measurements of the positron spectra  by PAMELA and AMS are very powerful to constrain models which favour DM
annihilation into $e^+e^-$ or $\tau^+\tau^-$ pairs, this is not the case in our models, we will therefore not consider these signatures.

\section{Results for the $\Z_{3}$ model}
\label{sec:Z:3}

To explore the phenomenology of the model, we perform a random scan over the parameter space. The masses and the cubic term are generated with uniform distribution in the ranges $1~\text{GeV} < M_{x_1}<1000~\text{GeV}$, $1~\text{GeV}<M_{x_2}, M_{H^\pm} < 2000~\text{GeV}$, $0~\text{GeV} < \mu''_S < 3500~\text{GeV}$, and $124~\text{GeV}< M_h <127~\text{GeV}$. The range of the mixing angle $\theta$ between the neutral components of $H_2$ and $S$ is $0 \leqslant \theta \leqslant \pi/2$ and we choose $M_{x_{1}} < M_{x_{2}}$ without loss of generality. In practice, we take the mixing angle in the range $0 \leqslant \theta \leqslant 0.06$ with uniform distribution. This guarantees that the DM candidate $x_1$ is dominantly singlet-like and so does not lead to a too large direct detection rate.

The quartic couplings are generated with triangular distribution (with the mode at zero) in the ranges allowed by perturbativity:\footnote{Since the mixing is very small, we ignore it in the derivation of the perturbativity bounds.}
\begin{equation}
\begin{aligned}
  \abs{\lambda_{1}} &< \frac{2 \pi}{3}, & \abs{\lambda_{2}} &< \pi, & \abs{\lambda_{3}} &< 4 \pi,
  & \abs{\lambda_{4}} &< 4 \sqrt{2} \pi, & \abs{\lambda_{3} + \lambda_{4}} &< 4 \pi, 
  \\
  \abs{\lambda_{S}} &< \pi, & \abs{\lambda_{S1}} &< 4 \pi, 
  & \abs{\lambda_{S2}} &< 4 \pi, & \abs{\lambda_{S12}} &< 4 \pi,
\end{aligned}
\end{equation}
except for $\lambda_{4}$ from \eqref{eq:Z3:def:lambda:4} and $\lambda_{1}$ from \eqref{eq:Z3:def:lambda:1}, the perturbativity of which is subsequently checked.

We then apply the unitarity, vacuum stability and globality bounds, the upper limit on the Higgs invisible decays, the 3$\sigma$ range for the DM relic density and the 3$\sigma$ range for the electroweak precision parameters $S$ and $T$. We present results for points that satisfy this set of constraints. We thten compute the Higgs diphoton signal strength as well as the DM  direct detection rate and self-annihilation cross section relevant for indirect detection.

To emphasize the role of semi-annihilation on the DM properties in  the model, we will  present our results with a colour code characterising the fraction  of semi-annihilation defined as
\begin{equation}
\alpha=\frac{1}{2} \frac{v\sigma^{x x\to x^* X}}{v\sigma^{x x^*\to XX}+ \frac{1}{2}
v\sigma^{x x\to x^*X }},
\end{equation}
where $x$ stands for $x_1,x_2,H^+$ and $X$ for any SM particle.
The dominant DM annihilation processes lead to gauge boson  and Higgs pairs while
the dominant semi-annihilation process is generally $x_1 x_1\to x_1 h$ which is is relevant for $M_{x_{1}} > M_{h}$. 
Other semi-annihilation processes  such as  $x_1 x_1 \to x_2 Z$, $x_1 x_1 \to x_2 h$ or $x_1 x_1 \to H^{\pm} W^\mp$
can dominate when $M_{x_2,H^{\pm}}< 2 M_{x_1}$. When the singlet-doublet mixing is small these processes depend on $\lambda_{S12}$ and/or $\mu''_S$.  For small mass differences between $x_1$ and $x_2$~(and/or $H^\pm$) coannihilation occurs and 
 new semi-annihilation processes  become possible, for example $x_1 x_2\to Z x_1, h x_1$ or even $x_1 H^+ \to x_1 W^+$.

\begin{figure}[h!]
\begin{center}
  \includegraphics[height=0.29\textheight]{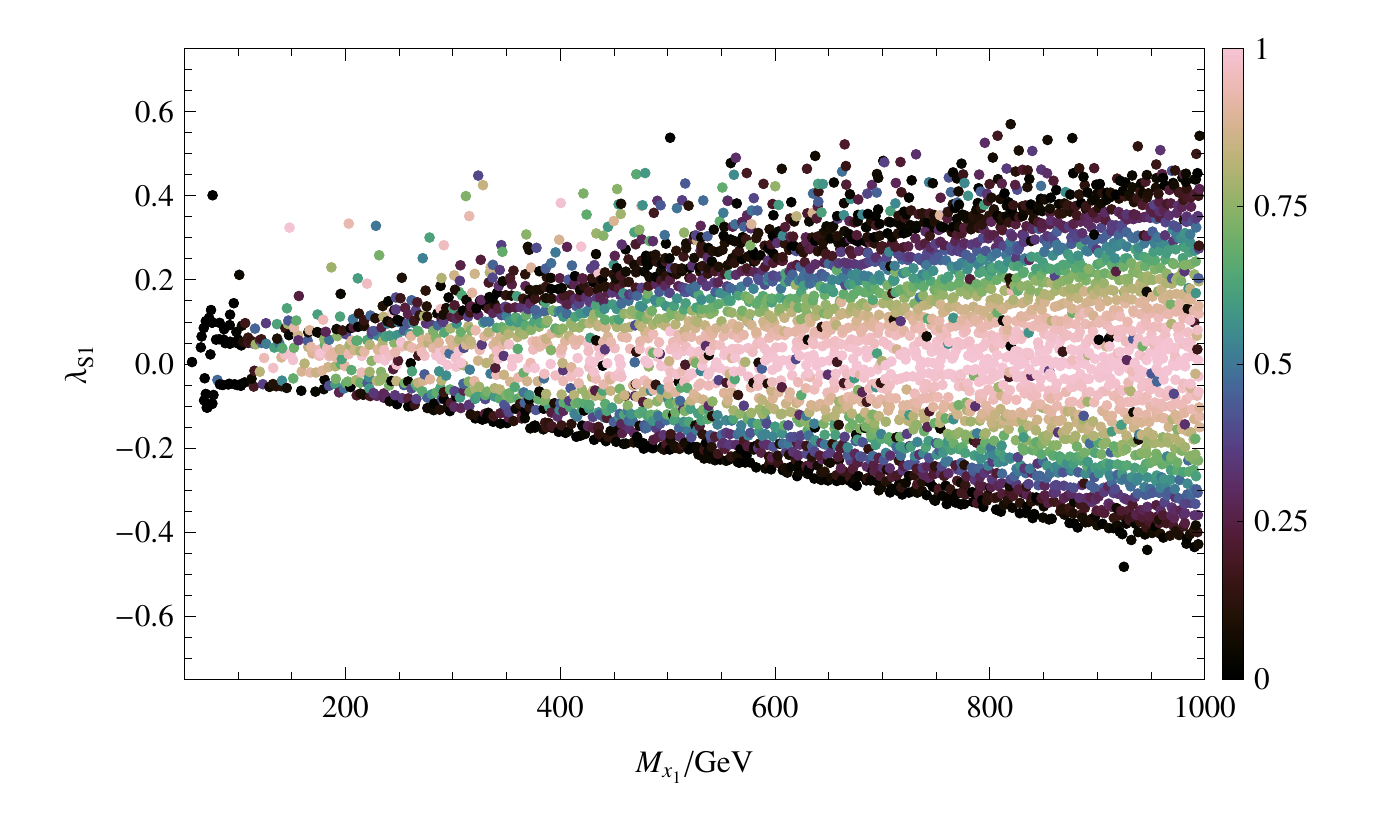}
  \includegraphics[height=0.29\textheight]{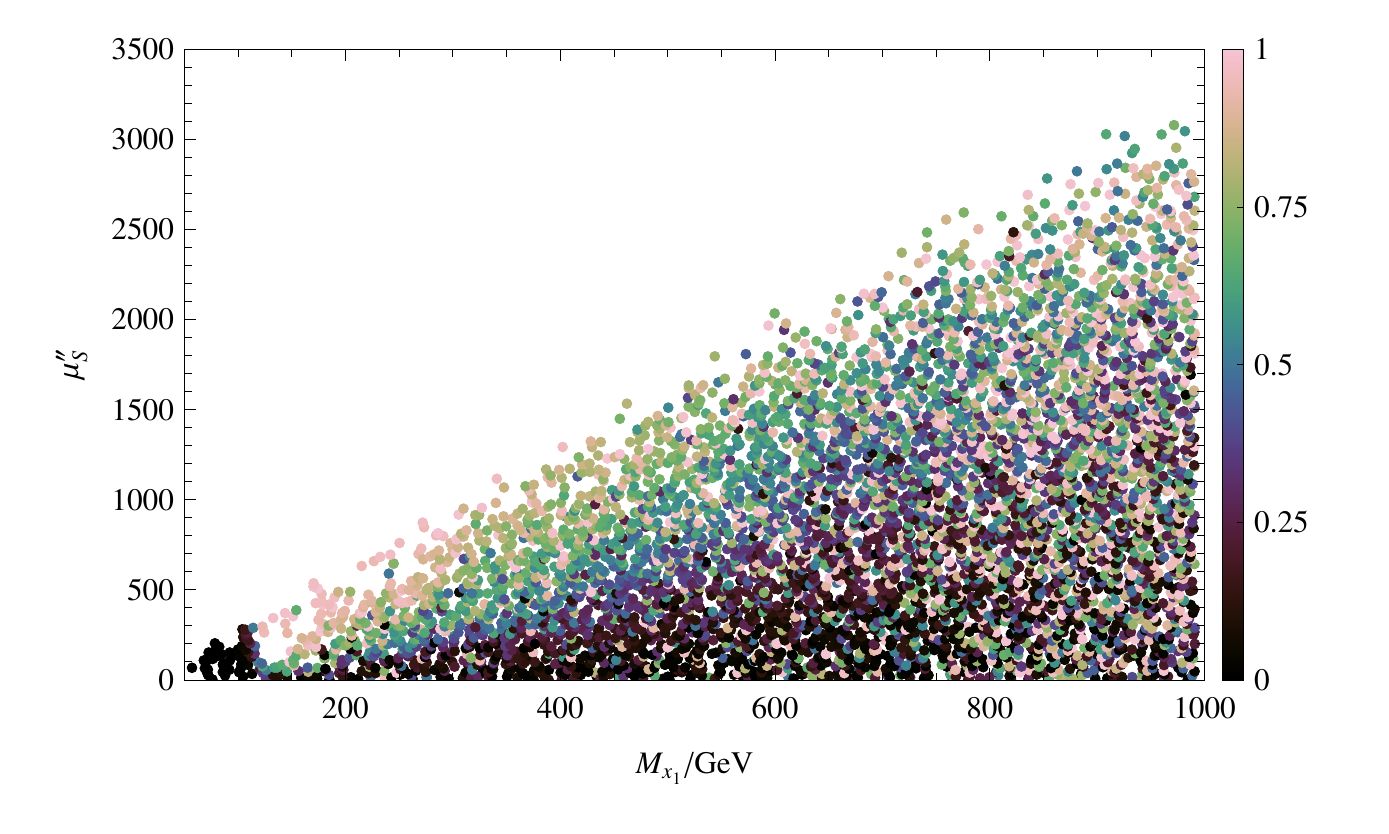}
  \includegraphics[height=0.29\textheight]{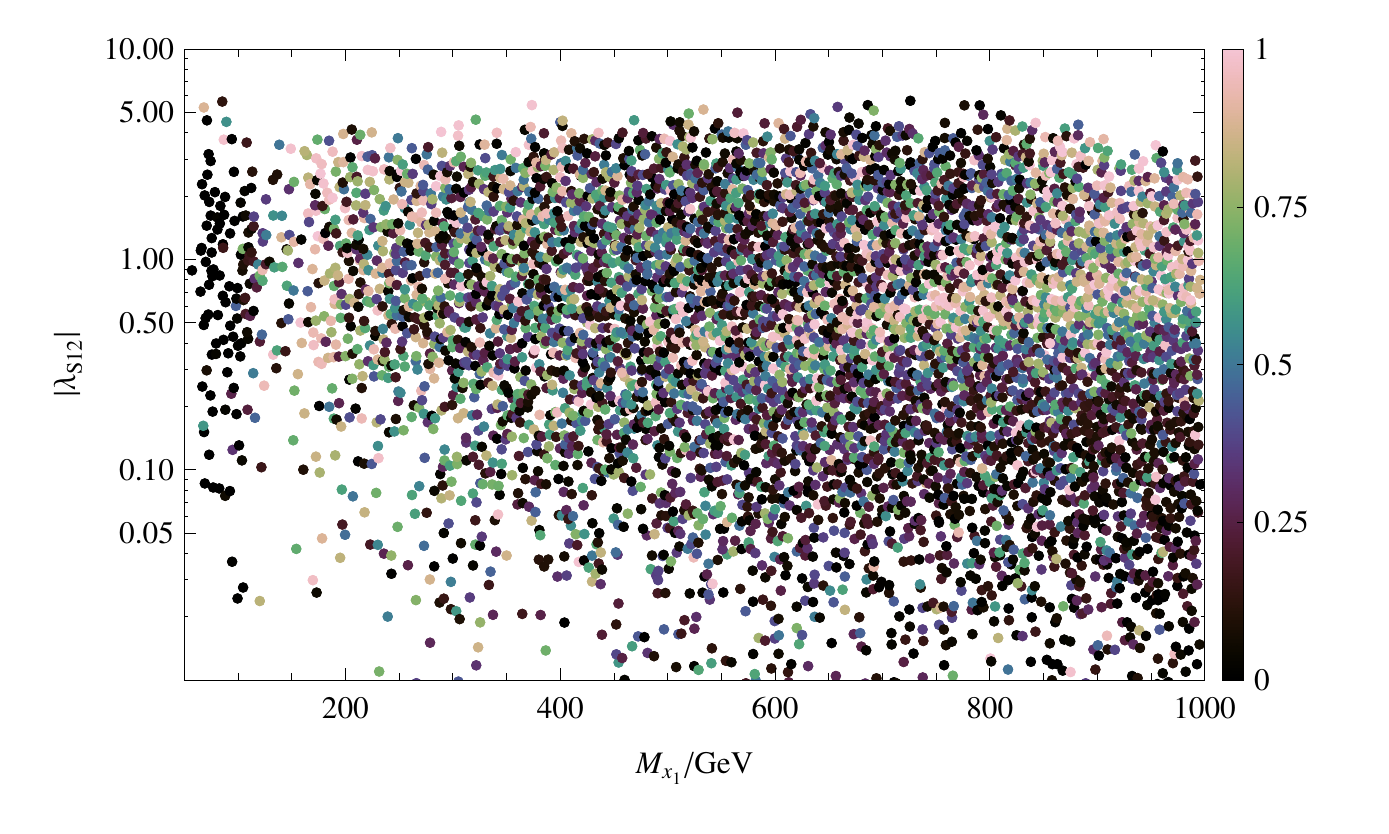}
\caption{Top: The Higgs-singlet coupling $\lambda_{S1}$ that determines the DM annihilation cross section vs. $M_{x_1}$. Middle and bottom: The parameters that bring about semi-annihilation, $\mu''_S$ and $|\lambda_{S12}|$, vs. $M_{x_1}$. The colour code shows the fraction of semi-annihilation $\alpha$.}
\label{fig:Z3:couplings:vs:Mx1}
\end{center}
\end{figure}

In figure~\ref{fig:Z3:couplings:vs:Mx1} we show the regions allowed by our  set of constraints   in the planes $\lambda_{S1}$, $\mu''_S$ and $\lambda_{S12}$ vs. $M_{x_1}$.
Basically it is possible to satisfy the Planck constraint for any mass of DM, while the upper bound on the Higgs invisible decay rules out all DM masses below $\approx 50~$GeV. The same range of masses are also incompatible with the upper limit on the direct detection rate as will be discussed in Section~\ref{sec:z3_dm}. Similarly to the case of the pure $\Z_3$ singlet DM \cite{Belanger:2012zr}, the relic density constraint determines the range of allowed values for $\lambda_{S1}/M_{x1}$, the combination of parameters that control DM annihilation, while smaller values of $\lambda_{S1}$ are possible when semi-annihilation is important.  Significant contributions from semi-annihilation is associated with large values for  $\mu''_S$ and/or $\lambda_{S12}$, 
the former contributing to the semi-annihilation process $x_1 x_1 \to x_1^* h$ (which is more important for relatively light DM masses), while the latter to processes with the dominantly doublet Higgs in the final state, these occur only if $M_{x_2,H^\pm}+M_{\rm SM}<2M_{x_1}$ where SM refers to the scalar or gauge boson produced in the semi-annihilation process. The requirement that the SM vacuum be the global one constrains the possible maximum value of $\mu''_S$. A few benchmarks satisfying these constraints are described in Appendix~\ref{sec:benchmarks}.

\subsection{Higgs and electroweak precision parameters}

\begin{figure}[tb]
\begin{center}
  \includegraphics[scale=0.7]{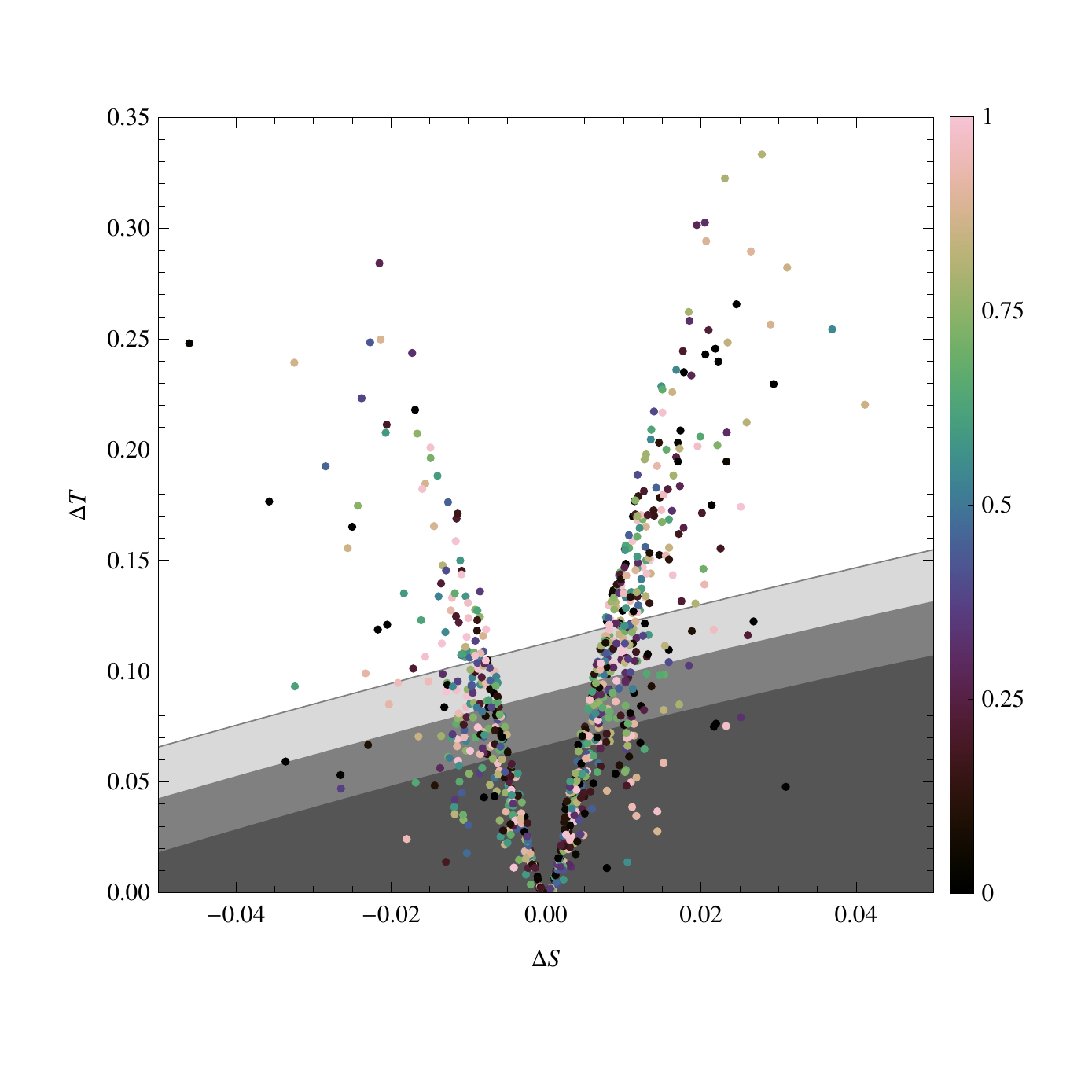}
  \vspace{-1cm}
\caption{Electroweak precision parameters $\Delta T$ vs. $\Delta S$  for the $\Z_3$ model. The grey regions show the 1, 2, and 3 $\sigma$ bounds \cite{Baak:2012kk}. The colour code shows the fraction of semi-annihilation $\alpha$.}
\label{fig:Z3:T:vs:S}
\end{center}
\end{figure}

Figure~\ref{fig:Z3:T:vs:S} shows the results of the electroweak precision parameters $S$ and $T$. Due to the fact that the allowed mixing angle is very small, the parameters virtually do not depend on the mass of the singlet-like DM. The constraint on the $T$ parameter  basically imposes $\abs{M_{H^{\pm}} - M_{x_{2}}} \lesssim 120$~GeV on  the mass splitting of $H^{\pm}$ with the doublet-like neutral scalar $x_{2}$. Therefore although it restricts the parameter space there is no direct correlation with other observables, in particular the spin-independent direct detection cross section $\sigma_{\text{SI}}$. In the following figures we impose the $3\sigma$ constraint from the $S$ and $T$ parameters.  

\begin{figure}[hbt]
\begin{center}
  \includegraphics{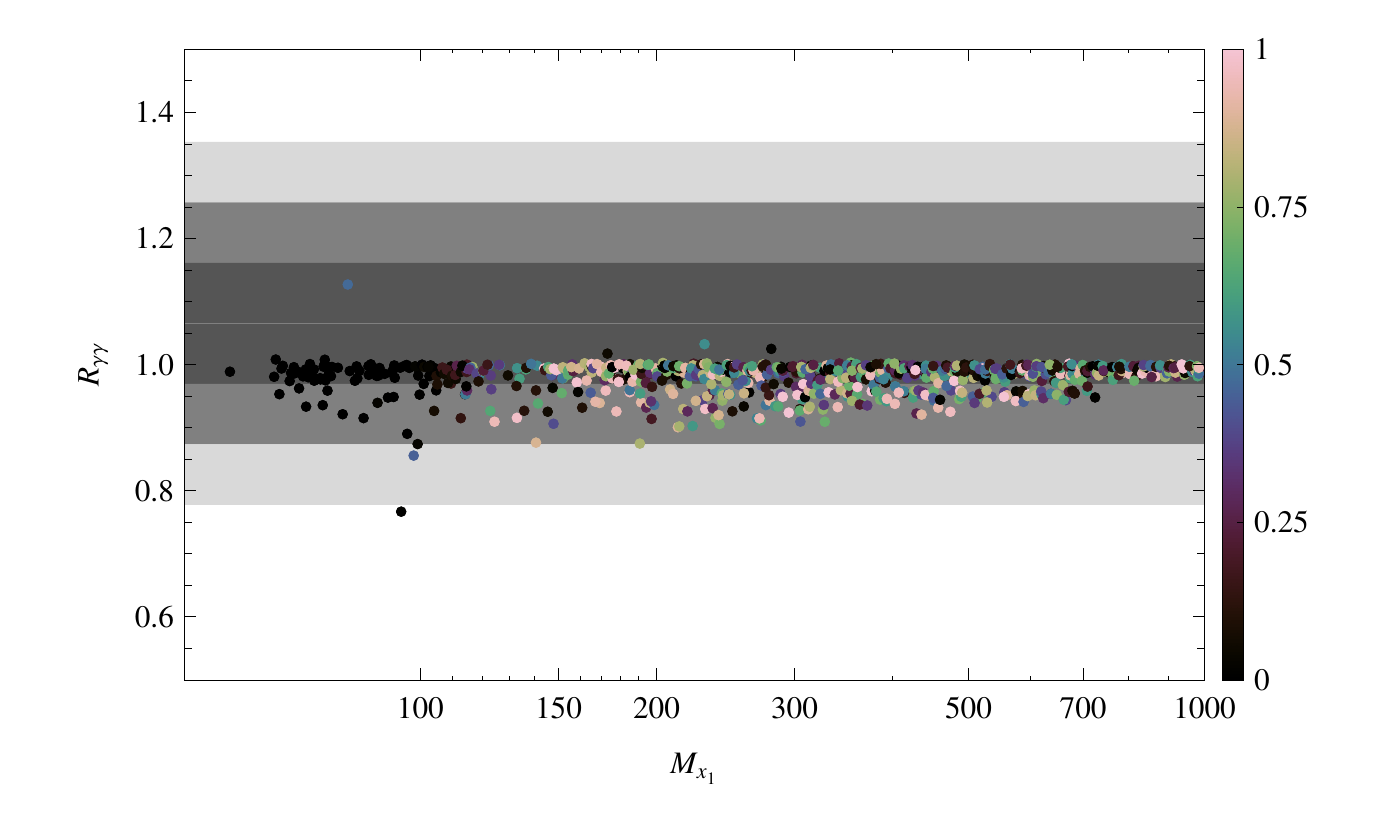}
  \vspace{-0.8cm}
\caption{The $h \to \gamma \gamma$ rate for the $\Z_3$ model normalised to the SM. The thick grey line is the central value from a combined fit of collider data, the coloured bands show 1, 2, and 3 $\sigma$. The colour code shows the fraction of semi-annihilation $\alpha$.}
\label{fig:Z3:Rgg:vs:Mx1}
\end{center}
\end{figure}

Figure~\ref{fig:Z3:Rgg:vs:Mx1} shows the $h \to \gamma \gamma$ signal strength (normalised to the SM) vs. the DM mass. The rate is systematically below the current average, and generally within the $2\sigma$ error band. Nevertheless there can be larger deviations from the SM for low DM masses.  Note that here the constraint from invisible Higgs width has been applied.

\subsection{Dark matter observables}
\label{sec:z3_dm}

\begin{figure}[htb]
\begin{center}
  \includegraphics{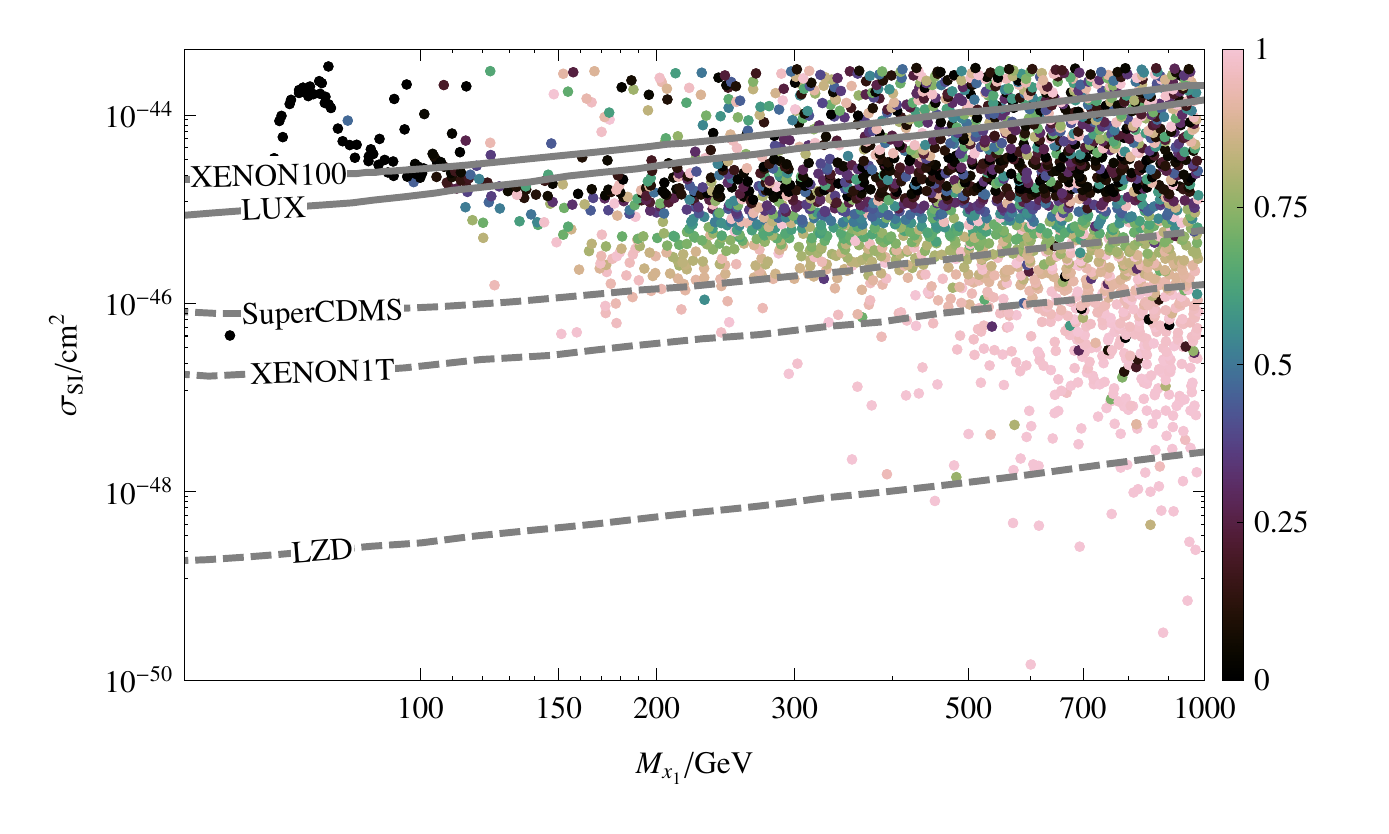}
\caption{Spin-independent direct detection cross section on xenon, $\sigma_{\text{SI}}$ vs. $M_{x_{1}}$ for the $\Z_3$ model. The solid grey lines are the XENON100 (2012) \cite{Aprile:2012nq} and LUX \cite{Akerib2013111} exclusion limits at $90\%$~C.L. and the dashed grey lines are the projected $90\%$~C.L. exclusion limits for  SuperCDMS(SNOLAB) \cite{supercdms}, XENON1T \cite{Aprile:2012zx} or  LZD \cite{2011arXiv1110.0103M}. The colour code shows the fraction of semi-annihilation $\alpha$.}
\label{fig:Z3:sigma:SI:vs:Mx1}
\end{center}
\end{figure}

In figure~\ref{fig:Z3:sigma:SI:vs:Mx1} we show the results of the DM spin-independent cross section $\sigma_\text{SI}$ vs. the DM mass $M_{x_1}$. The colour variation from black to pink (black to light grey) shows the fraction of semi-annihilation. The solid grey lines are the XENON100 (2012) \cite{Aprile:2012nq} and LUX \cite{Akerib2013111} exclusion limits at $90\%$~C.L. and the dashed grey lines are the projected $90\%$~C.L. exclusion limits for  SuperCDMS(SNOLAB) \cite{supercdms}, XENON1T \cite{Aprile:2012zx} or LZD \cite{2011arXiv1110.0103M} (which will be at the limit of liquid xenon based experiments).

The current LUX upper limit severely constrains  $M_{x_1}<120$~GeV  as well as points with a large doublet singlet mixing. 
As expected scenarios dominated by semi-annihilation lead to suppressed cross sections. 
Note also the dark points with small semi-annihilation but low $\sigma_\text{SI}$ at high DM masses. They correspond to co-annihilation -- that is, the relic density is dominated by self-annihilation of either $x_2$ or $H^{\pm}$ and the cross section for annihilation of $x_1$ is small, resulting in a small cross section with nucleons as well. Some of the points will remain out of reach of Xenon1T.

We also investigate the indirect DM signatures in this model.  The annihilation channels for DM in the Galaxy are, as in the early Universe,  often dominated by  the  $bb,\tau\tau, WW,ZZ$ channels, with some contribution from the $hh$ channel. The main new feature is the possibility of semi-annihilation with final states such as $x_1 Z,x_1 h$ or even  $x_2 Z$, $x_2 h$ or $ H^\mp W^\pm$.

\begin{figure}[htb]
\begin{center}
  \includegraphics{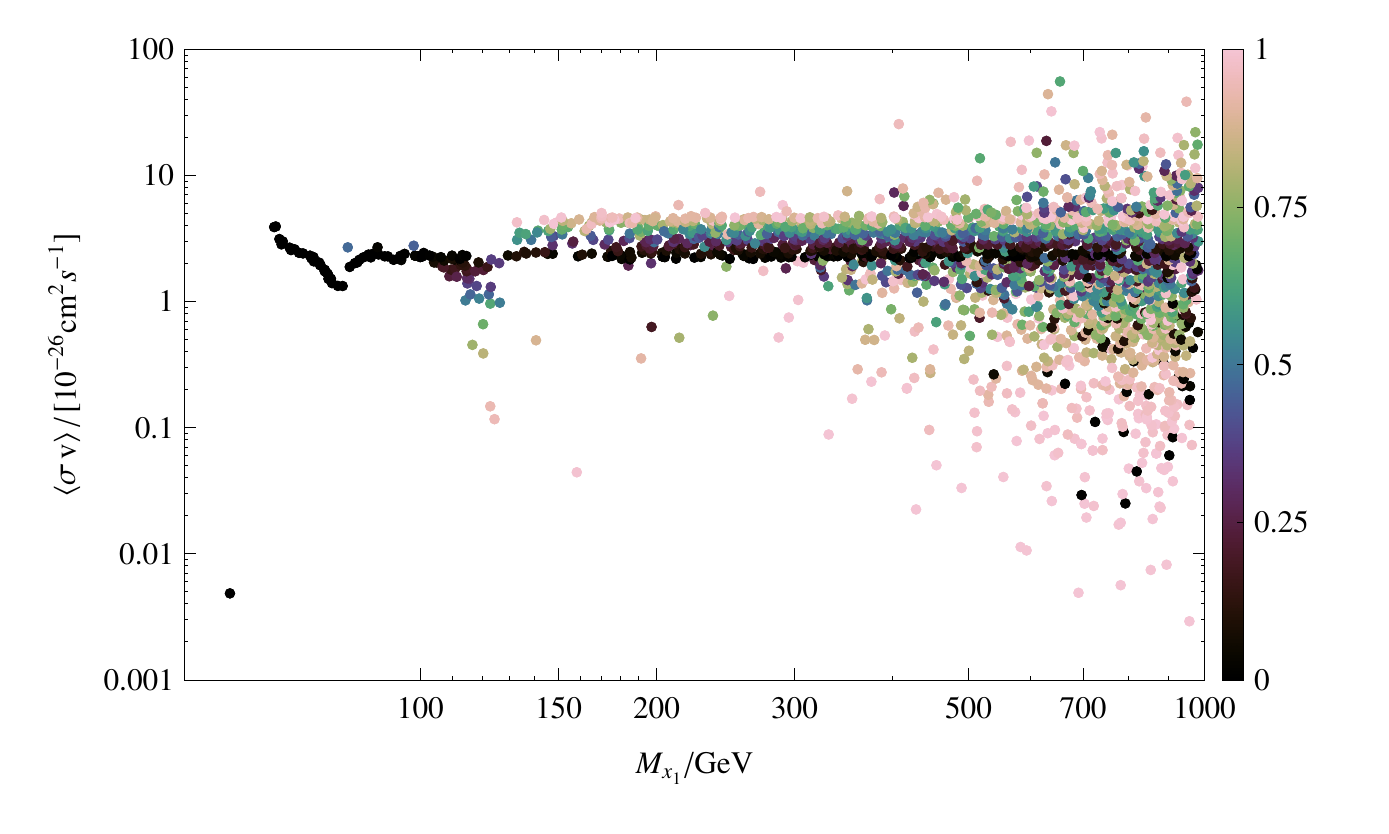}
\caption{Thermally averaged annihilation cross section $\langle \sigma v \rangle$ vs. $M_{x_{1}}$ for the $\Z_3$ model. The colour code shows the fraction of semi-annihilation $\alpha$.}
\label{fig:Z3:sigma:v:vs:Mx1}
\end{center}
\end{figure}
 
In figure~\ref{fig:Z3:sigma:v:vs:Mx1} we show the results of the DM annihilation cross section, $\langle \sigma v \rangle$ for $v\approx 10^{-3} c$, the quantity relevant for indirect detection.
We find that generally $\langle \sigma v \rangle \approx 3\, (5)\times 10^{-26}~{\rm cm}^3/{\rm s}$ when $\alpha\rightarrow 0\,(1)$,   although the predictions can span several orders of magnitude. In particular $\langle \sigma v \rangle$ is suppressed when  coannihilation dominates (indeed coannihilation is relevant only for the computation of the relic density), while  kinematic effects can lead to either a large enhancement or suppression of the (semi-)annihilation cross section.
This occurs  when $M_{x_1}\approx M_{x_2}/2$.  The cross section can be enhanced by more than one order of magnitude because of the near resonant $x_2$ exchange in the $s$-channel.  Large suppression of  $\langle \sigma v \rangle$  can also be found when the thermal annihilation relevant for the relic density benefits from a resonance effect while the cross section at small velocities in the galaxy does not for kinematical reason. 

The largest  values of $\langle \sigma v \rangle$ can potentially lead to a strong enhancement of the anti-matter flux, in particular of the anti-proton flux. 
To ascertain the viability of all our scenarios in a quantitative way, we have performed a $\chi^2$ fit to the data measured by PAMELA for energies between $10-100$ GeV. To avoid large solar modulation effects we ignore the data at lower energies. 
For the background we assume the analytical parametrisation in Ref.~\cite{Donato:2003xg} and determine the 95\%C.L. allowed region by imposing that $\chi^2=\chi^2_{\rm background}+4$. Despite the enhancement in the annihilation cross-section, we find that only a handful of points are constrained for the MED propagation parameters~\cite{Donato:2003xg}. This is because semi-annihilation channels have only one SM particle in the final state decaying into anti-matter, hence a flux reduced by a factor 2 and a softer anti-matter spectra with a shape similar to the one for DM annihilation into a pair of SM particles (here typically $Z$ or $h$). Furthermore the largest $\langle \sigma v \rangle$ are found for DM masses above a few hundred GeV where indirect detection have a reduced sensitivity. Note that  for MIN propagation parameters our results are always compatible with the background only hypothesis.

\subsection{Renormalisation group running}
\label{sec:Z3:RGE}

\begin{figure}[htb]
\begin{center}
  \includegraphics[width=0.33\textwidth]{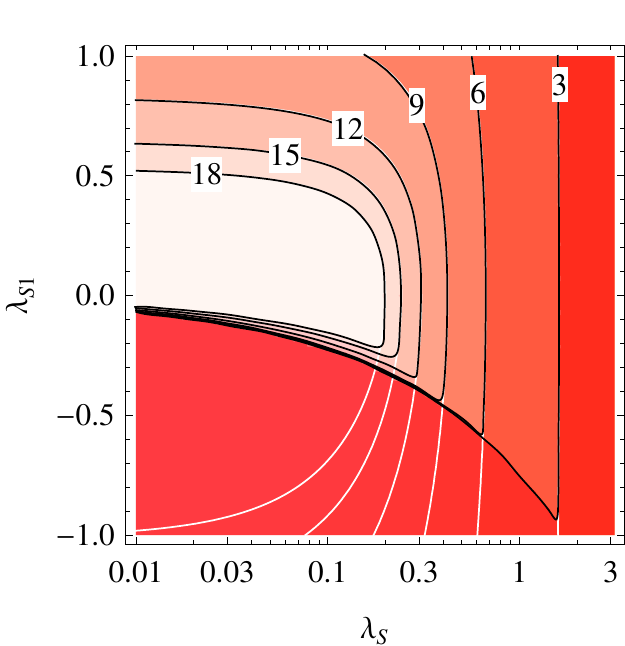}~
  \includegraphics[width=0.33\textwidth]{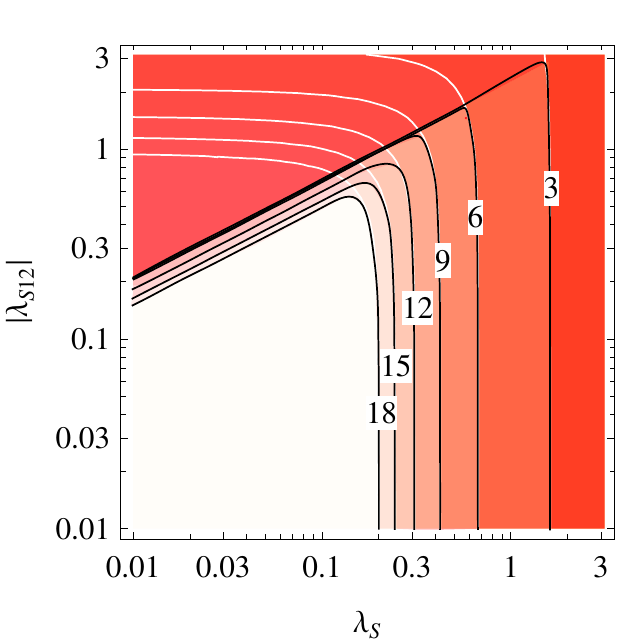}~
  \includegraphics[width=0.33\textwidth]{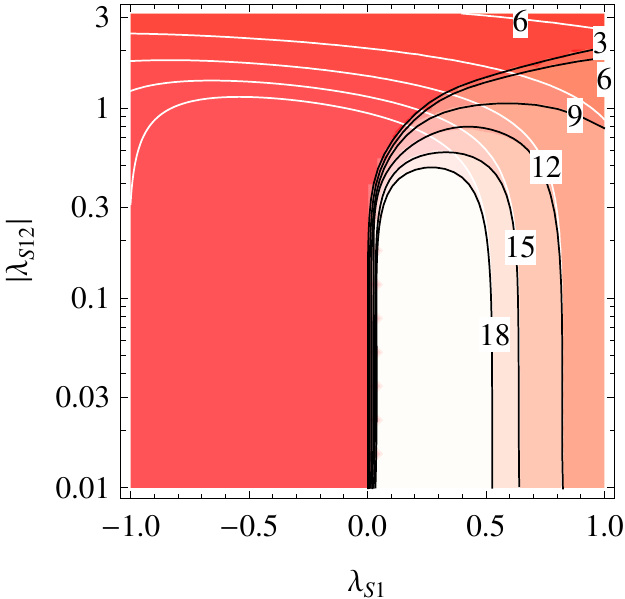}
  \vspace{0.2cm}
\caption{Vacuum stability and perturbativity bounds for the $\Z_3$ model. The values of the couplings are set at the EW scale. The black contour lines show the logarithm $\log_{10}(\Lambda/1~\text{GeV})$ of the combined bound, while the white lines show the bound from perturbativity only. All the other couplings are set to zero, except for $\lambda_3$ (and for $\lambda_{S2} = \lambda_{3}$ for the right panel) which is chosen as to maximise the bound.}
\label{fig:Z3:RGE:planes}
\end{center}
\end{figure}

The interaction couplings depend on the energy scale via renormalisation group equations (RGEs). Due to the RGE running, above some scale $\Lambda$ the model may become non-perturbative or the scalar potential may not be bounded from below. At energies greater than $\Lambda$, new physics, for example a Grand Unified Theory, is expected to appear to ensure that the full theory is perturbative and stable up to the Planck scale.

Since we are interested in the influence of semi-annihilation on the relic density and direct detection, we show in figure~\ref{fig:Z3:RGE:planes} the bounds on the $\lambda_{S1}$ vs. $\lambda_{S}$, $\lambda_{S12}$ vs. $\lambda_{S}$ and $\lambda_{S12}$ vs. $\lambda_{S1}$ planes at the EW scale.%
\footnote{ Note that $\beta_{\lambda_{S12}} \propto \lambda_{S12}$ since a non-zero $\lambda_{S12}$ means hard breaking of a global $U(1)$. If $\lambda_{S12}$ is set to zero at the EW scale, it will not be generated by radiative corrections at any other scale.} 
All the other couplings are set to zero with the exception of, obviously, $\lambda_{1}$,  and $0 \leqslant \lambda_3, \lambda_{S2} \leqslant 0.6$ which are chosen to roughly maximise the scale of vacuum stability of the model in each point (e.g. prevent the Higgs quartic coupling from running to negative values). We use the SM two-loop $\beta$-functions for the gauge couplings and the top Yukawa coupling and the one-loop $\beta$-functions \label{eq:RGEs:scalar} for the scalar quartic couplings given in Appendix~\ref{se:z3:z4:1-loop:beta:functions}. 

The white contour lines show the logarithm of the scale $\log_{10} (\Lambda$/1 GeV) where perturbativity is lost, while the black contour lines show the combined bound from loss of perturbativity and vacuum stability. 

A large fraction of semi-annihilation is associated with small values of $\lambda_{S1}$,  figure~\ref{fig:Z3:RGE:planes} shows that the model can then be valid up to the GUT scale provided $\lambda_S$ and $\lambda_{S12}$ are not too large. In fact when semi-annihilation into doublet final states is dominant (recall that this requires  $x_2$  not to be  too heavy compared to $x_1$), the value of $\abs{\lambda_{S12}}$ that produces maximal semi-annihilation is about 0.5, therefore  the model can be valid up to the GUT or  Planck scale.
For light enough DM, large semi-annihilation rather arises from the cubic $\mu''_S$ term in which case a large value of $\lambda_S$ is needed in order to for the SM vacuum to be the global minimum of the potential (see \cite{Belanger:2012zr}) and perturbativity will be lost close to the TeV scale.

\section{Results for the $\Z_{4}$ model}
\label{sec:Z:4}

We perform a random scan over the parameter space. The masses are generated with uniform distribution in the ranges $1~\text{GeV} < M_{S}, M_{H^{0}}, M_{A^{0}}<1000~\text{GeV}$, $1~\text{GeV}< M_{H^\pm} < 2000~\text{GeV}$, and the Higgs mass $124~\text{GeV}< M_h <127~\text{GeV}$. We consider only the cases when $M_{H^{0},A^{0}} < 2 M_{S}$, since otherwise the neutral doublet would decay before the freeze-out of $S$ and the situation would therefore be analogous to the one particle DM model. Here and below, $M_{H^{0},A^{0}} \equiv \min(M_{H^{0}}, M_{A^{0}})$ stands for the mass of the lighter neutral component of the doublet $H_{2}$.
The quartic couplings are generated with triangular distribution (with the mode at zero) in the ranges allowed by perturbativity:
\begin{equation}
\begin{aligned}
  \abs{\lambda_{1}} &< \frac{2 \pi}{3},\hspace{-2mm} 
  & \abs{\lambda_{2}} &< \frac{2 \pi}{3},\hspace{-2mm} 
  & \abs{\lambda_{3}} &< 4 \pi,\hspace{-2mm}
  & \abs{\lambda_{4}} &< 4 \pi,\hspace{-2mm}
  & \abs{\lambda_{5}} &< 2 \pi,\hspace{-2mm}
  \\
  \abs{\lambda_{3} + \lambda_{4}} &< 4 \pi,\hspace{-2mm} 
  & \abs{\lambda_3 + \lambda_4 \pm \lambda_5} &< 4 \pi,\hspace{-2mm} 
  & \abs{\lambda_4 \pm \lambda_5} &< 4 \pi,\hspace{-2mm}
  & \abs{\lambda_{S}} &< \pi,\hspace{-2mm} 
  & \abs{\lambda'_{S}} &< \frac{\pi}{3},\hspace{-2mm} 
  \\
  \abs{\lambda_{S1}} &< 4 \pi,\hspace{-2mm} 
  & \abs{\lambda_{S2}} &< 4 \pi,\hspace{-2mm} 
  & \abs{\lambda_{S12}} &< 4 \pi,\hspace{-2mm}
  & \abs{\lambda_{S21}} &< 4 \pi,\hspace{-2mm}
  &\abs{\lambda_{S12} - \lambda_{S21}} &< 2 \pi,\hspace{-2mm}
\end{aligned}
\end{equation}
except for $\lambda_{1}$, $\lambda_{4}$ and $\lambda_{5}$ which are derived from the free parameters, \eqref{eq:Z4:def:lambda:1},  \eqref{eq:Z4:def:lambda:4} and \eqref{eq:Z4:def:lambda:5}, for these the perturbativity condition is subsequently checked.

We then apply the unitarity, vacuum stability and globality bounds, the upper limit on the Higgs invisible decays, the 3$\sigma$ range for the DM relic density and the 3$\sigma$ range for the electroweak precision parameters $S$ and $T$. We present results for points that satisfy this set of constraints. We compute the Higgs diphoton signal strength as well as the DM  direct detection rate and self-annihilation cross section relevant for indirect detection.

\subsection{Higgs and electroweak precision parameters}

\begin{figure}[htb]
\begin{center}
  \includegraphics[scale=0.62]{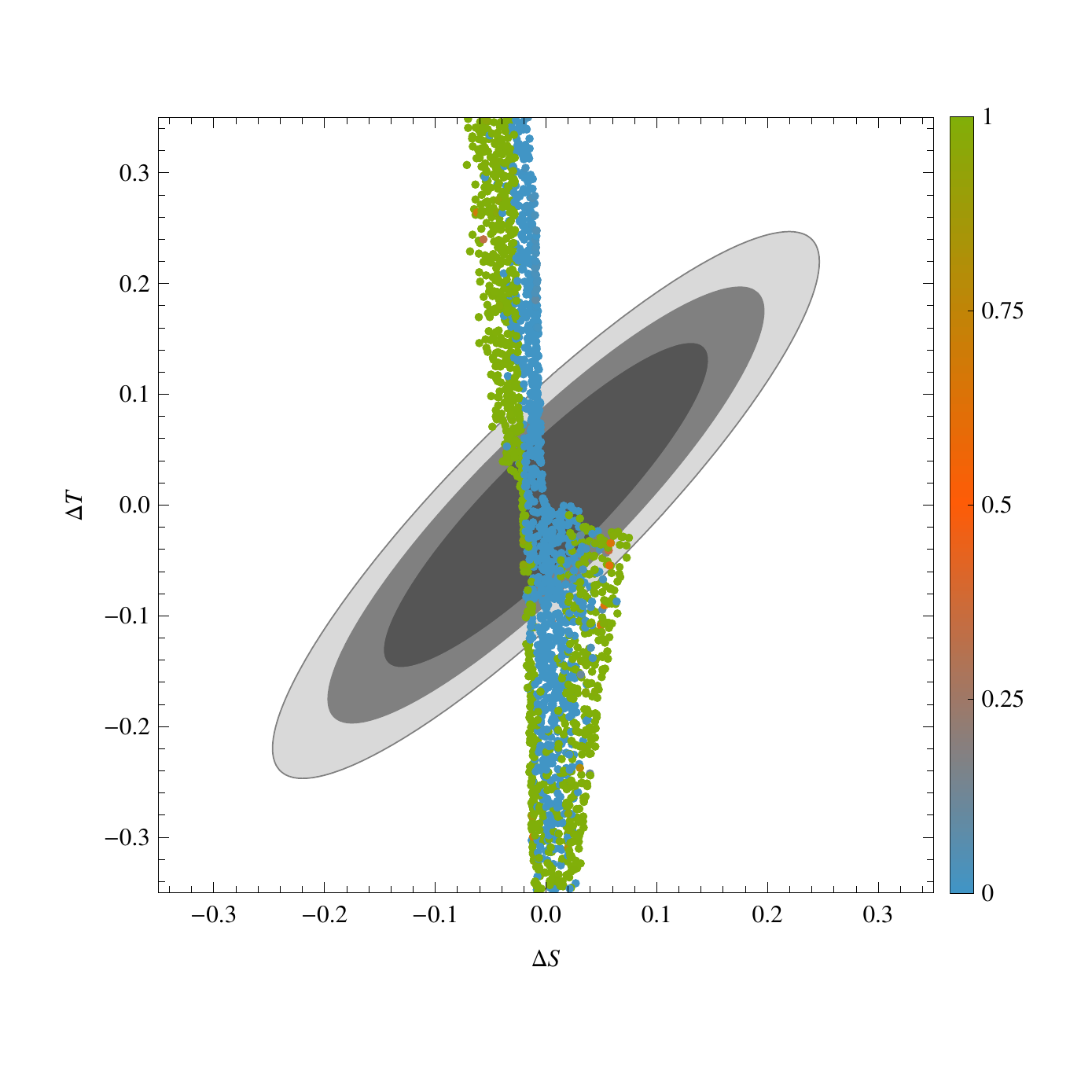}
  \vspace{-0.8cm}
\caption{Electroweak precision parameters $\Delta T$ vs. $\Delta S$ for the $\Z_4$ model. We show the 1, 2, and 3 $\sigma$ contours \cite{Baak:2012kk}. The colour code shows $\Omega_{2}/(\Omega_{1} + \Omega_{2})$. 
}
\label{fig:Z4:T:vs:S}
\end{center}
\end{figure}

\begin{figure}[hb]
\begin{center}
  \includegraphics{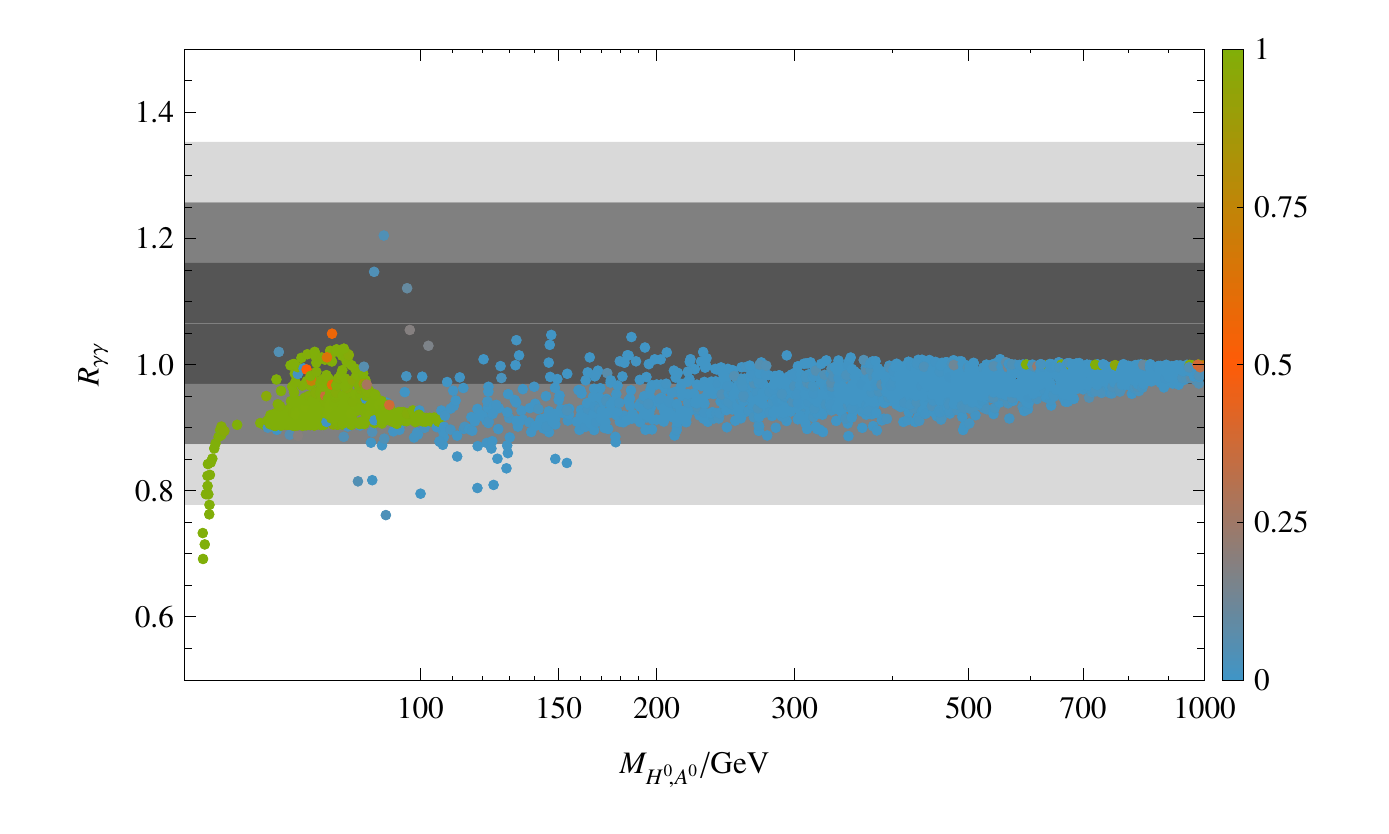}
\caption{The $h \to \gamma \gamma$ rate vs. $M_{H^{0},A^{0}}$ for the $\Z_4$ model normalised to the SM. The thick grey line is the central value from a combined fit of collider data, the coloured bands show 1, 2, and 3 $\sigma$. The colour code shows $\Omega_{2}/(\Omega_{1} + \Omega_{2})$.}
\label{fig:Z4:Rgg:vs:MH0A0}
\end{center}
\end{figure}

Figure~\ref{fig:Z4:T:vs:S} shows the results of the electroweak precision parameters $S$ and $T$. Since there is no mixing between the singlet and doublet components, the $S$ and $T$ parameters depend only on the latter. They tend to restrict the three mass splittings between $H^{0}$, $A^{0}$ and $H^{\pm}$, but cancellations in the parameters can occur for specific combinations of masses. Restricting $S$ and $T$ to the $3 \sigma$ range, however, does not directly exclude any specific region on e.g. the direct detection plots.

Figure~\ref{fig:Z4:Rgg:vs:MH0A0} shows the $h \to \gamma \gamma$ signal strength (normalised to the SM) vs. $M_{H^{0},A^{0}}$. For higher masses the rate is between $0.9 - 1.0$, but higher or lower rates are possible for doublet masses below $150$~GeV. The allowed rates are similar to those of the inert doublet model (see \cite{Swiezewska:2012eh}).

\subsection{Dark matter observables}

We find that most of the time the relic density is dominated by one or the other of the  dark sector.
First note that for the doublet DM -- as in the inert doublet model -- annihilation channels into gauge bosons are efficient and the relic density is typically too small unless the DM mass is below the mass of $W$ or above 500 GeV. 
For the singlet component the relic density constraint can be satisfied over the full mass range \cite{Belanger:2012vp}
when ignoring the interactions between the two dark sectors. 
 The interactions between the two dark sectors influence this picture. 

\begin{figure}[htbp]
\begin{center}
\includegraphics[width=0.6\textwidth]{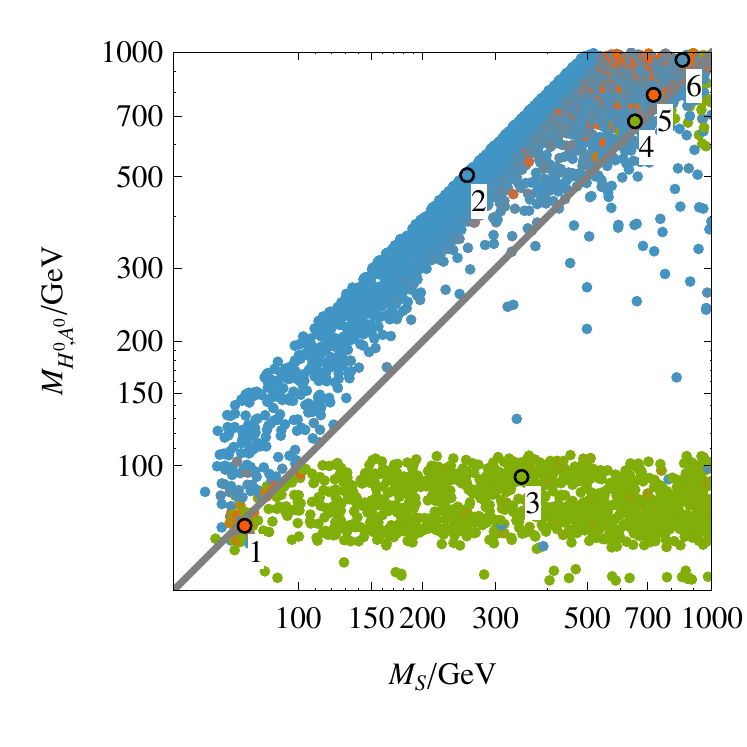}
\vspace{-1cm}
\caption{The allowed points on the $M_{H^{0},A^{0}}$ vs. $M_{S}$ plane. The colour code shows $\Omega_{2}/(\Omega_{1} + \Omega_{2})$.}
\label{fig:mass}
\end{center}
\end{figure}

\begin{figure}[htb]
\begin{center}
   \includegraphics{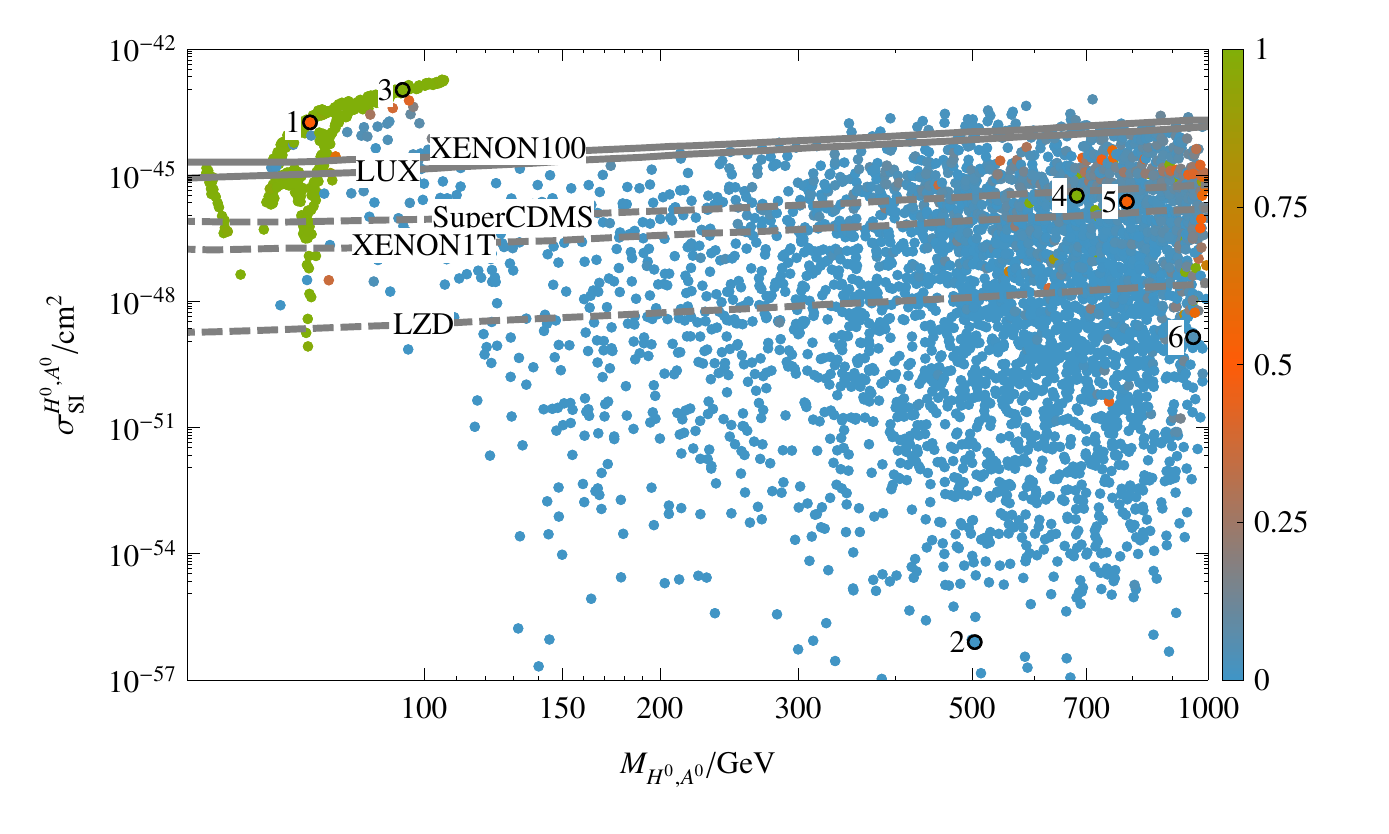}\\
    \includegraphics{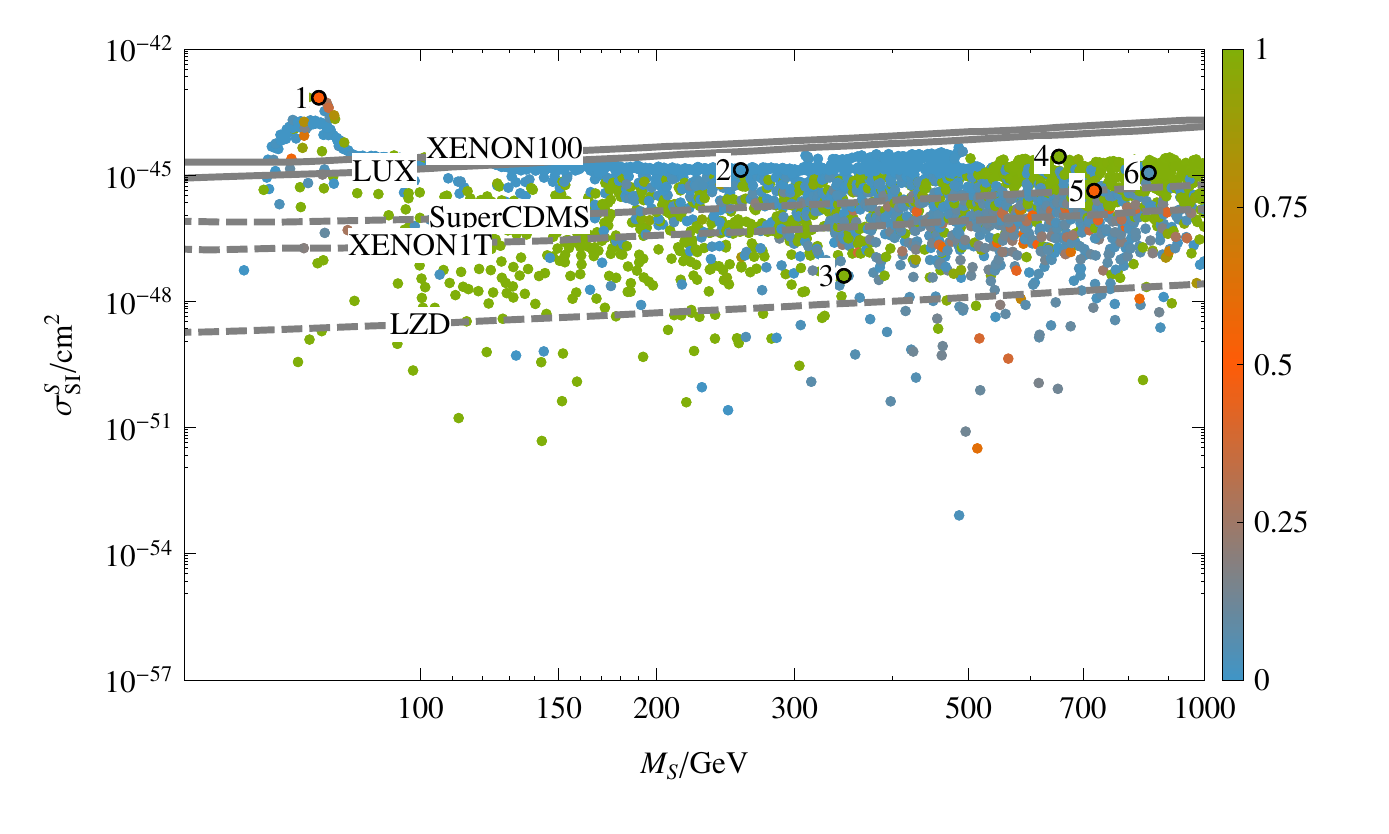}
\caption{Normalised spin independent cross section on xenon for $H^{0},A^{0}$ (top) and $S$ (bottom). The solid grey lines are the XENON100 (2012) \cite{Aprile:2012nq} and LUX \cite{Akerib2013111} exclusion limits at $90\%$~C.L. and the dashed grey lines are the projected $90\%$~C.L. exclusion limits for  SuperCDMS(SNOLAB) \cite{supercdms}, XENON1T \cite{Aprile:2012zx} or  LZD \cite{2011arXiv1110.0103M}. The colour code shows $\Omega_{2}/(\Omega_{1} + \Omega_{2})$.}
\label{fig:Z4:SHA}
\end{center}
\end{figure}

Figure~\ref{fig:mass} shows the allowed region in the $M_{H^{0},A^{0}} - M_{S}$ plane, 
the color code indicate which is the dominant component from blue (singlet dominated) to orange (equal contribution) to green (doublet dominated). In the different regions we have indicated some benchmark points, numbered 1 to 6, which are also shown in plots below and described in Appendix~\ref{sec:benchmarks}.  
Typically the lighter component is dominant.

When $S$ is the heavier DM component, both DM conversion and semi-annihilation, notably $S S \rightarrow H h$, lead to a decrease in $Y_1$ (see \eqref{z4eq1}), the DM is therefore usually dominated by $H^{0},A^{0}$, and falls into two regions one with $M_{H^{0},A^{0}} < 100$~GeV the other with $M_{H^{0},A^{0}} > 500$~GeV. These regions correspond roughly to the medium and high mass ranges of the inert doublet model. However in some cases when the relic abundance of $S$ falls in the measured range, then the subdominant DM component, $H^{0},A^{0}$ can have any mass, see the points below the diagonal in figure~\ref{fig:mass}.

When $H^{0},A^{0}$ are heavier than $S$,  both self-interactions and semi-annihilations tend to  reduce $Y_2$, thus the relic density is typically dominated by the singlet. The lion's share of the points are quite similar to the model with a single complex singlet. As a result the doublet mass can span the whole range up to $M_{H^{0},A^{0}}< 2M_S$. 
Note that for both components to contribute significantly (orange points in figure~\ref{fig:mass}), their mass must be of the same order, therefore they are restricted to the region
 $M_S,M_{H^{0},A^{0}}< 80$~GeV and to the region between $300$~GeV and $1000$~GeV, otherwise the doublet component would be sub-dominant. 

Annihilation of $S$ depends on the $\lambda_{S1}$ coupling and annihilation of $H^{0}, A^{0}$ respectively on $\lambda_{L,R} = \lambda_{3} + \lambda_{4} \pm \lambda_{5}$. Dark matter conversion is mediated by the $\lambda_{S2}$ coupling and the semi-annihilation couplings for $H^{0}, A^{0}$ are $\lambda_{ShH^{0}, ShA^{0}} = \lambda_{S12} \pm \lambda_{S21}$. In general, if the lighter component dominates, the absolute value of its annihilation coupling is less than $0.5$, its (semi-)annihilation couplings and $\lambda_{S2}$ have a large range, while for the heavier component to dominate, the latter must also be within $0.5$.

In figure~\ref{fig:Z4:SHA}  we show the predictions for the SI cross section for each DM component independently. The cross section has been rescaled by a factor $\Omega_i/(\Omega_1+\Omega_2)$ to take into account the relative abundance of each DM component.
On top,  the SI cross section on protons for the doublet component is displayed. As in the inert doublet model most of the points where the DM mass is near the electroweak scale have a large cross section -- exceeding the current bounds -- this is so even when the doublet is a subdominant DM component (blue points). Only a few points around $M_Z/2$ or $M_h/2$  remain below the current bound. 
For heavier DM the current limits are generally satisfied. It is intriguing that we also find scenarios where the doublet leads to a cross section detectable by the next generation of detectors even if it forms a subdominant DM component. This means that direct detection experiments could detect two DM components of different masses in this model.

In figure~\ref{fig:Z4:SHA} (bottom) the SI cross section on xenon for the singlet component is displayed. As in the singlet DM model, the cross section exceeds the current limit for masses below $\approx 100$~GeV except for a few cases where $M_S \approx M_h/2$. For higher masses the predictions are below the current limit but mostly in a range accessible by future detectors such as  SuperCDMS or XENON1T. The lowest cross sections can be related to  strong semi-annihilation in the singlet sector or simply to the fact that the abundance of the singlet component is small (recall that points in green are doublet-dominated). 
Comparing both figures one can see that points where both component have a comparable abundance typically lead to similar cross sections, see e.g. points 1 and 5.

\begin{figure}[htb]
\begin{center}
  \includegraphics{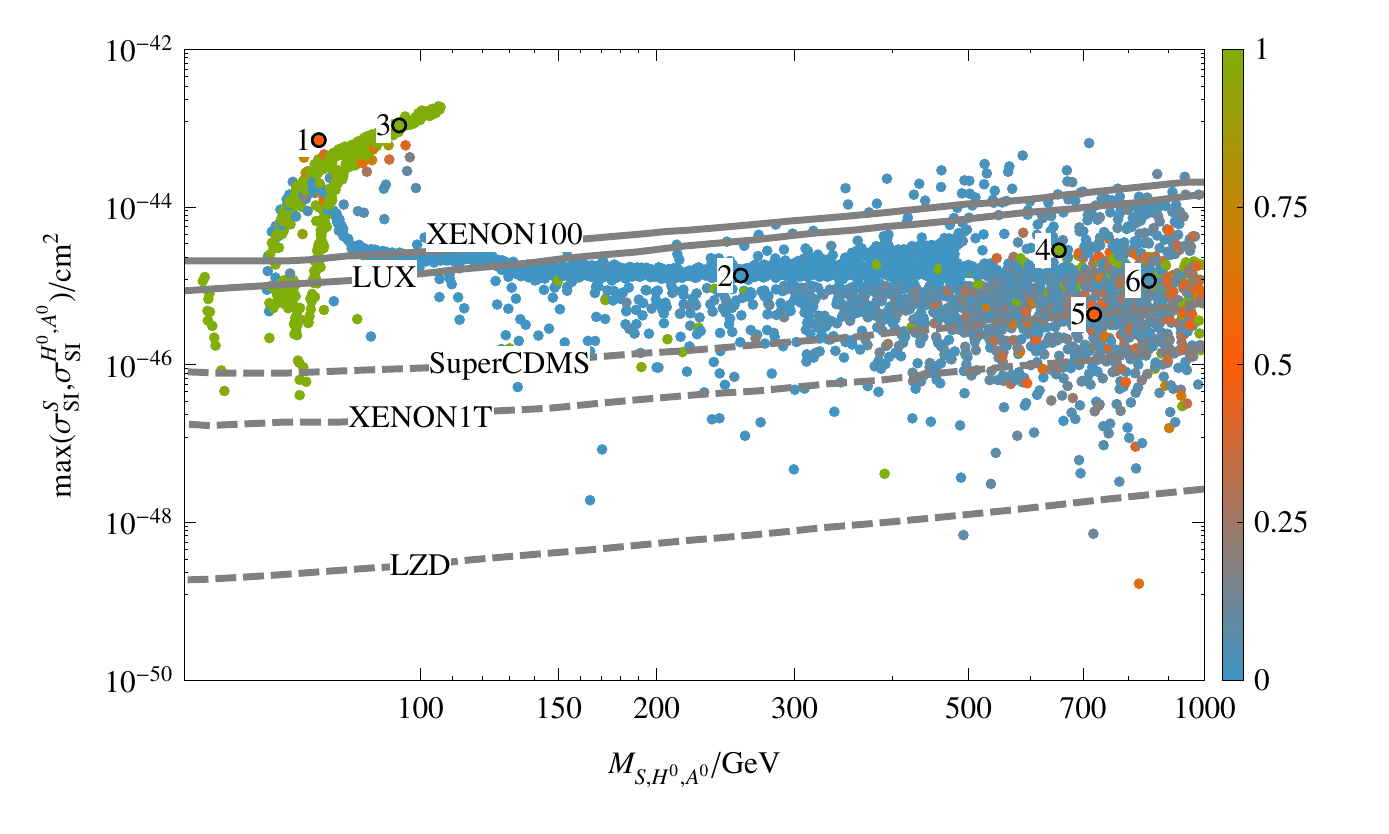}
\caption{Dominant spin independent cross section on xenon for either $S$ or  $H^{0},A^{0}$. The solid grey lines are the XENON100 (2012) \cite{Aprile:2012nq} and LUX \cite{Akerib2013111} exclusion limits at $90\%$~C.L. and the dashed grey lines are the projected $90\%$~C.L. exclusion limits for  SuperCDMS(SNOLAB) \cite{supercdms}, XENON1T \cite{Aprile:2012zx} or  LZD \cite{2011arXiv1110.0103M}. The colour code shows $\Omega_{2}/(\Omega_{1} + \Omega_{2})$.}
\label{fig:Z4:dominant}
\end{center}
\end{figure}

These two figures might give the impression that for many input parameters, DM would escape direct detection. However,
figure~\ref{fig:Z4:dominant}  which displays for each point in parameter space the value of the largest $\sigma_{\rm SI}^S$ or $\sigma_{\rm SI}^{H^{0},A^{0}}$, clearly shows that most of the model parameter space  leads to a signal that could be detected in the near future. Only some points remain out of reach of the projected sensitivity of Xenon1T or even LZD. 

In this model the $\langle \sigma v\rangle$ relevant for indirect detection can be very large for either DM candidate, however after rescaling by the fraction of DM density for each component, we obtain $\langle \sigma v\rangle\approx 3\times 10^{-26}~{\rm cm}^3/{\rm s}$. Therefore the limits from PAMELA on antiprotons are easily satisfied.

\subsection{Renormalisation group running}
\label{sec:Z4:RGE}

\begin{figure}[htb]
\begin{center}
  \includegraphics[width=0.33\textwidth]{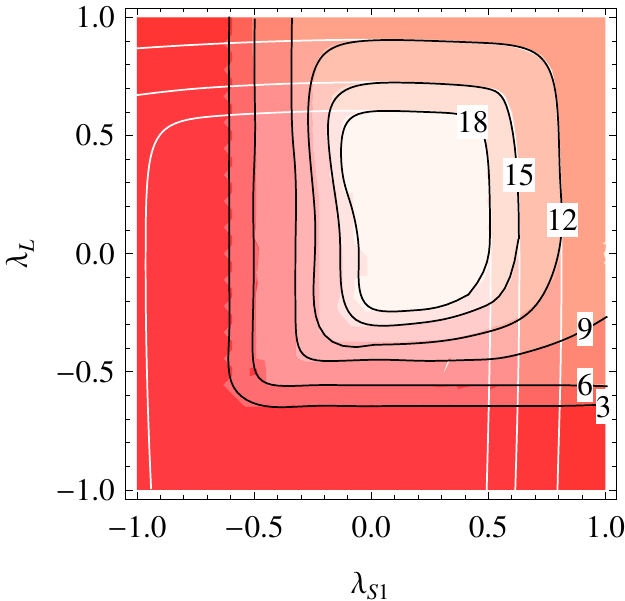}~
  \includegraphics[width=0.33\textwidth]{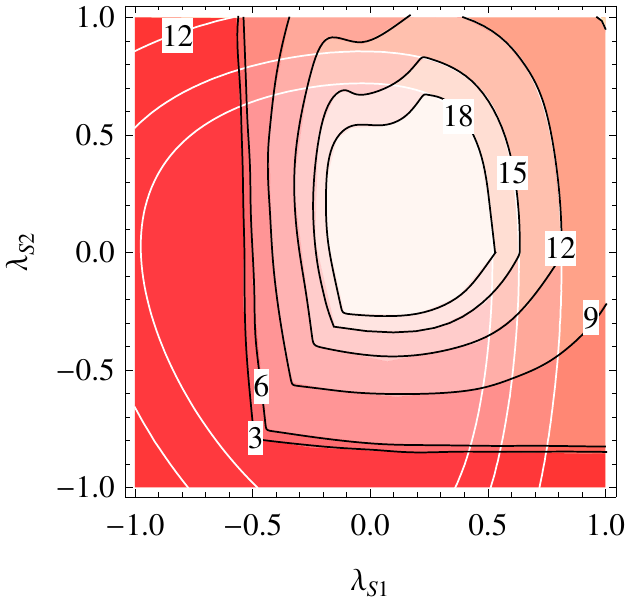}~
  \includegraphics[width=0.33\textwidth]{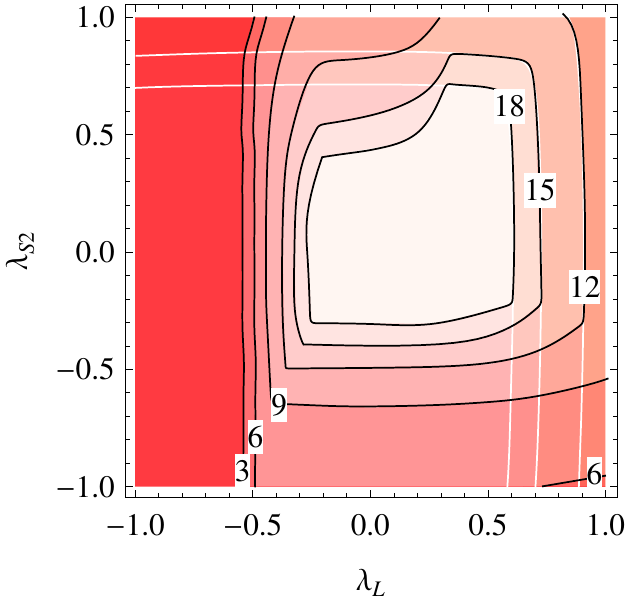} \\
    \includegraphics[width=0.33\textwidth]{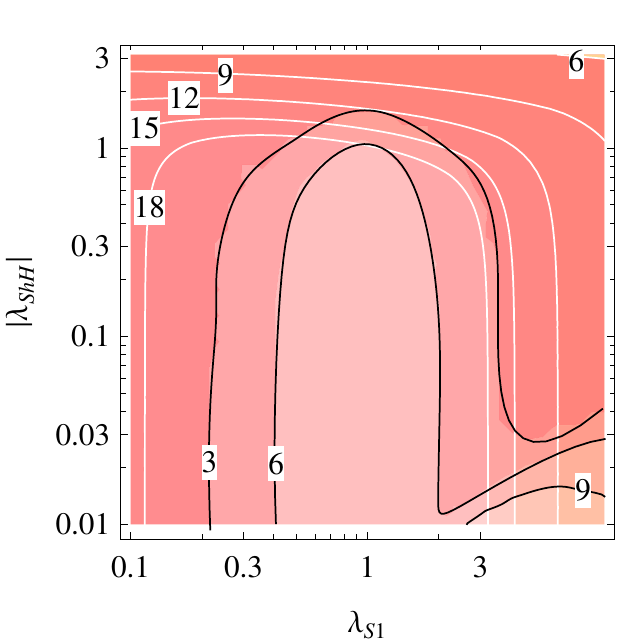}~
  \includegraphics[width=0.33\textwidth]{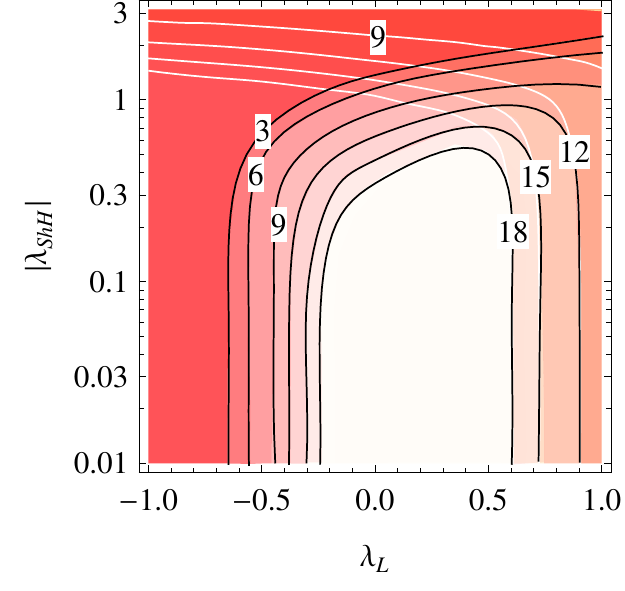}~
  \\
\caption{Vacuum stability and perturbativity bounds for the $\Z_4$ model. The values of the couplings are set at the EW scale. The black contour lines show the logarithm $\log_{10} (\Lambda/1~\text{GeV})$ of the combined bound, while the white lines show the bound from perturbativity only. }
\label{fig:Z4:RGE:planes}
\end{center}
\end{figure}

Without loss of generality, we consider here only $H^{0}$ as the doublet component of DM, because the RGEs are symmetric under the exchange of $\lambda_{L}$ with $\lambda_{R}$ and $\lambda_{ShH^{0}}$ with $\lambda_{ShA^{0}}$. There are four relevant parameters: $\lambda_{S1}$, $\lambda_{L}$, $\lambda_{S2}$ and $\lambda_{ShH^{0}}$. We show the bounds on various parameter planes at the EW scale in figure~\ref{fig:Z4:RGE:planes}.  The values of other couplings are set to zero, except $\lambda_{1}$, and $\lambda_{2}$, $\lambda_{3}$ and $\lambda_{S}$ which are chosen to roughly maximise the scale of vacuum stability of the model. 

Each subplot can be thought of as relevant to a limiting case where only two of the processes of annihilation of $S$ and $H^{0}$, DM conversion and semi-annihilation are relevant.

The top left panel corresponds to the bounds in the $\lambda_{L}$ vs. $\lambda_{S1}$ plane in the special case where interaction between the two sectors is negligible, but independent annihilation of $S$ and $H^{0}$ ensures the correct relic density.

The top middle panel corresponds to the bounds in the $\lambda_{S2}$ vs. $\lambda_{S1}$ plance in the case where the singlet is lighter and $H^{0}$ decays to the SM only via DM conversion to $S$. Similarly, the panel at top right corresponds to the bounds in the $\lambda_{S2}$ vs. $\lambda_{L}$ in the case where the doublet component is lighter and $S$ decays to the SM only via DM conversion to $H^{0}$.

The bottom left panel shows the bounds in the $\lambda_{ShH^{0}}$ vs. $\lambda_{S1}$ plane for the case where $S$ is lighter and $H^{0}$ decays into it via semi-annihilation; the bottom right panel shows the bounds in the $\lambda_{ShH^{0}}$ vs. $\lambda_{S1}$ plane for the similar case where $H^{0}$ is lighter and $S$ decays into it via semi-annihilation.

For these special cases, except if $S$ is lighter and $H^{0}$ decays into it via semi-annihilation (bottom left), one can find points with realistic relic density where the model is valid up to the GUT scale. However, for generic points where all the quartic couplings are of $\mathcal{O}(1)$, for most of the points the model loses perturbativity at about the TeV scale.

\section{Conclusions}
\label{sec:concl}

We have explored the phenomenology of an
inert doublet and complex scalar dark matter model stabilized by $\Z_N$ symmetries, with explicit investigation of the $\Z_3$ and $\Z_4$ cases.
The new feature of these models as compared to the $\Z_2$ case  is the possibility of semi-annihilation and dark matter conversion.
This has important consequences for all dark matter observables.

In the $\Z_3$ model, semi-annihilation processes, e.g. $x_1 x_1\to x_1 h$, can give the dominant contribution to
the relic abundance through the cubic ($\mu''_{S} S^3$) or quartic ($\lambda_{S12} S^2 H_1^\dagger H_2$) terms in the scalar potential.
 This means that  the $\lambda_{S1}$ parameter which sets the coupling of DM to the Higgs and thus  the direct detection cross section is not uniquely determined by the relic density constraint as occurs in the $\Z_2$ model. Large semi-annihilation is therefore associated with suppressed direct detection rate. While the bulk of the points will be testable by ton-scale detectors, it is possible to satisfy the constraints
from vacuum stability and globality of the minimum of the potential with very small values of $\lambda_{S1}$ -- hence to escape all future searches, in particular when the DM is near the TeV range. The direct detection limits from LUX almost completely rule out the region where  dark matter masses are below 120 GeV since for kinematic reasons the  semi-annihilation does not play an important r\^{o}le (the Higgs cannot be produced in the final state).
In this model, because there can be resonance enhancement of the annihilation cross section in the Galaxy when the mass of the DM is tuned to be half the mass of the doublet Higgs,  the indirect detection cross section can be much enhanced as compared to the canonical value.
The semi-annihilation processes will however lead to softer spectra since the DM particle in the final state drains part of the energy of the reaction. Furthermore we have shown that the model can be perturbative up to the GUT scale  even with a large fraction of semi-annihilation.

Enlarging the symmetry to $\Z_4$ entails  two dark sectors, hence two dark matter candidates: a singlet and a doublet.
In this case both  semi-annihilation and  dark matter conversion significantly affects the dark matter phenomenology of the model.
While this model shares many characteristics of the inert doublet model especially when interactions between the two dark sectors can be ignored, the presence of the singlet dark matter candidate means that the doublet DM could only contribute to a fraction of the relic density (and vice versa). This  means in particular that the doublet DM can have any mass instead of being   confined to be at the electroweak scale or heavier than 500 GeV as in the inert doublet model. 

We found that for the sub-dominant  dark matter component, it is possible to have a detectable signal in future direct detection experiments even after taking into account the fraction of each component in the DM density. This occurs in
particular when the sub-dominant component is the doublet since  it typically has a large direct detection rate.
Furthermore in some cases a detectable signal in future ton-scale experiments is predicted for each DM component, opening up the exciting possibility of discovering two DM particles.

\newpage

\appendix

\section{Benchmarks}
\label{sec:benchmarks}

\begin{table}[!htb]
\begin{center}
\caption{\label{tab:z3} Benchmarks for the $\Z_3$ model. All masses and dimensionful parameters are in GeV.}
\vspace{0.5cm}
\begin{tabular}{cccc}
       \hline
 $M_{x_1}$& 225.6 & 398.6 &763.4 \\
  $M_{x_2}$& 615.4 & 655.5 &1518.0 \\ 
   $M_{h}$& 124.8 & 125.5 &124.5 \\ 
    $M_{H^+}$& 731.8& 642.2 &143.0 \\ \hline
     $\theta$& 0.0253& 0.0291 &0.0103 \\ 
      $\mu''_S$& 271.1 & 329.5 &793.2 \\ 
       $\lambda_2$& 1.588 & 2.263 &1.400 \\ 
        $\lambda_3$& 5.446 & 1.615 &-0.1986 \\ 
         $\lambda_S$& 1.917 & 0.7888 &1.280 \\ 
          $\lambda_{S1}$& $-7.188\times 10^{-3}$ & $7.374\times 10^{-2}$ &$-4.874\times 10^{-2}$ \\ 
           $\lambda_{S2}$& $-1.803$ & $7.715$ & $-2.231$ \\ 
            $\lambda_{S12}$& $-2.6285$ & $-0.3063$ & $-0.4525$ \\ \hline
             $\Delta S$   &  $-9.28\times 10^{-3}$   &      $1.01\times 10^{-3}$    &  $3.17\times 10^{-3}$   \\
            $\Delta T$   &  $6.09\times 10^{-2}$   &    $7.72\times 10^{-4}$      &  $3.50\times 10^{-2}$   \\
                     $R_{\gamma\gamma}$& 0.969&0.988 & 1.00\\\hline
                  $\Omega h^2$ &   0.1172  &   0.1182       & 0.1203    \\
                      $\alpha$ &  0.982   &  0.819   & 0.976 \\
                  $\sigma_{\rm SI}^p$ (pb)   & $5.1\times 10^{-11}$    &   $2.8\times 10^{-10}$         &  $4.5\times 10^{-11}$     \\
                 $\sigma_{\rm SI}^n$  (pb) & $3.1\times 10^{-9}$    &   $5.6\times 10^{-9}$         &   $1.3\times 10^{-10}$    \\
                 $\sigma_{SI}^{\rm Xe}$ (pb)   &  $9.8\times 10^{-10}$   &  $1.9\times 10^{-9}$          &  $7.1\times 10^{-11}$     \\           
              $\sigma v (\rm{cm}^3/\rm{s})$   &   $4.3\times 10^{-26}$  &    $3.8\times 10^{-26}$      & $1.0\times 10^{-25} $   \\\hline
    \end{tabular}
\end{center}
\end{table}

\begin{table}[!htb]
\begin{center}
\caption{\label{tab:z3} Benchmarks for the $\Z_4$ model. All masses and dimensionful parameters are in GeV.}
\vspace{0.5cm}
\begin{tabular}{ccccccc}
No. & $1$ & $2$ & $3$ & $4$ & $5$ & $6$ \\
       \hline 
$M_S$ & $74.2$ & $256.1$ & $346.8$ & $652.0$ & $723.1$ & $849.2$ \\
 $M_{H^0}$ & $211.5$ & $608.6$ & $94.0$ & $680.1$ & $803.7$ & $957.5$ \\
 $M_{A^0}$ & $71.6$ & $504.0$ & $389.8$ & $683.8$ & $788.0$ & $983.1$ \\
 $M_{H^\pm}$ & $100.9$ & $562.7$ & $385.7$ & $683.2$ & $791.1$ & $989.3$ \\
$\lambda_2$ & $0.2207$ & $0.8534$ & $0.8984$ & $0.9948$ & $0.1942$ & $0.7090$ \\
$\lambda_3$ & $0.0203$ & $2.077$ & $4.275$ & $6.378\! \! \times \! \! 10^{-3}$ & $-0.01815$ & $2.026$ \\
$\lambda_{S}$ & $0.9007$ & $1.012$ & $2.174$ & $1.782$ & $1.022$ & $1.392$ \\
$\lambda_{S1}$ & $0.2980$ & $-0.1007$ & $-0.1387$ & $8.752$ & $0.2366$ & $0.3149$ \\
$\lambda_{S2}$ & $10.01$ & $-0.8182$ & $1.453$ & $6.211$ & $1.174$ & $7.691$ \\
$\lambda_{S12}$ & $-0.3375$ & $3.692$ & $-3.489$ & $0.6821$ & $0.4821$ & $-0.1650$ \\
$\lambda_{S21}$ & $0.7413$ & $-1.179$ & $-1.136$ & $0.2624$ & $0.04920$ & $-0.3180$ \\
\hline
 $\Delta S$ & $0.0201$ & $-5.57 \! \! \times \! \!  10^{-4}$ & $-0.0193$ & $-0.0159$ & $2.74\! \! \times \! \!10^{-4}$ & $-1.06\! \! \times \! \!10^{-3}$ \\
 $\Delta T$ & $-0.0261$ & $-0.0241$ & $-9.65\! \! \times \! \! 10^{-3}$ & $-1.53 \! \! \times \! \! 10^{-4}$ & $-3.51 \! \! \times \! \!10^{-4}$ & $1.78 \! \! \times \! \!10^{-3}$ \\
 $R_{\gamma\gamma}$ & $0.992$ & $0.980$ & $0.913$ & $1.00$ & $1.00$ & $0.994$ \\
 \hline
$\Omega_1 h^2$ & $0.06175$ & $0.1130$ & $3.664\! \! \times \! \! 10^{-4}$ & $2.071\! \! \times \! \! 10^{-4}$ & $0.0597$ & $0.1187$ \\
$\Omega_2 h^2$ & $0.06165$ & $1.38\! \! \times \! \! 10^{-10}$ & $0.1209$ & 
   $0.1163$ & $0.06737$ & $5.536 \times 10^{-3}$ \\
$\sigma_{\rm SI,1}^{\rm Xe}$ (pb) & $1.4\! \! \times \! \! 10^{-7}$ & $1.3\! \! \times \! \! 10^{-9}$ & $1.4\! \! \times \! \! 10^{-9}$ &
$1.6\! \! \times \! \! 10^{-6}$ & $9.2\! \! \times \! \! 10^{-10}$ & $1.2\! \! \times \! \! 10^{-9}$ \\
$\sigma_{\rm SI,2}^{\rm Xe}$ (pb) & $3.7\! \! \times \! \! 10^{-8}$ & $6.5\! \! \times \! \! 10^{-12}$ & $1.1\! \! \times \! \! 10^{-7}$ &
$3.3\! \! \times \! \! 10^{-10}$ & $4.5\! \! \times \! \! 10^{-10}$ & $3.1\! \! \times \! \! 10^{-12}$ \\
              $\sigma_1 v(\rm{cm}^3/\rm{s}) $ & $1.6\! \! \times \! \! 10^{-25}$ & $2.2\! \! \times \! \! 10^{-26}$ & $8.6\! \! \times \! \! 10^{-24}$ & $1.8\! \! \times \! \! 10^{-23}$ & $4.9\! \! \times \! \! 10^{-26}$ & $3.0\! \! \times \! \! 10^{-26}$ \\
              $\sigma_2 v (\rm{cm}^3/\rm{s})$ & $5.0\! \! \times \! \! 10^{-26}$ & $3.9\! \! \times \! \! 10^{-24}$ & $2.5\! \! \times \! \! 10^{-26}$ &
$4.5\! \! \times \! \! 10^{-26}$ & $7.2\! \! \times \! \! 10^{-26}$ & $7.0\! \! \times \! \! 10^{-25}$ \\\hline
    \end{tabular}
\end{center}
\end{table}

\newpage
\section{One-loop $\beta$-functions}
\label{se:z3:z4:1-loop:beta:functions}

We present the $\beta$-functions for the $\Z_{4}$ potential \eqref{eq:pot:3:Z:4}. The $\beta$-functions for the $\Z_{3}$ potential \eqref{eq:V:Z:3} can be obtained by setting $\lambda'_S = \lambda_5 = \lambda_{S21} = 0$.

The $\beta$-functions for the quartic couplings are
\begin{equation}
\begin{split}
    \beta_{\lambda_{1}} &= 
    24 \lambda_{1}^{2} + 2 \lambda_{3}^{2} + 2 \lambda_{3} \lambda_{4}
    + \lambda_{4}^{2} + \lambda_{5}^{2} + \lambda_{S1}^{2} - 6 y_{t}^4 \\
    &+ \frac{3}{8}(3g^4 + g^{\prime 4} +2g^2 g^{\prime 2}) - 3 (3g^2
    + g^{\prime 2}-4y_{t}^2) \lambda_1,
    \\
    \beta_{\lambda_{2}} &= 
    24 \lambda_{2}^{2} + 2 \lambda_{3}^{2} + 2 \lambda_{3} \lambda_{4}
    + \lambda_{4}^{2} + \lambda_{5}^{2} + \lambda_{S2}^{2} \\
    &+ \frac{3}{8}(3g^4 + g^{\prime 4} +2g^2 g^{\prime 2})
    - 3 (3g^2 +g^{\prime 2}) \lambda_2,
    \\
    \beta_{\lambda_{3}} &= 
    4 (\lambda_{1} + \lambda_{2}) (3\lambda_{3} +  \lambda_{4})
    + 4 \lambda_{3}^{2} + 2 \lambda_{4}^{2} + 2 \lambda_{5}^{2} + 2 \lambda_{S1} \lambda_{S2} \\
    & 
    + \frac{3}{4}(3g^4 + g^{\prime 4} -2g^2 g^{\prime 2})
    - 3 (3g^2 +g^{\prime 2}-2y_{t}^2) \lambda_3,
    \\
    \beta_{\lambda_{4}} &= 
    4(\lambda_{1} + \lambda_{2}) \lambda_{4} + 8 \lambda_{3} \lambda_{4}
    + 4 \lambda_{4}^{2} + 8 \lambda_{5}^{2} \\
    &+ \lambda_{S12}^2 + \lambda_{S21}^2 + 3g^2 g^{\prime 2} - 3 (3g^2 + g^{\prime 2}-2y_{t}^2) \lambda_4,
    \\
    \beta_{\lambda_{5}} &= 
    4 (\lambda_{1} + \lambda_{2} + 2 \lambda_{3}
    + 3 \lambda_{4}) \lambda_{5} \\ 
    &+ 2 \lambda_{S12} \lambda_{S21} - 3 (3g^2 +g^{\prime 2}-2y_{t}^2) \lambda_5,
    \\
    \beta_{\lambda_{S}} &= 
    20 \lambda_{S}^{2} + 36 \lambda_{S}^{\prime 2} + 2 \lambda_{S1}^{2} +
    2 \lambda_{S2}^{2} + \lambda_{S12}^2 + \lambda_{S21}^2,
    \\
    \beta_{\lambda'_{S}} &= 
    24 \lambda_{S} \lambda'_{S} + 2 \lambda_{S12} \lambda_{S21},
    \\
    \beta_{\lambda_{S1}} &= 
    4 (3 \lambda_{1} + 2 \lambda_{S}  + \lambda_{S1}) \lambda_{S1}
    + (4 \lambda_{3} + 2  \lambda_{4}) \lambda_{S2} \\
    &+ 2 (\lambda_{S12}^2 + \lambda_{S21}^2) 
    - \frac{3}{2} (3 g^{2} + g^{\prime 2} - 4 y_{t}^{2}) \lambda_{S1},
    \\
    \beta_{\lambda_{S2}} &= 
    4 (3 \lambda_{2} + 2 \lambda_{S} + \lambda_{S2}) \lambda_{S2}
    +  (4 \lambda_{3} + 2 \lambda_{4}) \lambda_{S1} \\
    & + 2 (\lambda_{S12}^2 + \lambda_{S21}^2) - \frac{3}{2} (3 g^{2} + g^{\prime 2}) \lambda_{S2},
    \\
    \beta_{\lambda_{S12}} &= 
    2 (\lambda_{3} + 2 \lambda_{4} + 2 \lambda_{S} + 2 \lambda_{S1} + 2 \lambda_{S2}) \lambda_{S12} \\
    &+ 6 (\lambda_{5} + 2 \lambda'_{S}) \lambda_{S21} 
    - \frac{3}{2} \left(3 g^2 + g^{\prime 2} - 2 y_t^2 \right) \lambda_{S12},
    \\
    \beta_{\lambda_{S21}} &= 
    2 (\lambda_{3} + 2 \lambda_{4} + 2 \lambda_{S} + 2 \lambda_{S1} + 2 \lambda_{S2}) \lambda_{S21} \\
    &+ 6 (\lambda_{5} + 2 \lambda'_{S}) \lambda_{S12} 
    - \frac{3}{2} \left(3 g^2 + g^{\prime 2} - 2 y_t^2 \right) \lambda_{S21}.
   \\
\end{split}
\label{eq:RGEs:scalar} 
\end{equation}

\section*{Acknowledgements}
K.K. would like to thank for hospitality the LPSC institute in Grenoble where part of this work was completed. This work was supported in part by the French ANR, Project DMAstroLHC,  ANR-12-BS05-0006, by Estonian grants IUT23-6, CERN+, MTT8, MTT60, MJD140, ESF8943, by the European Union through the European Regional Development Fund by the CoE program, and by the project 3.2.0304.11-0313 Estonian Scientific Computing Infrastructure (ETAIS). A.P. was supported by the Russian foundation for Basic Research, grant RFBR-10-02-01443-a. The work of A.P. and G.B. was supported in part by the LIA-TCAP of CNRS.

\bibliography{SIID_full_ZN}

\providecommand{\href}[2]{#2}\begingroup\raggedright\begin{thebibliography}{100}

\bibitem{Ade:2013zuv}
{\bf Planck} Collaboration, P.~Ade {\em et.~al.}, {\it {Planck 2013 results.
  XVI. Cosmological parameters}},
  \href{http://xxx.lanl.gov/abs/1303.5076}{{\tt arXiv:1303.5076}}.

\bibitem{Jungman:1995df}
G.~Jungman, M.~Kamionkowski, and K.~Griest, {\it {Supersymmetric dark matter}},
   {\em Phys.Rept.} {\bf 267} (1996) 195--373,
  [\href{http://xxx.lanl.gov/abs/hep-ph/9506380}{{\tt hep-ph/9506380}}].

\bibitem{Bertone:2004pz}
G.~Bertone, D.~Hooper, and J.~Silk, {\it {Particle dark matter: Evidence,
  candidates and constraints}},  {\em Phys.Rept.} {\bf 405} (2005) 279--390,
  [\href{http://xxx.lanl.gov/abs/hep-ph/0404175}{{\tt hep-ph/0404175}}].

\bibitem{Bergstrom:2000pn}
L.~Bergstrom, {\it {Nonbaryonic dark matter: Observational evidence and
  detection methods}},  {\em Rept.Prog.Phys.} {\bf 63} (2000) 793,
  [\href{http://xxx.lanl.gov/abs/hep-ph/0002126}{{\tt hep-ph/0002126}}].

\bibitem{Nilles:1983ge}
H.~P. Nilles, {\it {Supersymmetry, Supergravity and Particle Physics}},  {\em
  Phys.Rept.} {\bf 110} (1984) 1--162.

\bibitem{Haber:1984rc}
H.~E. Haber and G.~L. Kane, {\it {The Search for Supersymmetry: Probing Physics
  Beyond the Standard Model}},  {\em Phys.Rept.} {\bf 117} (1985) 75--263.

\bibitem{Aprile:2011hi}
{\bf XENON100} Collaboration, E.~Aprile {\em et.~al.}, {\it {Dark Matter
  Results from 100 Live Days of XENON100 Data}},  {\em Phys.Rev.Lett.} {\bf
  107} (2011) 131302, [\href{http://xxx.lanl.gov/abs/1104.2549}{{\tt
  arXiv:1104.2549}}].

\bibitem{Aprile:2012nq}
{\bf XENON100} Collaboration, E.~Aprile {\em et.~al.}, {\it {Dark Matter
  Results from 225 Live Days of XENON100 Data}},  {\em Phys.Rev.Lett.} {\bf
  109} (2012) 181301, [\href{http://xxx.lanl.gov/abs/1207.5988}{{\tt
  arXiv:1207.5988}}].

\bibitem{Akerib2013111}
{\bf LUX} Collaboration, D.~Akerib {\em et.~al.}, {\it {The Large Underground
  Xenon (LUX) experiment}},  {\em Nuclear Instruments and Methods in Physics
  Research Section A: Accelerators, Spectrometers, Detectors and Associated
  Equipment} {\bf 704} (2013), no.~0 111 -- 126.

\bibitem{Cirelli:2010xx}
M.~Cirelli, G.~Corcella, A.~Hektor, G.~Hutsi, M.~Kadastik, {\em et.~al.}, {\it
  {PPPC 4 DM ID: A Poor Particle Physicist Cookbook for Dark Matter Indirect
  Detection}},  {\em JCAP} {\bf 1103} (2011) 051,
  [\href{http://xxx.lanl.gov/abs/1012.4515}{{\tt arXiv:1012.4515}}].

\bibitem{Weinberg:2013aya}
D.~H. Weinberg, J.~S. Bullock, F.~Governato, R.~K. de~Naray, and A.~H.~G.
  Peter, {\it {Cold dark matter: controversies on small scales}},
  \href{http://xxx.lanl.gov/abs/1306.0913}{{\tt arXiv:1306.0913}}.

\bibitem{Aad:2012tfa}
{\bf ATLAS} Collaboration, G.~Aad {\em et.~al.}, {\it {Observation of a new
  particle in the search for the Standard Model Higgs boson with the ATLAS
  detector at the LHC}},  {\em Phys.Lett.} {\bf B716} (2012) 1--29,
  [\href{http://xxx.lanl.gov/abs/1207.7214}{{\tt arXiv:1207.7214}}].

\bibitem{Chatrchyan:2012ufa}
{\bf CMS} Collaboration, S.~Chatrchyan {\em et.~al.}, {\it {Observation of a
  new boson at a mass of 125 GeV with the CMS experiment at the LHC}},  {\em
  Phys.Lett.} {\bf B716} (2012) 30--61,
  [\href{http://xxx.lanl.gov/abs/1207.7235}{{\tt arXiv:1207.7235}}].

\bibitem{McDonald:1993ex}
J.~McDonald, {\it {Gauge Singlet Scalars as Cold Dark Matter}},  {\em Phys.
  Rev.} {\bf D50} (1994) 3637--3649,
  [\href{http://xxx.lanl.gov/abs/hep-ph/0702143}{{\tt hep-ph/0702143}}].

\bibitem{Barger:2007im}
V.~Barger, P.~Langacker, M.~McCaskey, M.~J. Ramsey-Musolf, and G.~Shaughnessy,
  {\it {LHC Phenomenology of an Extended Standard Model with a Real Scalar
  Singlet}},  {\em Phys. Rev.} {\bf D77} (2008) 035005,
  [\href{http://xxx.lanl.gov/abs/0706.4311}{{\tt arXiv:0706.4311}}].

\bibitem{Barger:2008jx}
V.~Barger, P.~Langacker, M.~McCaskey, M.~Ramsey-Musolf, and G.~Shaughnessy,
  {\it {Complex Singlet Extension of the Standard Model}},  {\em Phys. Rev.}
  {\bf D79} (2009) 015018, [\href{http://xxx.lanl.gov/abs/0811.0393}{{\tt
  arXiv:0811.0393}}].

\bibitem{Burgess:2000yq}
C.~P. Burgess, M.~Pospelov, and T.~ter Veldhuis, {\it {The minimal model of
  nonbaryonic dark matter: A singlet scalar}},  {\em Nucl. Phys.} {\bf B619}
  (2001) 709--728, [\href{http://xxx.lanl.gov/abs/hep-ph/0011335}{{\tt
  hep-ph/0011335}}].

\bibitem{Gonderinger:2009jp}
M.~Gonderinger, Y.~Li, H.~Patel, and M.~J. Ramsey-Musolf, {\it {Vacuum
  Stability, Perturbativity, and Scalar Singlet Dark Matter}},  {\em JHEP} {\bf
  1001} (2010) 053, [\href{http://xxx.lanl.gov/abs/0910.3167}{{\tt
  arXiv:0910.3167}}].

\bibitem{Cai:2011kb}
Y.~Cai, X.-G. He, and B.~Ren, {\it {Low Mass Dark Matter and Invisible Higgs
  Width In Darkon Models}},  {\em Phys.Rev.} {\bf D83} (2011) 083524,
  [\href{http://xxx.lanl.gov/abs/1102.1522}{{\tt arXiv:1102.1522}}].

\bibitem{Deshpande:1977rw}
N.~G. Deshpande and E.~Ma, {\it {Pattern of Symmetry Breaking with Two Higgs
  Doublets}},  {\em Phys.Rev.} {\bf D18} (1978) 2574.

\bibitem{Ma:2006km}
E.~Ma, {\it {Verifiable radiative seesaw mechanism of neutrino mass and dark
  matter}},  {\em Phys.Rev.} {\bf D73} (2006) 077301,
  [\href{http://xxx.lanl.gov/abs/hep-ph/0601225}{{\tt hep-ph/0601225}}].

\bibitem{Barbieri:2006dq}
R.~Barbieri, L.~J. Hall, and V.~S. Rychkov, {\it {Improved naturalness with a
  heavy Higgs: An Alternative road to LHC physics}},  {\em Phys.Rev.} {\bf D74}
  (2006) 015007, [\href{http://xxx.lanl.gov/abs/hep-ph/0603188}{{\tt
  hep-ph/0603188}}].

\bibitem{LopezHonorez:2006gr}
L.~Lopez~Honorez, E.~Nezri, J.~F. Oliver, and M.~H. Tytgat, {\it {The Inert
  Doublet Model: An Archetype for Dark Matter}},  {\em JCAP} {\bf 0702} (2007)
  028, [\href{http://xxx.lanl.gov/abs/hep-ph/0612275}{{\tt hep-ph/0612275}}].

\bibitem{Belanger:2012zr}
G.~Belanger, K.~Kannike, A.~Pukhov, and M.~Raidal, {\it {$Z_3$ Scalar Singlet
  Dark Matter}},  {\em JCAP} {\bf 1301} (2013) 022,
  [\href{http://xxx.lanl.gov/abs/1211.1014}{{\tt arXiv:1211.1014}}].

\bibitem{Belanger:2012vp}
G.~Belanger, K.~Kannike, A.~Pukhov, and M.~Raidal, {\it {Impact of
  semi-annihilations on dark matter phenomenology - an example of $Z_N$
  symmetric scalar dark matter}},  {\em JCAP} {\bf 1204} (2012) 010,
  [\href{http://xxx.lanl.gov/abs/1202.2962}{{\tt arXiv:1202.2962}}].

\bibitem{Bezrukov:2012sa}
F.~Bezrukov, M.~Y. Kalmykov, B.~A. Kniehl, and M.~Shaposhnikov, {\it {Higgs
  Boson Mass and New Physics}},  {\em JHEP} {\bf 1210} (2012) 140,
  [\href{http://xxx.lanl.gov/abs/1205.2893}{{\tt arXiv:1205.2893}}].

\bibitem{Degrassi:2012ry}
G.~Degrassi, S.~Di~Vita, J.~Elias-Miro, J.~R. Espinosa, G.~F. Giudice, {\em
  et.~al.}, {\it {Higgs mass and vacuum stability in the Standard Model at
  NNLO}},  {\em JHEP} {\bf 1208} (2012) 098,
  [\href{http://xxx.lanl.gov/abs/1205.6497}{{\tt arXiv:1205.6497}}].

\bibitem{Buttazzo:2013uya}
D.~Buttazzo, G.~Degrassi, P.~P. Giardino, G.~F. Giudice, F.~Sala, {\em
  et.~al.}, {\it {Investigating the near-criticality of the Higgs boson}},
  {\em JHEP} {\bf 1312} (2013) 089,
  [\href{http://xxx.lanl.gov/abs/1307.3536}{{\tt arXiv:1307.3536}}].

\bibitem{Masina:2012tz}
I.~Masina, {\it {Higgs boson and top quark masses as tests of electroweak
  vacuum stability}},  {\em Phys.Rev.} {\bf D87} (2013), no.~5 053001,
  [\href{http://xxx.lanl.gov/abs/1209.0393}{{\tt arXiv:1209.0393}}].

\bibitem{Patt:2006fw}
B.~Patt and F.~Wilczek, {\it {Higgs-field portal into hidden sectors}},
  \href{http://xxx.lanl.gov/abs/hep-ph/0605188}{{\tt hep-ph/0605188}}.

\bibitem{Chu:2011be}
X.~Chu, T.~Hambye, and M.~H. Tytgat, {\it {The Four Basic Ways of Creating Dark
  Matter Through a Portal}},  {\em JCAP} {\bf 1205} (2012) 034,
  [\href{http://xxx.lanl.gov/abs/1112.0493}{{\tt arXiv:1112.0493}}].

\bibitem{Djouadi:2011aa}
A.~Djouadi, O.~Lebedev, Y.~Mambrini, and J.~Quevillon, {\it {Implications of
  LHC searches for Higgs--portal dark matter}},  {\em Phys.Lett.} {\bf B709}
  (2012) 65--69, [\href{http://xxx.lanl.gov/abs/1112.3299}{{\tt
  arXiv:1112.3299}}].

\bibitem{Djouadi:2012zc}
A.~Djouadi, A.~Falkowski, Y.~Mambrini, and J.~Quevillon, {\it {Direct Detection
  of Higgs-Portal Dark Matter at the LHC}},  {\em Eur.Phys.J.} {\bf C73} (2013)
  2455, [\href{http://xxx.lanl.gov/abs/1205.3169}{{\tt arXiv:1205.3169}}].

\bibitem{Kadastik:2011aa}
M.~Kadastik, K.~Kannike, A.~Racioppi, and M.~Raidal, {\it {Implications of the
  125 GeV Higgs boson for scalar dark matter and for the CMSSM phenomenology}},
   {\em JHEP} {\bf 1205} (2012) 061,
  [\href{http://xxx.lanl.gov/abs/1112.3647}{{\tt arXiv:1112.3647}}].

\bibitem{Chen:2012faa}
C.-S. Chen and Y.~Tang, {\it {Vacuum stability, neutrinos, and dark matter}},
  {\em JHEP} {\bf 1204} (2012) 019,
  [\href{http://xxx.lanl.gov/abs/1202.5717}{{\tt arXiv:1202.5717}}].

\bibitem{Cheung:2012nb}
C.~Cheung, M.~Papucci, and K.~M. Zurek, {\it {Higgs and Dark Matter Hints of an
  Oasis in the Desert}},  \href{http://xxx.lanl.gov/abs/1203.5106}{{\tt
  arXiv:1203.5106}}.

\bibitem{Gonderinger:2012rd}
M.~Gonderinger, H.~Lim, and M.~J. Ramsey-Musolf, {\it {Complex Scalar Singlet
  Dark Matter: Vacuum Stability and Phenomenology}},  {\em Phys.Rev.} {\bf D86}
  (2012) 043511, [\href{http://xxx.lanl.gov/abs/1202.1316}{{\tt
  arXiv:1202.1316}}].

\bibitem{Chao:2012mx}
W.~Chao, M.~Gonderinger, and M.~J. Ramsey-Musolf, {\it {Higgs Vacuum Stability,
  Neutrino Mass, and Dark Matter}},  {\em Phys.Rev.} {\bf D86} (2012) 113017,
  [\href{http://xxx.lanl.gov/abs/1210.0491}{{\tt arXiv:1210.0491}}].

\bibitem{Gabrielli:2013hma}
E.~Gabrielli, M.~Heikinheimo, K.~Kannike, A.~Racioppi, M.~Raidal, {\em
  et.~al.}, {\it {Towards Completing the Standard Model: Vacuum Stability, EWSB
  and Dark Matter}},  {\em Phys.Rev.} {\bf D89} (2014) 015017,
  [\href{http://xxx.lanl.gov/abs/1309.6632}{{\tt arXiv:1309.6632}}].

\bibitem{Lebedev:2012zw}
O.~Lebedev, {\it {On Stability of the Electroweak Vacuum and the Higgs
  Portal}},  {\em Eur.Phys.J.} {\bf C72} (2012) 2058,
  [\href{http://xxx.lanl.gov/abs/1203.0156}{{\tt arXiv:1203.0156}}].

\bibitem{EliasMiro:2012ay}
J.~Elias-Miro, J.~R. Espinosa, G.~F. Giudice, H.~M. Lee, and A.~Strumia, {\it
  {Stabilization of the Electroweak Vacuum by a Scalar Threshold Effect}},
  {\em JHEP} {\bf 1206} (2012) 031,
  [\href{http://xxx.lanl.gov/abs/1203.0237}{{\tt arXiv:1203.0237}}].

\bibitem{Hambye:2013dgv}
T.~Hambye and A.~Strumia, {\it {Dynamical generation of the weak and Dark
  Matter scale}},  {\em Phys.Rev.} {\bf D88} (2013) 055022,
  [\href{http://xxx.lanl.gov/abs/1306.2329}{{\tt arXiv:1306.2329}}].

\bibitem{Batell:2010bp}
B.~Batell, {\it {Dark Discrete Gauge Symmetries}},  {\em Phys.Rev.} {\bf D83}
  (2011) 035006, [\href{http://xxx.lanl.gov/abs/1007.0045}{{\tt
  arXiv:1007.0045}}].

\bibitem{DeMontigny:1993gy}
M.~De~Montigny and M.~Masip, {\it {Discrete gauge symmetries in supersymmetric
  grand unified models}},  {\em Phys. Rev.} {\bf D49} (1994) 3734--3740,
  [\href{http://xxx.lanl.gov/abs/hep-ph/9309312}{{\tt hep-ph/9309312}}].

\bibitem{Agashe:2010gt}
K.~Agashe, D.~Kim, M.~Toharia, and D.~G. Walker, {\it {Distinguishing Dark
  Matter Stabilization Symmetries Using Multiple Kinematic Edges and Cusps}},
  {\em Phys.Rev.} {\bf D82} (2010) 015007,
  [\href{http://xxx.lanl.gov/abs/1003.0899}{{\tt arXiv:1003.0899}}].

\bibitem{DEramo:2010ep}
F.~D'Eramo and J.~Thaler, {\it {Semi-annihilation of Dark Matter}},  {\em JHEP}
  {\bf 1006} (2010) 109, [\href{http://xxx.lanl.gov/abs/1003.5912}{{\tt
  arXiv:1003.5912}}].

\bibitem{Martin:1992mq}
S.~P. Martin, {\it {Some simple criteria for gauged R-parity}},  {\em
  Phys.Rev.} {\bf D46} (1992) 2769--2772,
  [\href{http://xxx.lanl.gov/abs/hep-ph/9207218}{{\tt hep-ph/9207218}}].

\bibitem{Agashe:2010tu}
K.~Agashe, D.~Kim, D.~G. Walker, and L.~Zhu, {\it {Using $M_{T2}$ to
  Distinguish Dark Matter Stabilization Symmetries}},  {\em Phys.Rev.} {\bf
  D84} (2011) 055020, [\href{http://xxx.lanl.gov/abs/1012.4460}{{\tt
  arXiv:1012.4460}}].

\bibitem{Ma:2007gq}
E.~Ma, {\it {Z(3) Dark Matter and Two-Loop Neutrino Mass}},  {\em Phys.Lett.}
  {\bf B662} (2008) 49--52, [\href{http://xxx.lanl.gov/abs/0708.3371}{{\tt
  arXiv:0708.3371}}].

\bibitem{Ivanov:2012hc}
I.~Ivanov and V.~Keus, {\it {$Z_p$ scalar dark matter from multi-Higgs-doublet
  models}},  {\em Phys.Rev.} {\bf D86} (2012) 016004,
  [\href{http://xxx.lanl.gov/abs/1203.3426}{{\tt arXiv:1203.3426}}].

\bibitem{Lovrekovic:2012bz}
I.~Lovrekovic, {\it {Dark Matter from Q4 Extension of Standard Model}},
  \href{http://xxx.lanl.gov/abs/1212.1145}{{\tt arXiv:1212.1145}}.

\bibitem{DEramo:2012rr}
F.~D'Eramo, M.~McCullough, and J.~Thaler, {\it {Multiple Gamma Lines from
  Semi-Annihilation}},  {\em JCAP} {\bf 1304} (2013) 030,
  [\href{http://xxx.lanl.gov/abs/1210.7817}{{\tt arXiv:1210.7817}}].

\bibitem{Ko:2014nha}
P.~Ko and Y.~Tang, {\it {Self-interacting scalar dark matter with local $Z_{3}$
  symmetry}},  \href{http://xxx.lanl.gov/abs/1402.6449}{{\tt arXiv:1402.6449}}.

\bibitem{Hambye:2009fg}
T.~Hambye and M.~H. Tytgat, {\it {Confined hidden vector dark matter}},  {\em
  Phys.Lett.} {\bf B683} (2010) 39--41,
  [\href{http://xxx.lanl.gov/abs/0907.1007}{{\tt arXiv:0907.1007}}].

\bibitem{Hambye:2008bq}
T.~Hambye, {\it {Hidden vector dark matter}},  {\em JHEP} {\bf 0901} (2009)
  028, [\href{http://xxx.lanl.gov/abs/0811.0172}{{\tt arXiv:0811.0172}}].

\bibitem{Arina:2009uq}
C.~Arina, T.~Hambye, A.~Ibarra, and C.~Weniger, {\it {Intense Gamma-Ray Lines
  from Hidden Vector Dark Matter Decay}},  {\em JCAP} {\bf 1003} (2010) 024,
  [\href{http://xxx.lanl.gov/abs/0912.4496}{{\tt arXiv:0912.4496}}].

\bibitem{Krauss:1988zc}
L.~M. Krauss and F.~Wilczek, {\it {Discrete Gauge Symmetry in Continuum
  Theories}},  {\em Phys.Rev.Lett.} {\bf 62} (1989) 1221.

\bibitem{Liu:2011aa}
Z.-P. Liu, Y.-L. Wu, and Y.-F. Zhou, {\it {Enhancement of dark matter relic
  density from the late time dark matter conversions}},  {\em Eur.Phys.J.} {\bf
  C71} (2011) 1749, [\href{http://xxx.lanl.gov/abs/1101.4148}{{\tt
  arXiv:1101.4148}}].

\bibitem{Belanger:2011ww}
G.~Belanger and J.-C. Park, {\it {Assisted freeze-out}},
  \href{http://xxx.lanl.gov/abs/1112.4491}{{\tt arXiv:1112.4491}}.

\bibitem{Adulpravitchai:2011ei}
A.~Adulpravitchai, B.~Batell, and J.~Pradler, {\it {Non-Abelian Discrete Dark
  Matter}},  {\em Phys.Lett.} {\bf B700} (2011) 207--216,
  [\href{http://xxx.lanl.gov/abs/1103.3053}{{\tt arXiv:1103.3053}}].

\bibitem{Kadastik:2009cu}
M.~Kadastik, K.~Kannike, and M.~Raidal, {\it {Dark Matter as the signal of
  Grand Unification}},  {\em Phys.Rev.} {\bf D80} (2009) 085020,
  [\href{http://xxx.lanl.gov/abs/0907.1894}{{\tt arXiv:0907.1894}}].

\bibitem{Kadastik:2009dj}
M.~Kadastik, K.~Kannike, and M.~Raidal, {\it {Matter parity as the origin of
  scalar Dark Matter}},  {\em Phys.Rev.} {\bf D81} (2010) 015002,
  [\href{http://xxx.lanl.gov/abs/0903.2475}{{\tt arXiv:0903.2475}}].

\bibitem{Kadastik:2009ca}
M.~Kadastik, K.~Kannike, A.~Racioppi, and M.~Raidal, {\it {EWSB from the soft
  portal into Dark Matter and prediction for direct detection}},  {\em
  Phys.Rev.Lett.} {\bf 104} (2010) 201301,
  [\href{http://xxx.lanl.gov/abs/0912.2729}{{\tt arXiv:0912.2729}}].

\bibitem{Kadastik:2009gx}
M.~Kadastik, K.~Kannike, A.~Racioppi, and M.~Raidal, {\it {Implications of the
  CDMS result on Dark Matter and LHC physics}},  {\em Phys.Lett.} {\bf B694}
  (2010) 242--245, [\href{http://xxx.lanl.gov/abs/0912.3797}{{\tt
  arXiv:0912.3797}}].

\bibitem{Huitu:2010uc}
K.~Huitu, K.~Kannike, A.~Racioppi, and M.~Raidal, {\it {Long-lived charged
  Higgs at LHC as a probe of scalar Dark Matter}},  {\em JHEP} {\bf 1101}
  (2011) 010, [\href{http://xxx.lanl.gov/abs/1005.4409}{{\tt
  arXiv:1005.4409}}].

\bibitem{Cohen:2011ec}
T.~Cohen, J.~Kearney, A.~Pierce, and D.~Tucker-Smith, {\it {Singlet-Doublet
  Dark Matter}},  {\em Phys.Rev.} {\bf D85} (2012) 075003,
  [\href{http://xxx.lanl.gov/abs/1109.2604}{{\tt arXiv:1109.2604}}].

\bibitem{Biswas:2013nn}
A.~Biswas, D.~Majumdar, A.~Sil, and P.~Bhattacharjee, {\it {Two Component Dark
  Matter : A Possible Explanation of 130 GeV $\gamma$-Ray Line from the
  Galactic Centre}},  \href{http://xxx.lanl.gov/abs/1301.3668}{{\tt
  arXiv:1301.3668}}.

\bibitem{Pati:1974yy}
J.~C. Pati and A.~Salam, {\it {Lepton Number as the Fourth Color}},  {\em
  Phys.Rev.} {\bf D10} (1974) 275--289.

\bibitem{Georgi:1974sy}
H.~Georgi and S.~Glashow, {\it {Unity of All Elementary Particle Forces}},
  {\em Phys.Rev.Lett.} {\bf 32} (1974) 438--441.

\bibitem{Fritzsch:1974nn}
H.~Fritzsch and P.~Minkowski, {\it {Unified Interactions of Leptons and
  Hadrons}},  {\em Annals Phys.} {\bf 93} (1975) 193--266.

\bibitem{Barr:1981qv}
S.~M. Barr, {\it {A New Symmetry Breaking Pattern for SO(10) and Proton
  Decay}},  {\em Phys.Lett.} {\bf B112} (1982) 219.

\bibitem{Derendinger:1983aj}
J.~Derendinger, J.~E. Kim, and D.~V. Nanopoulos, {\it {Anti-SU(5)}},  {\em
  Phys.Lett.} {\bf B139} (1984) 170.

\bibitem{Barr:1988yj}
S.~M. Barr, {\it {Some Comments on Flipped SU(5) X U(1) and Flipped Unification
  in General}},  {\em Phys.Rev.} {\bf D40} (1989) 2457.

\bibitem{Dreiner:2005rd}
H.~K. Dreiner, C.~Luhn, and M.~Thormeier, {\it {What is the discrete gauge
  symmetry of the MSSM?}},  {\em Phys.Rev.} {\bf D73} (2006) 075007,
  [\href{http://xxx.lanl.gov/abs/hep-ph/0512163}{{\tt hep-ph/0512163}}].

\bibitem{Lerner:2009xg}
R.~N. Lerner and J.~McDonald, {\it {Gauge singlet scalar as inflaton and
  thermal relic dark matter}},  {\em Phys.Rev.} {\bf D80} (2009) 123507,
  [\href{http://xxx.lanl.gov/abs/0909.0520}{{\tt arXiv:0909.0520}}].

\bibitem{Kanemura:1993hm}
S.~Kanemura, T.~Kubota, and E.~Takasugi, {\it {Lee-Quigg-Thacker bounds for
  Higgs boson masses in a two doublet model}},  {\em Phys.Lett.} {\bf B313}
  (1993) 155--160, [\href{http://xxx.lanl.gov/abs/hep-ph/9303263}{{\tt
  hep-ph/9303263}}].

\bibitem{Akeroyd:2000wc}
A.~G. Akeroyd, A.~Arhrib, and E.-M. Naimi, {\it {Note on tree level unitarity
  in the general two Higgs doublet model}},  {\em Phys.Lett.} {\bf B490} (2000)
  119--124, [\href{http://xxx.lanl.gov/abs/hep-ph/0006035}{{\tt
  hep-ph/0006035}}].

\bibitem{Ginzburg:2003fe}
I.~Ginzburg and I.~Ivanov, {\it {Tree level unitarity constraints in the 2HDM
  with CP violation}},  \href{http://xxx.lanl.gov/abs/hep-ph/0312374}{{\tt
  hep-ph/0312374}}.

\bibitem{Ginzburg:2005dt}
I.~Ginzburg and I.~Ivanov, {\it {Tree-level unitarity constraints in the most
  general 2HDM}},  {\em Phys.Rev.} {\bf D72} (2005) 115010,
  [\href{http://xxx.lanl.gov/abs/hep-ph/0508020}{{\tt hep-ph/0508020}}].

\bibitem{samuelson}
{Samuelson, Paul A.}, {\it {How Deviant Can You Be?}},  {\em {Journal of the
  American Statistical Association}} {\bf 63} (Dec., 1968) 1522--1525.

\bibitem{Kannike:2012pe}
K.~Kannike, {\it {Vacuum Stability Conditions From Copositivity Criteria}},
  {\em Eur.Phys.J.} {\bf C72} (2012) 2093,
  [\href{http://xxx.lanl.gov/abs/1205.3781}{{\tt arXiv:1205.3781}}].

\bibitem{Ginzburg:2004vp}
I.~F. Ginzburg and M.~Krawczyk, {\it {Symmetries of two Higgs doublet model and
  CP violation}},  {\em Phys.Rev.} {\bf D72} (2005) 115013,
  [\href{http://xxx.lanl.gov/abs/hep-ph/0408011}{{\tt hep-ph/0408011}}].

\bibitem{Branco:2011iw}
G.~Branco, P.~Ferreira, L.~Lavoura, M.~Rebelo, M.~Sher, {\em et.~al.}, {\it
  {Theory and phenomenology of two-Higgs-doublet models}},  {\em Phys.Rept.}
  {\bf 516} (2012) 1--102, [\href{http://xxx.lanl.gov/abs/1106.0034}{{\tt
  arXiv:1106.0034}}].

\bibitem{Ginzburg:2010wa}
I.~Ginzburg, K.~Kanishev, M.~Krawczyk, and D.~Sokolowska, {\it {Evolution of
  Universe to the present inert phase}},  {\em Phys.Rev.} {\bf D82} (2010)
  123533, [\href{http://xxx.lanl.gov/abs/1009.4593}{{\tt arXiv:1009.4593}}].

\bibitem{Ferreira:2004yd}
P.~Ferreira, R.~Santos, and A.~Barroso, {\it {Stability of the tree-level
  vacuum in two Higgs doublet models against charge or CP spontaneous
  violation}},  {\em Phys.Lett.} {\bf B603} (2004) 219--229,
  [\href{http://xxx.lanl.gov/abs/hep-ph/0406231}{{\tt hep-ph/0406231}}].

\bibitem{Barroso:2005sm}
A.~Barroso, P.~Ferreira, and R.~Santos, {\it {Charge and CP symmetry breaking
  in two Higgs doublet models}},  {\em Phys.Lett.} {\bf B632} (2006) 684--687,
  [\href{http://xxx.lanl.gov/abs/hep-ph/0507224}{{\tt hep-ph/0507224}}].

\bibitem{Maniatis:2006fs}
M.~Maniatis, A.~von Manteuffel, O.~Nachtmann, and F.~Nagel, {\it {Stability and
  symmetry breaking in the general two-Higgs-doublet model}},  {\em
  Eur.Phys.J.} {\bf C48} (2006) 805--823,
  [\href{http://xxx.lanl.gov/abs/hep-ph/0605184}{{\tt hep-ph/0605184}}].

\bibitem{Ivanov:2006yq}
I.~Ivanov, {\it {Minkowski space structure of the Higgs potential in 2HDM}},
  {\em Phys.Rev.} {\bf D75} (2007) 035001,
  [\href{http://xxx.lanl.gov/abs/hep-ph/0609018}{{\tt hep-ph/0609018}}].

\bibitem{Ivanov:2007de}
I.~P. Ivanov, {\it {Minkowski space structure of the Higgs potential in 2HDM.
  II. Minima, symmetries, and topology}},  {\em Phys.Rev.} {\bf D77} (2008)
  015017, [\href{http://xxx.lanl.gov/abs/0710.3490}{{\tt arXiv:0710.3490}}].

\bibitem{DBLP:journals/toms/Verschelde99}
J.~Verschelde, {\it {Algorithm 795: PHCpack: a general-purpose solver for
  polynomial systems by homotopy continuation}},  {\em ACM Trans. Math. Softw.}
  {\bf 25} (1999), no.~2 251--276.

\bibitem{Baak:2012kk}
M.~Baak, M.~Goebel, J.~Haller, A.~Hoecker, D.~Kennedy, {\em et.~al.}, {\it {The
  Electroweak Fit of the Standard Model after the Discovery of a New Boson at
  the LHC}},  {\em Eur.Phys.J.} {\bf C72} (2012) 2205,
  [\href{http://xxx.lanl.gov/abs/1209.2716}{{\tt arXiv:1209.2716}}].

\bibitem{Grimus:2008nb}
W.~Grimus, L.~Lavoura, O.~Ogreid, and P.~Osland, {\it {The Oblique parameters
  in multi-Higgs-doublet models}},  {\em Nucl.Phys.} {\bf B801} (2008) 81--96,
  [\href{http://xxx.lanl.gov/abs/0802.4353}{{\tt arXiv:0802.4353}}].

\bibitem{Grimus:2007if}
W.~Grimus, L.~Lavoura, O.~Ogreid, and P.~Osland, {\it {A Precision constraint
  on multi-Higgs-doublet models}},  {\em J.Phys.G} {\bf G35} (2008) 075001,
  [\href{http://xxx.lanl.gov/abs/0711.4022}{{\tt arXiv:0711.4022}}].

\bibitem{Gustafsson:2007pc}
M.~Gustafsson, E.~Lundstrom, L.~Bergstrom, and J.~Edsjo, {\it {Significant
  Gamma Lines from Inert Higgs Dark Matter}},  {\em Phys.Rev.Lett.} {\bf 99}
  (2007) 041301, [\href{http://xxx.lanl.gov/abs/astro-ph/0703512}{{\tt
  astro-ph/0703512}}].

\bibitem{Cao:2007rm}
Q.-H. Cao, E.~Ma, and G.~Rajasekaran, {\it {Observing the Dark Scalar Doublet
  and its Impact on the Standard-Model Higgs Boson at Colliders}},  {\em
  Phys.Rev.} {\bf D76} (2007) 095011,
  [\href{http://xxx.lanl.gov/abs/0708.2939}{{\tt arXiv:0708.2939}}].

\bibitem{Pierce:2007ut}
A.~Pierce and J.~Thaler, {\it {Natural Dark Matter from an Unnatural Higgs
  Boson and New Colored Particles at the TeV Scale}},  {\em JHEP} {\bf 0708}
  (2007) 026, [\href{http://xxx.lanl.gov/abs/hep-ph/0703056}{{\tt
  hep-ph/0703056}}].

\bibitem{Lundstrom:2008ai}
E.~Lundstrom, M.~Gustafsson, and J.~Edsjo, {\it {The Inert Doublet Model and
  LEP II Limits}},  {\em Phys.Rev.} {\bf D79} (2009) 035013,
  [\href{http://xxx.lanl.gov/abs/0810.3924}{{\tt arXiv:0810.3924}}].

\bibitem{Arhrib:2012ia}
A.~Arhrib, R.~Benbrik, and N.~Gaur, {\it {$H\to \gamma \gamma$ in Inert Higgs
  Doublet Model}},  {\em Phys.Rev.} {\bf D85} (2012) 095021,
  [\href{http://xxx.lanl.gov/abs/1201.2644}{{\tt arXiv:1201.2644}}].

\bibitem{Posch:2010hx}
P.~Posch, {\it {Enhancement of $h \to \gamma \gamma$ in the Two Higgs Doublet
  Model Type I}},  {\em Phys.Lett.} {\bf B696} (2011) 447--453,
  [\href{http://xxx.lanl.gov/abs/1001.1759}{{\tt arXiv:1001.1759}}].

\bibitem{Borah:2012pu}
D.~Borah and J.~M. Cline, {\it {Inert Doublet Dark Matter with Strong
  Electroweak Phase Transition}},  {\em Phys.Rev.} {\bf D86} (2012) 055001,
  [\href{http://xxx.lanl.gov/abs/1204.4722}{{\tt arXiv:1204.4722}}].

\bibitem{Swiezewska:2012eh}
B.~Swiezewska and M.~Krawczyk, {\it {Diphoton rate in the Inert Doublet Model
  with a 125 GeV Higgs boson}},  \href{http://xxx.lanl.gov/abs/1212.4100}{{\tt
  arXiv:1212.4100}}.

\bibitem{Aaltonen:2013kxa}
{\bf CDF, D0} Collaboration, T.~Aaltonen {\em et.~al.}, {\it {Higgs Boson
  Studies at the Tevatron}},  {\em Phys.Rev.} {\bf D88} (2013) 052014,
  [\href{http://xxx.lanl.gov/abs/1303.6346}{{\tt arXiv:1303.6346}}].

\bibitem{Aad:2013wqa}
{\bf ATLAS} Collaboration, G.~Aad {\em et.~al.}, {\it {Measurements of Higgs
  boson production and couplings in diboson final states with the ATLAS
  detector at the LHC}},  {\em Phys.Lett.} {\bf B726} (2013) 88--119,
  [\href{http://xxx.lanl.gov/abs/1307.1427}{{\tt arXiv:1307.1427}}].

\bibitem{ATLAS-CONF-2013-108}
{\it {Evidence for Higgs Boson Decays to the $\tau^+\tau^-$ Final State with
  the ATLAS Detector}},  Tech. Rep. ATLAS-CONF-2013-108, CERN, Geneva, Nov,
  2013.

\bibitem{ATLAS-CONF-2013-079}
{\it {Search for the $bb$ decay of the Standard Model Higgs boson in associated
  $W/ZH$ production with the ATLAS detector}},  Tech. Rep. ATLAS-CONF-2013-079,
  CERN, Geneva, Jul, 2013.

\bibitem{ATLAS-CONF-2013-010}
{\bf {ATLAS}} Collaboration, {\it {Search for a Standard Model Higgs boson in
  $H\rightarrow \mu\mu$ decays with the ATLAS detector}},  Tech. Rep.
  ATLAS-CONF-2013-010, CERN, Geneva, Mar, 2013.

\bibitem{ATLAS-CONF-2013-009}
{\bf {ATLAS}} Collaboration, {\it {Search for the Standard Model Higgs boson in
  the $H \rightarrow Z\gamma$ decay mode with pp collisions at $\sqrt{s} =$ 7
  and 8 TeV}},  Tech. Rep. ATLAS-CONF-2013-009, CERN, Geneva, Mar, 2013.
\newblock ATLAS-CONF-2013-009.

\bibitem{ATLAS-CONF-2013-014}
{\bf {ATLAS}} Collaboration, {\it {Combined measurements of the mass and signal
  strength of the Higgs-like boson with the ATLAS detector using up to 25
  fb$^{-1}$ of proton-proton collision data}},  Tech. Rep. ATLAS-CONF-2013-014,
  CERN, Geneva, Mar, 2013.
\newblock ATLAS-CONF-2013-014.

\bibitem{cms:2013:bosonic}
{\bf CMS} Collaboration, G.~Gomez-Ceballos. Talk at the Moriond 2013 EW
  session.

\bibitem{cms:2013:comb:1}
{\bf CMS} Collaboration, M.~Shen. Talk at the Moriond 2013 EW session.

\bibitem{cms:2013:comb:2}
{\bf CMS} Collaboration, B.~Mansoulie. Talk at the Moriond 2013 EW session.

\bibitem{cms:2013:fermions}
{\bf CMS} Collaboration, V.~Dutta. Talk at the Moriond 2013 EW session.

\bibitem{CMS-PAS-HIG-12-050}
{\bf {CMS}} Collaboration, {\it {Higgs to tau tau (MSSM) (HCP)}}, .
  CMS-PAS-HIG-12-050.

\bibitem{CMS-PAS-HIG-13-001}
{\bf {CMS}} Collaboration, {\it {Updated measurements of the Higgs boson at 125
  GeV in the two photon decay channel}},  Tech. Rep. CMS-PAS-HIG-13-001, CERN,
  Geneva, 2013.
\newblock CMS-PAS-HIG-13-001.

\bibitem{Chatrchyan:2013mxa}
{\bf CMS Collaboration} Collaboration, S.~Chatrchyan {\em et.~al.}, {\it
  {Measurement of the properties of a Higgs boson in the four-lepton final
  state}},  \href{http://xxx.lanl.gov/abs/1312.5353}{{\tt arXiv:1312.5353}}.

\bibitem{Chatrchyan:2013iaa}
{\bf CMS Collaboration} Collaboration, S.~Chatrchyan {\em et.~al.}, {\it
  {Measurement of Higgs boson production and properties in the WW decay channel
  with leptonic final states}},  \href{http://xxx.lanl.gov/abs/1312.1129}{{\tt
  arXiv:1312.1129}}.

\bibitem{CMS-PAS-HIG-13-004}
{\bf {CMS}} Collaboration, {\it {Search for the Standard-Model Higgs boson
  decaying to tau pairs in proton-proton collisions at sqrt(s) = 7 and 8 TeV}},
   Tech. Rep. CMS-PAS-HIG-13-004, CERN, Geneva, 2013.
\newblock CMS-PAS-HIG-13-004.

\bibitem{cms:2013:tautau:update}
{\bf CMS} Collaboration, M.~V. Acosta. Talk at CERN.

\bibitem{Chatrchyan:2013zna}
{\bf CMS} Collaboration, S.~Chatrchyan {\em et.~al.}, {\it {Search for the
  standard model Higgs boson produced in association with a W or a Z boson and
  decaying to bottom quarks}},  \href{http://xxx.lanl.gov/abs/1310.3687}{{\tt
  arXiv:1310.3687}}.

\bibitem{CMS-PAS-HIG-13-006}
{\bf {CMS}} Collaboration, {\it {Search for the standard model Higgs boson in
  the $Z$ boson plus a photon channel in pp collisions at sqrt(s) = 7 and
  8\,TeV}},  Tech. Rep. CMS-PAS-HIG-13-006, CERN, Geneva, 2013.
\newblock CMS-PAS-HIG-13-006.

\bibitem{CMS-PAS-HIG-13-009}
{\bf {CMS}} Collaboration, {\it {Search for SM Higgs in $WH$ to $WWW$ to $3l
  3\nu$}},  Tech. Rep. CMS-PAS-HIG-13-009, CERN, Geneva, 2013.
\newblock CMS-PAS-HIG-13-009.

\bibitem{Giardino:2013bma}
P.~P. Giardino, K.~Kannike, I.~Masina, M.~Raidal, and A.~Strumia, {\it {The
  universal Higgs fit}},  \href{http://xxx.lanl.gov/abs/1303.3570}{{\tt
  arXiv:1303.3570}}.

\bibitem{Belanger:2013xza}
G.~Belanger, B.~Dumont, U.~Ellwanger, J.~Gunion, and S.~Kraml, {\it {Global fit
  to Higgs signal strengths and couplings and implications for extended Higgs
  sectors}},  {\em Phys.Rev.} {\bf D88} (2013) 075008,
  [\href{http://xxx.lanl.gov/abs/1306.2941}{{\tt arXiv:1306.2941}}].

\bibitem{Hinshaw:2012aka}
{\bf WMAP} Collaboration, G.~Hinshaw {\em et.~al.}, {\it {Nine-Year Wilkinson
  Microwave Anisotropy Probe (WMAP) Observations: Cosmological Parameter
  Results}},  \href{http://xxx.lanl.gov/abs/1212.5226}{{\tt arXiv:1212.5226}}.

\bibitem{Belanger:2013oya}
G.~Belanger, F.~Boudjema, A.~Pukhov, and A.~Semenov, {\it {micrOMEGAs3.1 : a
  program for calculating dark matter observables}},
  \href{http://xxx.lanl.gov/abs/1305.0237}{{\tt arXiv:1305.0237}}.

\bibitem{micromegas:in:prep}
G.~Belanger, F.~Boudjema, and A.~Pukhov. In Preparation.

\bibitem{supercdms}
D.~Bauer. G2 Directorate Briefing.

\bibitem{Aprile:2012zx}
{\bf XENON1T} Collaboration, E.~Aprile, {\it {The XENON1T Dark Matter Search
  Experiment}},  \href{http://xxx.lanl.gov/abs/1206.6288}{{\tt
  arXiv:1206.6288}}.

\bibitem{2011arXiv1110.0103M}
D.~{Malling} {\em et.~al.}, {\it {After LUX: The LZ Program}},  {\em ArXiv
  e-prints} (Oct., 2011) [\href{http://xxx.lanl.gov/abs/1110.0103}{{\tt
  arXiv:1110.0103}}].

\bibitem{Ackermann:2013yva}
{\bf Fermi-LAT} Collaboration, M.~Ackermann {\em et.~al.}, {\it {Dark Matter
  Constraints from Observations of 25 Milky Way Satellite Galaxies with the
  Fermi Large Area Telescope}},  \href{http://xxx.lanl.gov/abs/1310.0828}{{\tt
  arXiv:1310.0828}}.

\bibitem{Ackermann:2011wa}
{\bf Fermi-LAT} Collaboration, M.~Ackermann {\em et.~al.}, {\it {Constraining
  Dark Matter Models from a Combined Analysis of Milky Way Satellites with the
  Fermi Large Area Telescope}},  {\em Phys.Rev.Lett.} {\bf 107} (2011) 241302,
  [\href{http://xxx.lanl.gov/abs/1108.3546}{{\tt arXiv:1108.3546}}].

\bibitem{Adriani:2010rc}
{\bf PAMELA} Collaboration, O.~Adriani {\em et.~al.}, {\it {PAMELA results on
  the cosmic-ray antiproton flux from 60 MeV to 180 GeV in kinetic energy}},
  {\em Phys.Rev.Lett.} {\bf 105} (2010) 121101,
  [\href{http://xxx.lanl.gov/abs/1007.0821}{{\tt arXiv:1007.0821}}].

\bibitem{Donato:2003xg}
F.~Donato, N.~Fornengo, D.~Maurin, and P.~Salati, {\it {Antiprotons in cosmic
  rays from neutralino annihilation}},  {\em Phys.Rev.} {\bf D69} (2004)
  063501, [\href{http://xxx.lanl.gov/abs/astro-ph/0306207}{{\tt
  astro-ph/0306207}}].

\bibitem{Cirelli:2013hv}
M.~Cirelli and G.~Giesen, {\it {Antiprotons from Dark Matter: Current
  constraints and future sensitivities}},  {\em JCAP} {\bf 1304} (2013) 015,
  [\href{http://xxx.lanl.gov/abs/1301.7079}{{\tt arXiv:1301.7079}}].

\end{thebibliography}\endgroup

\end{document}